# Synthesis of fast multiplication algorithms for arbitrary tensors

P. Dourbal [1]

[1] Dourbal Electric, Inc. 10 Schalk's Crossing Rd., St. 501-295, Princeton Jct., NJ 08536 paul@dourbalelectric.com

**Abstract**

A method of fast linear transform algorithm synthesis for an arbitrary tensor, matrix, or vector is proposed. The method is based on factorization of a tensor and using the factors for building computational structures performing fast tensor - vector multiplication on a computer or dedicated hardware platform.

**Introduction**

For many classes of linear transformations, methods of fast calculation already exist and are in use. These include transformations such as the Fourier transform, the transforms based on Vilenkin functions, Haar functions, complex rectangular functions, Walsh functions, sawtooth functions, and composite systems of basis functions[1]. Despite the variety of transformations that have fast algorithms, in many cases none of them is applicable to the situation at hand. Then it becomes necessary to search for a fast transformation method that emerges from the given situation. This problem has been described and solved for a wide variety of orthogonal transformations [2,3]. Until this time the method for performing fast linear transformations relied on finding some sparsely populated matrices (with the majority of elements equal to 0), whose product was the original transformation matrix. This approach uses symmetry and other properties of the system of basis functions of a linear transformation.

The present work examines the problem of performing fast linear transformations arising from arbitrary systems of functions – orthogonal or not orthogonal; symmetric or not symmetric. As in previous approaches [1-4], the factorization (or decomposition into factors) of an initial tensor or matrix is used, but there is no limitation on the rank of the factors. For example, two of the factors of any given tensor can be a tensor of higher rank and a vector, which, as it will be shown later in this work, have symmetrical and other properties essential for a fast transformation.

Additionally, the method proposed in the current work is applicable to transformations with symbolically as well as numerically expressed elements or coefficients.

**1. Factorization of a vector, matrix, or tensor.**

A vector, matrix, or tensor of arbitrary dimension can be expressed as the product of a matrix or tensor of dimension one higher than the original vector, matrix or tensor, henceforth termed the "commutator" and a vector containing all the unique nonzero elements of the original vector, matrix, or tensor, henceforth termed the "kernel". Here the commutator has the same number of elements as the original vector, matrix, or tensor in every dimension except the additional one; the number of elements in the additional dimension is equal to the length of the obtained vector. The number of unit elements in the commutator is equal to the number of nonzero elements in the original vector, matrix or tensor, and all other elements of the commutator are equal to zero. Moreover, the unit elements in the commutator are distributed such that the number of nonzero elements in the additional dimension does not exceed one. Thus, any vector, matrix, or tensor of any dimension may be expressed as the product of a commutator and a kernel.

For vectors, matrices or tensors whose elements are either real or purely imaginary numbers of arbitrary precision, the dimension of the kernel cannot exceed the greatest integer value less than the ratio of the difference between maximal and minimal elements of the original vector, matrix or tensor to the precision of the elements of the original vector, matrix or tensor.

For vectors, matrices or tensors whose elements are complex numbers of arbitrary precision, the dimension of the kernel cannot exceed the value of the product of the dimensions of the kernels obtained separately for the real and imaginary parts of the original vector, matrix or tensor.

For a matrix or tensor $[T]_{(N_1,N_2,...N_M)}$ of dimensions $(N_1, N_2, ... N_M)$ and finite dimension $N_M < \infty$ and finite dimension $K < \infty$ of the kernel, the rows (vectors) of the matrix or sub-tensors of the tensor $[T]_{(N_1,N_2,...N_M)}$ of dimensions $(N_1, N_2, ... N_{M-1})$ can all be unique only under the following condition:

$$\prod_{p=1}^{M-1} N_p \leq K^{N_M} \qquad (1)$$

## 2. Definition of method and process for factorization of a vector.

Suppose there exists a vector

$$[T]_N = \begin{bmatrix} t_1 \\ ... \\ t_n \\ ... \\ t_N \end{bmatrix} \qquad (2)$$

of length $N$ containing $L \leq N$ distinct nonzero elements. From this vector the kernel consisting of the vector

$$[U]_L = \begin{bmatrix} u_1 \\ ... \\ u_l \\ ... \\ u_L \end{bmatrix} \qquad (3)$$

is obtained by including the unique nonzero elements of $[T]_N$ in the vector $[U]_L$, in arbitrary order.

From the same vector $[T]_N$ the intermediate vector

$$[Y]_N = \begin{bmatrix} y_1 \\ \vdots \\ y_n \\ \vdots \\ y_N \end{bmatrix} \qquad (4)$$

is formed, with the same dimension $N$ as the vector $[T]_N$ and with each element equal either to zero or to the index of the element of the vector $[U]_L$ which is equal in value to this element of vector $[T]_N$. The vector $[Y]_N$ can be obtained by replacing every nonzero element $t_n$ of the vector $[T]_N$ by the index $l$ of the element $u_l$ of the vector $[U]_L$ that has the same value.

From the intermediate vector $[Y]_N$ the commutator

$$[Z]_{N,L} = \begin{bmatrix} z_{1,1} & \cdots & z_{1,L} \\ \vdots & z_{n,l} & \vdots \\ z_{N,1} & \cdots & z_{N,L} \end{bmatrix} \qquad (5)$$

is obtained by replacing every nonzero element $y_n$ of the vector $[Y]_N$ with a row vector of length L, with a single unit element in the position with index equal to the value of $y_n$ and L-1 zero elements in all other positions. The resulting commutator is represented as:

$$[Z]_{N,L} = \begin{bmatrix} \begin{cases} [0 \ldots 0]_L, \text{ for } y_1 = 0 \\ [0 \ldots 0]_{y_1-1} \; 1 \; [0 \ldots 0]_{L-y_1}, \text{ for } y_1 > 0 \end{cases} \\ \cdots \\ \begin{cases} [0 \ldots 0]_L, \text{ for } y_n = 0 \\ [0 \ldots 0]_{y_n-1} \; 1 \; [0 \ldots 0]_{L-y_n}, \text{ for } y_n > 0 \end{cases} \\ \cdots \\ \begin{cases} [0 \ldots 0]_L, \text{ for } y_N = 0 \\ [0 \ldots 0]_{y_N-1} \; 1 \; [0 \ldots 0]_{L-y_N}, \text{ for } y_N > 0 \end{cases} \end{bmatrix} \qquad (6)$$

The vector $[T]_N$ is factored as the product of the multiplication of the commutator $[Z]_{N,L}$ by the kernel $[U]_L$:

$$[T]_N = [Z]_{N,L} \cdot [U]_L = \begin{bmatrix} z_{1,1} & \cdots & z_{1,L} \\ \vdots & z_{n,l} & \vdots \\ z_{N,1} & \cdots & z_{N,L} \end{bmatrix} \cdot \begin{bmatrix} u_1 \\ \vdots \\ u_l \\ \vdots \\ u_L \end{bmatrix} \qquad (7)$$

### 3. Example of vector factorization.

The vector $[T]_N = \begin{bmatrix} t_1 \\ t_2 \\ t_3 \\ t_4 \\ t_5 \\ t_6 \\ t_7 \end{bmatrix} = \begin{bmatrix} 0 \\ 1 \\ 5 \\ 7 \\ 5 \\ 0 \\ 1 \end{bmatrix}$ of length $N = 7$ contains $L = 3$ distinct nonzero elements, 1, 5, and 7, which yield the kernel $[U]_L = \begin{bmatrix} u_1 \\ u_2 \\ u_3 \end{bmatrix} = \begin{bmatrix} 5 \\ 1 \\ 7 \end{bmatrix}$.

From the intermediate vector $[Y]_N = \begin{bmatrix} y_1 \\ y_2 \\ y_3 \\ y_4 \\ y_5 \\ y_6 \\ y_7 \end{bmatrix} = \begin{bmatrix} 0 \\ 2 \\ 1 \\ 3 \\ 1 \\ 0 \\ 2 \end{bmatrix}$ the commutator $[Z]_{N,L} = \begin{bmatrix} 0 & 0 & 0 \\ 0 & 1 & 0 \\ 1 & 0 & 0 \\ 0 & 0 & 1 \\ 1 & 0 & 0 \\ 0 & 0 & 0 \\ 0 & 1 & 0 \end{bmatrix}$ is obtained.

The factorization of the vector $[T]_N$ is the same as the product of the multiplication of the commutator $[Z]_{N,L}$ by the kernel $[U]_L$:

$$[T]_N = [Z]_{N,L} \cdot [U]_L = \begin{bmatrix} 0 & 0 & 0 \\ 0 & 1 & 0 \\ 1 & 0 & 0 \\ 0 & 0 & 1 \\ 1 & 0 & 0 \\ 0 & 0 & 0 \\ 0 & 1 & 0 \end{bmatrix} \cdot \begin{bmatrix} u_1 \\ u_2 \\ u_3 \end{bmatrix} = \begin{bmatrix} 0 & 0 & 0 \\ 0 & 1 & 0 \\ 1 & 0 & 0 \\ 0 & 0 & 1 \\ 1 & 0 & 0 \\ 0 & 0 & 0 \\ 0 & 1 & 0 \end{bmatrix} \cdot \begin{bmatrix} 5 \\ 1 \\ 7 \end{bmatrix} = \begin{bmatrix} 0 \\ 1 \\ 5 \\ 7 \\ 5 \\ 0 \\ 1 \end{bmatrix}$$

## 4. Definition of method and process for the factorization of a matrix.

Suppose that there exists the matrix

$$[T]_{M,N} = \begin{bmatrix} t_{1,1} & \cdots & t_{1,N} \\ \vdots & t_{m,n} & \vdots \\ t_{M,1} & \cdots & t_{M,N} \end{bmatrix} \tag{8}$$

of dimension M x $N$ and containing $L \leq M \cdot N$ distinct nonzero elements. This commutator yields the kernel consisting of the vector Из этой матрицы образуется ядро представляющее собой вектор

$$[U]_L = \begin{bmatrix} u_1 \\ \cdots \\ u_l \\ \cdots \\ u_L \end{bmatrix} \tag{9}$$

consisting of all the unique nonzero elements of the matrix $[T]_{M,N}$.

This same matrix $[T]_{M,N}$ is used to form a new intermediate matrix

$$[Y]_{M,N} = \begin{bmatrix} y_{1,1} & \cdots & y_{1,N} \\ \vdots & y_{m,n} & \vdots \\ y_{M,1} & \cdots & y_{M,N} \end{bmatrix} \tag{10}$$

of the same dimension M x $N$ as the matrix $[T]_{M,N}$ each of whose elements is either equal to zero or equal to the index of the element of the vector $[U]_L$, which is equal in value to this element of the matrix $[T]_{M,N}$. The matrix $[Y]_{M,N}$ can be obtained by replacing each nonzero element $t_{m,n}$ of the matrix $[T]_{M,N}$ by the index l of the equivalent element $u_l$ in the vector $[U]_L$.

From the resulting intermediate matrix $[Y]_{M,N}$ the commutator

$$[Z]_{M,N,L} = \{ z_{m,n,l} \mid m \in [1, M], n \in [1, N], l \in [1, L] \} \tag{11}$$

a tensor of rank 3, is obtained by replacing each nonzero element $y_{m,n}$ of the matrix $[Y]_{M,N}$ by the vector of length L with all elements equal to 0 if $y_{m,n} = 0$, or with a single unit element in the position corresponding to the nonzero value of $y_{m,n}$ and L-1 zero elements in all other positions.

The resulting commutator can be expressed as:

$$[Z]_{M,N,L} = \left\{ \begin{array}{l} [0 \ldots 0]_L, \text{ for } y_{m,n} = 0 \\ [0 \ldots 0]_{y_{m,n}-1} \ 1 \ [0 \ldots 0]_{L-y_{m,n}}, \text{ for } y_{m,n} > 0 \end{array} \middle| m \in [1, M], n \in [1, N] \right\} \tag{12}$$

The factorization of the matrix $[T]_{M,N}$ is equivalent to the convolution of the commutator $[Z]_{M,N,L}$ with the kernel $[U]_L$:

$$[T]_{M,N} = [Z]_{M,N,L} \cdot [U]_L = \left\{ \sum_{l=1}^{l=L} z_{m,n,l} \cdot u_l \ \middle| \ m \in [1, M], n \in [1, N] \right\} \tag{13}$$

## 5. Example of matrix factorization.

The matrix $[T]_{M,N} = \begin{bmatrix} t_{1,1} & t_{1,2} & t_{1,3} \\ t_{2,1} & t_{2,2} & t_{2,3} \\ t_{3,1} & t_{3,2} & t_{3,3} \\ t_{4,1} & t_{4,2} & t_{4,3} \end{bmatrix} = \begin{bmatrix} 2 & 5 & 2 \\ 3 & 0 & 9 \\ 0 & 7 & 0 \\ 9 & 2 & 3 \end{bmatrix}$ of dimension M x $N = 4$ x $3$ contains $L = 5$ distinct nonzero elements

2, 3, 5, 7, and 9 comprising the kernel $[U]_L = \begin{bmatrix} u_1 \\ u_2 \\ u_3 \\ u_4 \\ u_5 \end{bmatrix} = \begin{bmatrix} 2 \\ 3 \\ 5 \\ 7 \\ 9 \end{bmatrix}$.

From the intermediate matrix $[Y]_{M,N} = \begin{bmatrix} y_{1,1} & y_{1,2} & y_{1,3} \\ y_{2,1} & y_{2,2} & y_{2,3} \\ y_{3,1} & y_{3,2} & y_{3,3} \\ y_{4,1} & y_{4,2} & y_{4,3} \end{bmatrix} = \begin{bmatrix} 1 & 3 & 1 \\ 2 & 0 & 5 \\ 0 & 4 & 0 \\ 5 & 1 & 2 \end{bmatrix}$ the following commutator, a tensor of rank 3, is

obtained:

$$[Z]_{M,N,L} = \{ z_{m,n,l} \mid m \in [1,4], n \in [1,3], l \in [1,5] \} = \{ Z_{1,2,1} \ldots Z_{1,2,5} \ldots \ldots Z_{1,3,5} \ldots \ldots \ldots Z_{4,3,5} \}$$

$$= \begin{bmatrix} [Z_{1,1,1} \ldots Z_{1,1,5}][Z_{1,2,1} \ldots Z_{1,2,5}][Z_{1,3,1} \ldots Z_{1,3,5}] \\ [Z_{2,1,1} \ldots Z_{2,1,5}][Z_{2,2,1} \ldots Z_{2,2,5}][Z_{2,3,1} \ldots Z_{2,3,5}] \\ [Z_{3,1,1} \ldots Z_{3,1,5}][Z_{3,2,1} \ldots Z_{3,2,5}][Z_{3,3,1} \ldots Z_{3,3,5}] \\ [Z_{4,1,1} \ldots Z_{4,1,5}][Z_{4,2,1} \ldots Z_{4,2,5}][Z_{4,3,1} \ldots Z_{4,3,5}] \end{bmatrix} = \begin{bmatrix} [1\ 0\ 0\ 0\ 0][0\ 0\ 1\ 0\ 0][1\ 0\ 0\ 0\ 0] \\ [0\ 1\ 0\ 0\ 0][0\ 0\ 0\ 0\ 0][0\ 0\ 0\ 0\ 1] \\ [0\ 0\ 0\ 0\ 0][0\ 0\ 0\ 1\ 0][0\ 0\ 0\ 0\ 0] \\ [0\ 0\ 0\ 0\ 1][1\ 0\ 0\ 0\ 0][0\ 1\ 0\ 0\ 0] \end{bmatrix}$$

The matrix $[T]_{M,N}$ has the form of the convolution of the commutator $[Z]_{M,N,L}$ with the kernel $[U]_L$:

$$[T]_{M,N} = \begin{bmatrix} \sum_{l=1}^{l=5} z_{1,1,l} \cdot u_l & \sum_{l=1}^{l=5} z_{1,2,l} \cdot u_l & \sum_{l=1}^{l=5} z_{1,3,l} \cdot u_l \\ \sum_{l=1}^{l=5} z_{2,1,l} \cdot u_l & \sum_{l=1}^{l=5} z_{2,2,l} \cdot u_l & \sum_{l=1}^{l=5} z_{2,3,l} \cdot u_l \\ \sum_{l=1}^{l=5} z_{3,1,l} \cdot u_l & \sum_{l=1}^{l=5} z_{3,2,l} \cdot u_l & \sum_{l=1}^{l=5} z_{3,3,l} \cdot u_l \\ \sum_{l=1}^{l=5} z_{4,1,l} \cdot u_l & \sum_{l=1}^{l=5} z_{4,2,l} \cdot u_l & \sum_{l=1}^{l=5} z_{4,3,l} \cdot u_l \end{bmatrix} = \begin{bmatrix} [1\,0\,0\,0\,0][0\,0\,1\,0\,0][1\,0\,0\,0\,0] \\ [0\,1\,0\,0\,0][0\,0\,0\,0\,0][0\,0\,0\,0\,1] \\ [0\,0\,0\,0\,0][0\,0\,0\,1\,0][0\,0\,0\,0\,0] \\ [0\,0\,0\,0\,1][1\,0\,0\,0\,0][0\,1\,0\,0\,0] \end{bmatrix} \cdot \begin{bmatrix} u_1 \\ u_2 \\ u_3 \\ u_4 \\ u_5 \end{bmatrix}$$

$$= \begin{bmatrix} [1\,0\,0\,0\,0][0\,0\,1\,0\,0][1\,0\,0\,0\,0] \\ [0\,1\,0\,0\,0][0\,0\,0\,0\,0][0\,0\,0\,0\,1] \\ [0\,0\,0\,0\,0][0\,0\,0\,1\,0][0\,0\,0\,0\,0] \\ [0\,0\,0\,0\,1][1\,0\,0\,0\,0][0\,1\,0\,0\,0] \end{bmatrix} \cdot \begin{bmatrix} 2 \\ 3 \\ 5 \\ 7 \\ 9 \end{bmatrix} = \begin{bmatrix} 2 & 5 & 2 \\ 3 & 0 & 9 \\ 0 & 7 & 0 \\ 9 & 2 & 3 \end{bmatrix}$$

## 6. Definition of method and process for factorization of a matrix whose rows constitute all possible combinations of a finite set of elements.

Suppose that there exist finitely many distinct nonzero elements

$$E = \{e_1, e_2, \ldots, e_k\} \tag{14}$$

Then the matrix $[T]_{M,N}$, of dimension M x $N$ containing $L \leq M \cdot N$ distinct nonzero elements, whose rows constitute a complete set of the permutations of the elements of $E$ of length M will contain N columns and $M = k^N$ rows:

$$[T]_{k^N,N} = \begin{bmatrix} t_{1,1} & \cdots & t_{1,N} \\ \vdots & t_{m,n} & \vdots \\ t_{M,1} & \cdots & t_{M,N} \end{bmatrix} = \begin{bmatrix} e_1\,e_1\,e_1\,\cdots\,e_1 \\ e_2\,e_1\,e_1\,\cdots\,e_1 \\ \cdots \\ e_k\,e_1\,e_1\,\cdots\,e_1 \\ e_1\,e_2\,e_1\,\cdots\,e_1 \\ e_2\,e_2\,e_1\,\cdots\,e_1 \\ \cdots \\ e_k\,e_2\,e_1\,\cdots\,e_1 \\ \cdots \\ \cdots \\ e_1\,e_k\,e_k\,\cdots\,e_k \\ e_2\,e_k\,e_k\,\cdots\,e_k \\ \cdots \\ e_k\,e_k\,e_k\,\cdots\,e_k \end{bmatrix} = \left\{ e_{1+\text{floor}(\frac{v+m-1}{k(h+n-1) \text{mod } N} \text{ mod } k)} \middle| m \in [1, k^N], n \in [1, N] \right\} =$$

$$\begin{bmatrix} e_{1+\text{floor}(\frac{v}{k(n) \text{mod } N} \text{ mod } k)} & & e_{1+\text{floor}(\frac{v}{k(h+n-1) \text{mod } N} \text{ mod } k)} \\ \vdots & e_{1+\text{floor}(\frac{v+m-1}{k(h+n-1) \text{mod } N} \text{ mod } k)} & \vdots \\ e_{1+\text{floor}(\frac{v+k^N-1}{k(h) \text{mod } N} \text{ mod } k)} & \cdots & e_{1+\text{floor}(\frac{v+k^N-1}{k(h+N-1) \text{mod } N} \text{ mod } k)} \end{bmatrix} \tag{15}$$

M x $N L \leq M \cdot N$ From this matrix the kernel is obtained as the vector

$$M \times NL \leq M \cdot N [U]_L = \begin{bmatrix} u_1 \\ \ldots \\ u_l \\ \ldots \\ u_L \end{bmatrix} \quad (16)$$

consisting of all the distinct nonzero elements of the matrix $[T]_{M,N}$.

From the same matrix $[T]_{M,N}$ the intermediate matrix

$$[Y]_{M,N} = \begin{bmatrix} y_{1,1} & \cdots & y_{1,N} \\ \vdots & y_{m,n} & \vdots \\ y_{M,1} & \cdots & y_{M,N} \end{bmatrix} \quad (17)$$

is obtained, with the same dimensions M x $N$ as the matrix $[T]_{M,N}$ and with each element equal either to zero or to the index of that element of the vector $[U]_L$ which is equal in value to this element of the matrix $[T]_{M,N}$. The matrix $[Y]_{M,N}$ may be obtained by replacing each nonzero element $t_{m,n}$ of the matrix $[T]_{M,N}$ by the index $l$ of the equivalent element $u_l$ of the vector $[U]_L$.

From the resulting intermediate matrix $[Y]_{M,N}$ the commutator,

$$[Z]_{M,N,L} = \{ Z_{m,n,l} \,|\, m \in [1, M], n \in [1, N], l \in [1, L] \} \quad (18)$$

a tensor of rank 3, is obtained by replacing each nonzero element $y_{m,n}$ of the matrix $[Y]_{M,N}$ by the vector of length L, with all elements equal to 0 if $y_{m,n} = 0$, or with a single unit element in the position corresponding to the nonzero value of $y_{m,n}$ and L-1 elements equal to 0 in all other positions.

The resulting commutator may be written as:

$$[Z]_{M,N,L} = \left\{ \begin{cases} [0 \ldots 0]_L, \text{ for } y_{m,n} = 0 \\ [0 \ldots 0]_{y_{m,n}-1} \, 1 \, [0 \ldots 0]_{L-y_{m,n}}, \text{ for } y_{m,n} > 0 \end{cases} \,|\, m \in [1, M], n \in [1, N] \right\} \quad (19)$$

The factorization of the matrix $[T]_{M,N}$ is of the form of the convolution of the commutator $[Z]_{M,N,L}$ with the kernel $[U]_L$:

$$[T]_{M,N} = [Z]_{M,N,L} \cdot [U]_L = \left\{ \sum_{l=1}^{l=L} z_{m,n,l} \cdot u_l \,|\, m \in [1, M], n \in [1, N] \right\} \quad (20)$$

**7. Example of the factorization of a matrix whose rows constitute all possible combinations of a finite set of elements.**

The matrix $[T]_{M,N} = \begin{bmatrix} t_{1,1} & t_{1,2} & t_{1,3} \\ t_{2,1} & t_{2,2} & t_{2,3} \\ t_{3,1} & t_{3,2} & t_{3,3} \\ t_{4,1} & t_{4,2} & t_{4,3} \\ t_{5,1} & t_{5,2} & t_{5,3} \\ t_{6,1} & t_{6,2} & t_{6,3} \\ t_{7,1} & t_{7,2} & t_{7,3} \\ t_{8,1} & t_{8,2} & t_{8,3} \end{bmatrix} = \begin{bmatrix} 7 & 7 & 7 \\ 7 & 7 & 9 \\ 7 & 9 & 7 \\ 7 & 9 & 9 \\ 9 & 7 & 7 \\ 9 & 7 & 9 \\ 9 & 9 & 7 \\ 9 & 9 & 9 \end{bmatrix}$ of dimensions M x $N = 8$ x $3$ contains $L = 2$ distinct nonzero elements 7 and 9 constituting the kernel $[U]_L = \begin{bmatrix} u_1 \\ u_2 \end{bmatrix} = \begin{bmatrix} 7 \\ 9 \end{bmatrix}$.

From the intermediate matrix $[Y]_{M,N} = \begin{bmatrix} y_{1,1} & y_{1,2} & y_{1,3} \\ y_{2,1} & y_{2,2} & y_{2,3} \\ y_{3,1} & y_{3,2} & y_{3,3} \\ y_{4,1} & y_{4,2} & y_{4,3} \\ y_{5,1} & y_{5,2} & y_{5,3} \\ y_{6,1} & y_{6,2} & y_{6,3} \\ y_{7,1} & y_{7,2} & y_{7,3} \\ y_{8,1} & y_{8,2} & y_{8,3} \end{bmatrix} = \begin{bmatrix} 1 & 1 & 1 \\ 1 & 1 & 2 \\ 1 & 2 & 1 \\ 1 & 2 & 2 \\ 2 & 1 & 1 \\ 2 & 1 & 2 \\ 2 & 2 & 1 \\ 2 & 2 & 2 \end{bmatrix}$ the following commutator, a tensor of rank 3, is obtained:

$$[Z]_{M,N,L} = \{ Z_{m,n,l} \mid m \in [1,8], n \in [1,3], l \in [1,2]\} = \{Z_{1,2,1}\ Z_{1,2,2}\ \ldots\ldots\ Z_{1,3,2}\ \ldots\ldots\ldots\ Z_{8,3,2}\}$$

$$= \begin{bmatrix} [Z_{1,1,1}\ Z_{1,1,2}][Z_{1,2,1}\ Z_{1,2,2}][Z_{1,3,1}\ Z_{1,3,2}] \\ [Z_{2,1,1}\ Z_{2,1,2}][Z_{2,2,1}\ Z_{2,2,2}][Z_{2,3,1}\ Z_{2,3,2}] \\ [Z_{3,1,1}\ Z_{3,1,2}][Z_{3,2,1}\ Z_{3,2,2}][Z_{3,3,1}\ Z_{3,3,2}] \\ [Z_{4,1,1}\ Z_{4,1,2}][Z_{4,2,1}\ Z_{4,2,2}][Z_{4,3,1}\ Z_{4,3,2}] \\ [Z_{5,1,1}\ Z_{5,1,2}][Z_{5,2,1}\ Z_{5,2,2}][Z_{5,3,1}\ Z_{5,3,2}] \\ [Z_{61,1}\ Z_{6,1,2}][Z_{6,2,1}\ Z_{6,2,2}][Z_{6,3,1}\ Z_{6,3,2}] \\ [Z_{7,1,1}\ Z_{7,1,2}][Z_{7,2,1}\ Z_{7,2,2}][Z_{7,3,1}\ Z_{7,3,2}] \\ [Z_{8,1,1}\ Z_{8,1,2}][Z_{8,2,1}\ Z_{8,2,2}][Z_{8,3,1}\ Z_{8,3,2}] \end{bmatrix} = \begin{bmatrix} [1\ 0][1\ 0][1\ 0] \\ [1\ 0][1\ 0][0\ 1] \\ [1\ 0][0\ 1][1\ 0] \\ [1\ 0][0\ 1][0\ 1] \\ [0\ 1][1\ 0][1\ 0] \\ [0\ 1][1\ 0][0\ 1] \\ [0\ 1][0\ 1][1\ 0] \\ [0\ 1][0\ 1][0\ 1] \end{bmatrix}$$

The matrix $[T]_{M,N}$ is equal to the convolution of the commutator $[Z]_{M,N,L}$ and the kernel $[U]_L$:

$$[T]_{M,N} = \begin{bmatrix} \sum_{l=1}^{l=2} z_{1,1,l} \cdot u_l & \sum_{l=1}^{l=2} z_{1,2,l} \cdot u_l & \sum_{l=1}^{l=2} z_{1,3,l} \cdot u_l \\ \sum_{l=1}^{l=2} z_{2,1,l} \cdot u_l & \sum_{l=1}^{l=2} z_{2,2,l} \cdot u_l & \sum_{l=1}^{l=2} z_{2,3,l} \cdot u_l \\ \sum_{l=1}^{l=2} z_{3,1,l} \cdot u_l & \sum_{l=1}^{l=2} z_{3,2,l} \cdot u_l & \sum_{l=1}^{l=2} z_{3,3,l} \cdot u_l \\ \sum_{l=1}^{l=2} z_{4,1,l} \cdot u_l & \sum_{l=1}^{l=2} z_{4,2,l} \cdot u_l & \sum_{l=1}^{l=2} z_{4,3,l} \cdot u_l \\ \sum_{l=1}^{l=2} z_{5,1,l} \cdot u_l & \sum_{l=1}^{l=2} z_{5,2,l} \cdot u_l & \sum_{l=1}^{l=2} z_{5,3,l} \cdot u_l \\ \sum_{l=1}^{l=2} z_{6,1,l} \cdot u_l & \sum_{l=1}^{l=2} z_{6,2,l} \cdot u_l & \sum_{l=1}^{l=2} z_{6,3,l} \cdot u_l \\ \sum_{l=1}^{l=2} z_{7,1,l} \cdot u_l & \sum_{l=1}^{l=2} z_{7,2,l} \cdot u_l & \sum_{l=1}^{l=2} z_{7,3,l} \cdot u_l \\ \sum_{l=1}^{l=2} z_{8,1,l} \cdot u_l & \sum_{l=1}^{l=2} z_{8,2,l} \cdot u_l & \sum_{l=1}^{l=2} z_{8,3,l} \cdot u_l \end{bmatrix} = \begin{bmatrix} [1\ 0][1\ 0][1\ 0] \\ [1\ 0][1\ 0][0\ 1] \\ [1\ 0][0\ 1][1\ 0] \\ [1\ 0][0\ 1][0\ 1] \\ [0\ 1][1\ 0][1\ 0] \\ [0\ 1][1\ 0][0\ 1] \\ [0\ 1][0\ 1][1\ 0] \\ [0\ 1][0\ 1][0\ 1] \end{bmatrix} \cdot \begin{bmatrix} 7 \\ 9 \end{bmatrix} = \begin{bmatrix} 7 & 7 & 7 \\ 7 & 7 & 9 \\ 7 & 9 & 7 \\ 7 & 9 & 9 \\ 9 & 7 & 7 \\ 9 & 7 & 9 \\ 9 & 9 & 7 \\ 9 & 9 & 9 \end{bmatrix}$$

## 8. Definition of method and process for tensor factorization.

Suppose that there exists a tensor

$$[T]_{N_1,N_2,...,N_m,...,N_M} = \{ t_{n_1,n_2,...,n_m,...,n_M} | n_m \in [1, N_m], m \in [1, M] \} \quad (21)$$

of dimensions $\prod_{m=1}^{M} N_m$ containing $L \leq \prod_{m=1}^{M} N_m$ distinct nonzero elements. From this tensor the kernel consisting of the vector

$$[U]_L = \begin{bmatrix} u_1 \\ ... \\ u_l \\ ... \\ u_L \end{bmatrix} \quad (22)$$

is obtained, containing all the distinct nonzero elements of the tensor $[T]_{N_1,N_2,...,N_m,...,N_M}$.

From the same tensor $[T]_{N_1,N_2,...,N_m,...,N_M}$ a new intermediate tensor $[Y]_{N_1,N_2,...,N_m,...,N_M} = \{ y_{n_1,n_2,...,n_m,...,n_M} | n_m \in [1, N_m], m \in [1, M] \}$ (23) is generated, with the same dimensions $\prod_{m=1}^{M} N_m$ as the original tensor $[T]_{N_1,N_2,...,N_m,...,N_M}$ and with each element equal either to 0, or to the index of that element of the vector $[U]_L$ which has the same value as this element of the tensor $[T]_{N_1,N_2,...,N_m,...,N_M}$. The tensor $[Y]_{N_1,N_2,...,N_m,...,N_M}$ may be obtained by replacing each nonzero element $t_{n_1,n_2,...,n_m,...,n_M}$ of the tensor $[T]_{N_1,N_2,...,N_m,...,N_M}$ by the index $l$ of the equivalent element $u_l$ of the vector $[U]_L$.

From the resulting intermediate tensor $[Y]_{N_1,N_2,...,N_m,...,N_M}$ the commutator

$$[Z]_{N_1,N_2,...,N_m,...,N_M,L} = \{ z_{n_1,n_2,...,n_m,...,n_M,l} | n_m \in [1, N_m], m \in [1, M], l \in [1, L] \} \quad (24)$$

a tensor of rank M+1, is obtained by replacing every nonzero element $y_{n_1,n_2,...,n_m,...,n_M}$ of the tensor $[Y]_{N_1,N_2,...,N_m,...,N_M}$ by a vector of length L whose elements are all 0 if $y_{n_1,n_2,...,n_m,...,n_M} = 0$, or which has one unit element in the position corresponding to the nonzero value $y_{n_1,n_2,...,n_m,...,n_M}$ and L-1 nonzero elements in all other positions.

The resulting commutator may be represented as:

$$[Z]_{N_1,N_2,...,N_m,...,N_M,L} = \left\{ \begin{cases} [0 ... 0]_L, \text{ for } y_{n_1,n_2,...,n_m,...,n_M} = 0 \\ [0 ... 0]_{y_{n_1,n_2,...,n_m,...,n_M}-1} \; 1 \; [0 ... 0]_{L-y_{n_1,n_2,...,n_m,...,n_M}}, \text{ for } y_{n_1,n_2,...,n_m,...,n_M} > 0 \end{cases} \; \Big| \; n_m \in [1, N_m], \; m \in [1, M] \right\}$$

(25)

The factorization of the tensor $[T]_{N_1,N_2,...,N_m,...,N_M}$ has the form of the convolution of the commutator $[Z]_{N_1,N_2,...,N_m,...,N_M,L}$ with the kernel $[U]_L$:

$$[T]_{N_1,N_2,...,N_m,...,N_M} = [Z]_{N_1,N_2,...,N_m,...,N_M,L} \cdot [U]_L = \{ \sum_{l=1}^{l=L} z_{n_1,n_2,...,n_m,...,n_M,l} \cdot u_l \; | \; n_m \in [1, N_m], m \in [1, M] \}$$

(26)

**9. Definition of method and process of scalar multiplication or convolution of two vectors with factorization of one vector.**

Consider the operation of transposing the vector

$$[T]_N = \begin{bmatrix} t_1 \\ \ldots \\ t_n \\ \ldots \\ t_N \end{bmatrix} \qquad (27)$$

and multiplying it by the vector

$$[V]_N = \begin{bmatrix} v_1 \\ \ldots \\ v_n \\ \ldots \\ v_N \end{bmatrix}. \qquad (28)$$

Suppose that the vector $[T]_N$ of length $N$, containing $L \leq N$ distinct nonzero elements, is equal to the product of the commutator $[Z]_{N,L} = \begin{bmatrix} z_{1,1} & \cdots & z_{1,L} \\ \vdots & z_{n,l} & \vdots \\ z_{N,1} & \cdots & z_{N,L} \end{bmatrix}$ by the kernel $[U]_L = \begin{bmatrix} u_1 \\ \ldots \\ u_l \\ \ldots \\ u_L \end{bmatrix}$:

$$[T]_N = [Z]_{N,L} \cdot [U]_L = \begin{bmatrix} z_{1,1} & \cdots & z_{1,L} \\ \vdots & z_{n,l} & \vdots \\ z_{N,1} & \cdots & z_{N,L} \end{bmatrix} \cdot \begin{bmatrix} u_1 \\ \ldots \\ u_l \\ \ldots \\ u_L \end{bmatrix} \qquad (29)$$

Then the product of the transposed vector $[T]_N$ by the vector $[V]_N$ may be represented as:

$$[T]_N^t \cdot [V]_N = \left([Z]_{N,L} \cdot [U]_L\right)^t \cdot [V]_N = \sum_{n=1}^N v_n \cdot \sum_{l=1}^L z_{n,l} \cdot u_l = \sum_{n=1}^N \left(\sum_{l=1}^L z_{n,l} \cdot u_l\right) \cdot v_n = \sum_{n=1}^N \sum_{l=1}^L z_{n,l} \cdot u_l \cdot v_n = \sum_{n=1}^N \sum_{l=1}^L z_{n,l} \cdot (u_l \cdot v_n) \qquad (30)$$

In the above expression, the coefficient of the nested sum $(u_l \cdot v_n)$ is an element of the matrix which is obtained by multplication of the vector $[U]_L$ by the transposed vector $[V]_N$:

$$[P]_{L,N} = [U]_L \cdot [V]_N^t \qquad (31)$$

With this expression in mind, the product of the vectors $[T]_N$ and $[V]_N$ may be written in the form:

$$[T]_N^t \cdot [V]_N = \sum_{n=1}^N \sum_{l=1}^L z_{n,l} \cdot (u_l \cdot v_n) = \sum_{n=1}^N \sum_{l=1}^L z_{n,l} \cdot p_{l,n} = \sum_{n=1}^N [z_{n,1}, \ldots, z_{n,l}, \ldots, z_{n,L}] \cdot \begin{bmatrix} p_{1,n} \\ \ldots \\ p_{l,n} \\ \ldots \\ p_{L,n} \end{bmatrix} = tr([Z]_{N,L} \cdot [P]_{L,N}) \qquad (32)$$

Thus, the multiplication of a row vector by a column vector, or equivalently the multiplication of a covariant tensor of rank 1 by a contravariant tensor of rank 1 and length N, may take place in two steps. First the matrix is obtained containing each product of an element of the column vector and an element of the transposed row vector. Then, the trace of the matrix which is the product of the matrix obtained in the first step and the commutator of the transposed row vector is calculated. As a result, all the multiplications are carried out during the first step, and their maximum number is not the length N of the original vector, as in ordinary multiplication, but the product of the length N of the original vector and the number L of distinct nonzero elements of the row vector. Meanwhile, all additions are carried out during the second step, and their maximum number is equal to N-1. Thus the ratio of the number of operations using the decomposition into commutator and kernel to the number of operations using a method without such a decomposition is equal to $C_+ \leq \frac{N-1}{N-1} = 1$ for addition operations and $C_* \leq \frac{N \cdot L}{N} = L$ for multiplication operations. Naturally, multiplication by 0 or by 1 is not counted.

## 10. Example of scalar multiplication or convolution of vectors.

The vector $[T]_N = \begin{bmatrix} 2 \\ 3 \\ 4 \\ 2 \end{bmatrix}$ of length $N = 4$ is transposed and multiplied by the vector $[V]_N = \begin{bmatrix} 5 \\ 6 \\ 7 \\ 8 \end{bmatrix}$. The vector $[T]_N$, containing $L = 3$ distinct nonzero elements, is represented by the convolution of the commutator $[Z]_{N,L} = \begin{bmatrix} 1 & 0 & 0 \\ 0 & 1 & 0 \\ 0 & 0 & 1 \\ 1 & 0 & 0 \end{bmatrix}$ by the kernel $[U]_L = \begin{bmatrix} 2 \\ 3 \\ 4 \end{bmatrix}$:

$$[T]_N = [Z]_{N,L} \cdot [U]_L = \begin{bmatrix} z_{1,1} & \cdots & z_{1,L} \\ \vdots & z_{n,l} & \vdots \\ z_{N,1} & \cdots & z_{N,L} \end{bmatrix} \cdot \begin{bmatrix} u_1 \\ \cdots \\ u_l \\ \cdots \\ u_L \end{bmatrix} = \begin{bmatrix} 1 & 0 & 0 \\ 0 & 1 & 0 \\ 0 & 0 & 1 \\ 1 & 0 & 0 \end{bmatrix} \cdot \begin{bmatrix} 2 \\ 3 \\ 4 \end{bmatrix} = \begin{bmatrix} 2 \\ 3 \\ 4 \\ 2 \end{bmatrix}.$$

The product of the transposed vector $[T]_N$ by the vector $[V]_N$ may be obtained in two stages. First, the matrix is obtained that contains the product of each element of the column vector and each element of the transposed row vector:

$$[P]_{L,N} = [U]_L \cdot [V]_N^t = \begin{bmatrix} 2 \\ 3 \\ 4 \end{bmatrix} \cdot [5\ 6\ 7\ 8] = \begin{bmatrix} 10 & 12 & 14 & 16 \\ 15 & 18 & 21 & 24 \\ 20 & 24 & 28 & 32 \end{bmatrix}$$

At this stage all multiplication operations are carried out, of which there are $N \cdot L = 12$, and the coefficient $C_* = \frac{N \cdot L}{N} = \frac{4 \cdot 3}{4} = 3$.

Then the trace of the product of the commutator with the matrix $[P]_{L,N}$ is found.

$$[T]_N^t \cdot [V]_N = \text{tr}([Z]_{N,L} \cdot [P]_{L,N}) = \text{tr}\left(\begin{bmatrix}1\ 0\ 0\\0\ 1\ 0\\0\ 0\ 1\\1\ 0\ 0\end{bmatrix} \cdot \begin{bmatrix}10\ 12\ 14\ 16\\15\ 18\ 21\ 24\\20\ 24\ 28\ 32\end{bmatrix}\right) = \text{tr}\left(\begin{bmatrix}10\ 12\ 14\ 16\\15\ 18\ 21\ 24\\20\ 24\ 28\ 32\\10\ 12\ 14\ 16\end{bmatrix}\right) = 10 + 18 + 28 + 16$$

$$= 72 = \left(\begin{bmatrix}2\\3\\4\\2\end{bmatrix}\right)^t \cdot \begin{bmatrix}5\\6\\7\\8\end{bmatrix} = \left(\begin{bmatrix}1\ 0\ 0\\0\ 1\ 0\\0\ 0\ 1\\1\ 0\ 0\end{bmatrix} \cdot \begin{bmatrix}2\\3\\4\end{bmatrix}\right)^t \cdot \begin{bmatrix}5\\6\\7\\8\end{bmatrix}$$

At this stage all additions are carried out, of which there are $N - 1 = 3$, and the coefficient $C_+ = \frac{3-1}{3-1} = 1$.

## 11. Definition of method and process for recursive scalar multiplication or recursive convolution of vectors.

Now we examine a different problem – that of multiple consecutive (in other words, recursive) calculation of the product of a known and constant transposed vector by a series of vectors, each of which is formed from the preceding vector in the series by shifting each element in the preceding vector up by one position. In the case of a linear shift, the lowest position is filled by a new element. In the case of a ring shift, the lowest position is filled by the element occupying the highest position in the preceding vector in the series. This is the equivalent of multiplying the known and constant vector by a left-circulant or Toeplitz matrix.

The vector

$$[T]_N = \begin{bmatrix} t_1 \\ \dots \\ t_n \\ \dots \\ t_N \end{bmatrix} \tag{33}$$

is to be transposed and multiplied by the vector

$$[V]_N = \begin{bmatrix} v_1 \\ \dots \\ v_n \\ \dots \\ v_N \end{bmatrix} \tag{34}$$

and all of its circularly-shifted variants:

$$\{[V_1]_N, [V_2]_N, \dots, [V_{N-1}]_N\} = \left\{\begin{bmatrix} v_2 \\ \dots \\ \dots \\ v_N \\ v_1 \end{bmatrix}, \begin{bmatrix} v_3 \\ \dots \\ \dots \\ v_1 \\ v_2 \end{bmatrix}, \dots, \begin{bmatrix} v_N \\ v_1 \\ \dots \\ \dots \\ v_{N-1} \end{bmatrix}\right\}. \tag{35}$$

Suppose that the vector $[T]_N = \begin{bmatrix} t_1 \\ \ldots \\ t_n \\ \ldots \\ t_N \end{bmatrix}$ of length $N$, containing $L \leq N$ distinct nonzero elements, is represented as the product of the commutator $[Z]_{N,L} = \begin{bmatrix} z_{1,1} & \cdots & z_{1,L} \\ \vdots & z_{n,l} & \vdots \\ z_{N,1} & \cdots & z_{N,L} \end{bmatrix}$ by the kernel $[U]_L = \begin{bmatrix} u_1 \\ \ldots \\ u_l \\ \ldots \\ u_L \end{bmatrix}$.

First to be obtained is the result of multiplying the transposed vector $[T]_N$ by the vector $[V]_N$. The product of the vectors $[T]_N$ and $[V]_N$ may be represented as:

$$[T]_N^t \cdot [V]_N = \text{tr}([Z]_{N,L} \cdot [P]_{L,N}), \tag{36}$$

where $[P]_{L,N}$ is a matrix formed by the multiplication of the kernel $[U]_L$ by the transposed vector $[V]_N$:

$$[P]_{L,N} = [U]_L \cdot [V]_N^t = \begin{bmatrix} u_1 \\ \ldots \\ u_l \\ \ldots \\ u_L \end{bmatrix} \cdot [v_1 \ldots v_n \ldots v_N] = \begin{bmatrix} v_1 \cdot u_1 & \cdots & v_N \cdot u_1 \\ \vdots & v_n \cdot u_l & \vdots \\ v_1 \cdot u_L & \cdots & v_N \cdot u_L \end{bmatrix} \tag{37}$$

To obtain the second value - is the result of multiplying the transposed vector $[T]_N$ by the first shifted variant of the vector $[V]_N$, which is the vector

$$[V_1]_N = \begin{bmatrix} v_2 \\ \ldots \\ \ldots \\ v_N \\ v_1 \end{bmatrix} \tag{38}$$

- the new matrix $[P_1]_{L,N}$ is obtained:

$$[P_1]_{L,N} = [U]_L \cdot [V_1]_N^t = \begin{bmatrix} u_1 \\ \ldots \\ u_l \\ \ldots \\ u_L \end{bmatrix} \cdot [v_2 \ldots v_{n+1} \ldots v_N \; v_1] = \begin{bmatrix} v_2 \cdot u_1 & \cdots & v_N \cdot u_1 & v_1 \cdot u_1 \\ \vdots & \vdots & \vdots & \vdots \\ v_1 \cdot u_L & \cdots & v_N \cdot u_L & v_1 \cdot u_L \end{bmatrix}, \tag{39}$$

Clearly the matrix $[P_1]_{L,N}$ is a copy of the matrix $[P]_{L,N}$ cyclically shifted one position to the left. Thus the matrix $[P_1]_{L,N}$ is obtained from $[P]_{L,N}$ via multiplication by the shift matrix

$$[C_1]_{N,N} = \begin{bmatrix} 0 & 0 & 0 & \cdots & 0 & 0 & 1 \\ 1 & 0 & 0 & \cdots & 0 & 0 & 0 \\ 0 & 1 & 0 & \cdots & 0 & 0 & 0 \\ & & & \vdots & & & \\ 0 & 0 & 0 & \cdots & 0 & 1 & 0 \end{bmatrix} \tag{40}$$

of dimension N x N , which is itself a variant of a diagonal matrix cyclically shifted left by one position.

$$[P_1]_{L,N} = [P]_{L,N} \cdot [C_1]_{N,N} = \begin{bmatrix} v_1 \cdot u_1 & \cdots & & v_N \cdot u_1 \\ \vdots & & v_n \cdot u_l & \vdots \\ v_1 \cdot u_L & \cdots & & v_N \cdot u_L \end{bmatrix} \cdot \begin{bmatrix} 0 & 0 & 0 & \cdots & 0 & 0 & 1 \\ 1 & 0 & 0 & \cdots & 0 & 0 & 0 \\ 0 & 1 & 0 & \cdots & 0 & 0 & 0 \\ & & & \vdots & & & \\ 0 & 0 & 0 & \cdots & 0 & 1 & 0 \end{bmatrix} \tag{41}$$

To obtain the remaining values, products of the multiplication of the transposed vector $[T]_N$ by the second and subsequent circularly-shifted variants of the vector $[V]_N$, which are the vectors $[V_2]_N, [V_3]_N, \ldots, [V_{N-1}]_N$, it is necessary to obtain the matrices $[P_2]_{L,N}, [P_3]_{L,N}, \ldots, [P_{N-1}]_{L,N}$. As in the case just described, each subsequent matrix $[P_n]_{L,N}$ is obtained by cyclically shifting the matrix $[P_{n-1}]_{L,N}$ to the left, or equivalently by multiplying the matrix $[P_{n-1}]_{L,N}$ by the displacement matrix $[C_1]_{N,N}$, or by multiplying the matrix $[P]_{L,N}$ by the displacement matrix $[C_n]_{N,N}$, which is the n[th] power of the displacement matrix $[C_n]_{N,N}$:

$$[P_n]_{L,N} = [U]_L \cdot [V_n]_N^t = [P_{n-1}]_{L,N} \cdot [C_1]_{N,N} = [P_{n-k}]_{L,N} \cdot [C_1]_{N,N}^k = [P]_{L,N} \cdot [C_1]_{N,N}^n = [P]_{L,N} \cdot [C_n]_{N,N}, \tag{42}$$

where $n, k \in [0, N-1]$. For the sake of generality we take

$$[P]_{L,N} = [P_0]_{L,N}, \tag{43}$$

$$[V_0]_N = [V]_N, \tag{44}$$

and

$$[C_1]_{N,N}^n = [C]_{N,N}^n. \tag{45}$$

Thus the result of the cyclic multiplication can be represented by the vector

$$[R]_N = \begin{bmatrix} r_1 \\ \cdots \\ r_n \\ \cdots \\ r_N \end{bmatrix} = \begin{bmatrix} [T]_N^t \cdot [V_0]_N \\ \cdots \\ [T]_N^t \cdot [V_{n-1}]_N \\ \cdots \\ [T]_N^t \cdot [V_{N-1}]_N \end{bmatrix} = \begin{bmatrix} tr([Z]_{N,L} \cdot [P_0]_{L,N}) \\ \cdots \\ tr([Z]_{N,L} \cdot [P_{n-1}]_{L,N}) \\ \cdots \\ tr([Z]_{N,L} \cdot [P_{N-1}]_{L,N}) \end{bmatrix} = \begin{bmatrix} tr([Z]_{N,L} \cdot [P]_{L,N} \cdot [C]_{N,N}^0) \\ \cdots \\ tr([Z]_{N,L} \cdot [P]_{L,N} \cdot [C]_{N,N}^{n-1}) \\ \cdots \\ tr([Z]_{N,L} \cdot [P]_{L,N} \cdot [C]_{N,N}^{N-1}) \end{bmatrix} \tag{46}$$

Each element of the vector $[R]_N$ contains the same matrix $[P]_{L,N}$. From this it follows that to compute al N vector multiplications, it is only necessary to obtain this matrix once, which requires no more than N x L multiplication operations. Thus the ratio of the number of operations in the cyclical multiplication using the method of decomposition of the vector into a commutator and kernel to the number of operations for the same multiplication using a method without this decomposition is $C_+ \leq \frac{N \cdot (N-1)}{N \cdot (N-1)} = 1$ for addition and $C_* \leq \frac{N \cdot L}{N \cdot N} = \frac{L}{N} \leq 1$ for multiplication.

## 12. Example of recursive scalar multiplication or recursive convolution of vectors.

The vector $[T]_N = \begin{bmatrix} 2 \\ 3 \\ 4 \\ 2 \end{bmatrix}$ of length $N = 4$ is to be transposed and multiplied by the vector $[V]_N = \begin{bmatrix} 5 \\ 6 \\ 7 \\ 8 \end{bmatrix}$ and all of its circularly-shifted variants:

$$\{[V_1]_N, [V_2]_N, [V_3]_N\} = \left\{ \begin{bmatrix} 6 \\ 7 \\ 8 \\ 5 \end{bmatrix}, \begin{bmatrix} 7 \\ 8 \\ 5 \\ 6 \end{bmatrix}, \begin{bmatrix} 8 \\ 5 \\ 6 \\ 7 \end{bmatrix} \right\}.$$

The vector $[T]_N$, containing $L = 3$ distinct nonzero elements, is represented by the product of the commutator $[Z]_{N,L} = \begin{bmatrix} 1 & 0 & 0 \\ 0 & 1 & 0 \\ 0 & 0 & 1 \\ 1 & 0 & 0 \end{bmatrix}$ and the kernel $[U]_L = \begin{bmatrix} 2 \\ 3 \\ 4 \end{bmatrix}$ :

$$[T]_N = [Z]_{N,L} \cdot [U]_L = \begin{bmatrix} z_{1,1} & \cdots & z_{1,L} \\ \vdots & z_{n,l} & \vdots \\ z_{N,1} & \cdots & z_{N,L} \end{bmatrix} \cdot \begin{bmatrix} u_1 \\ \cdots \\ u_l \\ \cdots \\ u_L \end{bmatrix} = \begin{bmatrix} 1 & 0 & 0 \\ 0 & 1 & 0 \\ 0 & 0 & 1 \\ 1 & 0 & 0 \end{bmatrix} \cdot \begin{bmatrix} 2 \\ 3 \\ 4 \end{bmatrix} = \begin{bmatrix} 2 \\ 3 \\ 4 \\ 2 \end{bmatrix}.$$

The matrix is obtained which contains the products of each element of the column vector with each element of the transposed row vector:

$$[P]_{L,N} = [U]_L \cdot [V]_N^t = \begin{bmatrix} 2 \\ 3 \\ 4 \end{bmatrix} \cdot [5\ 6\ 7\ 8] = \begin{bmatrix} 10 & 12 & 14 & 16 \\ 15 & 18 & 21 & 24 \\ 20 & 24 & 28 & 32 \end{bmatrix}$$

In the same way the displacement matrix $[C_1]_{N,N} = [C]_{N,N} = \begin{bmatrix} 0 & 0 & 0 & 1 \\ 1 & 0 & 0 & 0 \\ 0 & 1 & 0 & 0 \\ 0 & 0 & 1 & 0 \end{bmatrix}$ with dimensions $N\text{x}N = 4 \text{ x } 4$ is obtained, and gives rise to the subsequent N-1=3 displacement matrices:

$$[C]_{N,N}^0 = \begin{bmatrix} 0 & 0 & 0 & 1 \\ 1 & 0 & 0 & 0 \\ 0 & 1 & 0 & 0 \\ 0 & 0 & 1 & 0 \end{bmatrix}^0 = \begin{bmatrix} 1 & 0 & 0 & 0 \\ 0 & 1 & 0 & 0 \\ 0 & 0 & 1 & 0 \\ 0 & 0 & 0 & 1 \end{bmatrix}, [C]_{N,N}^2 = \begin{bmatrix} 0 & 0 & 0 & 1 \\ 1 & 0 & 0 & 0 \\ 0 & 1 & 0 & 0 \\ 0 & 0 & 1 & 0 \end{bmatrix}^2 = \begin{bmatrix} 0 & 0 & 1 & 0 \\ 0 & 0 & 0 & 1 \\ 1 & 0 & 0 & 0 \\ 0 & 1 & 0 & 0 \end{bmatrix}, [C]_{N,N}^3 = \begin{bmatrix} 0 & 0 & 0 & 1 \\ 1 & 0 & 0 & 0 \\ 0 & 1 & 0 & 0 \\ 0 & 0 & 1 & 0 \end{bmatrix}^3 = \begin{bmatrix} 0 & 1 & 0 & 0 \\ 0 & 0 & 1 & 0 \\ 0 & 0 & 0 & 1 \\ 1 & 0 & 0 & 0 \end{bmatrix}$$

Then, the trace of the product of the commutator with the matrix $[P]_{L,N}$ is obtained.

$$[R]_N = \begin{bmatrix} r_1 \\ r_2 \\ r_3 \\ r_4 \end{bmatrix} = \begin{bmatrix} \text{tr}([Z]_{N,L} \cdot [P]_{L,N} \cdot [C]_{N,N}^0) \\ \text{tr}([Z]_{N,L} \cdot [P]_{L,N} \cdot [C]_{N,N}^1) \\ \text{tr}([Z]_{N,L} \cdot [P]_{L,N} \cdot [C]_{N,N}^2) \\ \text{tr}([Z]_{N,L} \cdot [P]_{L,N} \cdot [C]_{N,N}^3) \end{bmatrix} = \begin{bmatrix} \text{tr}\left( \begin{bmatrix} 1 & 0 & 0 \\ 0 & 1 & 0 \\ 0 & 0 & 1 \\ 1 & 0 & 0 \end{bmatrix} \cdot \begin{bmatrix} 10 & 12 & 14 & 16 \\ 15 & 18 & 21 & 24 \\ 20 & 24 & 28 & 32 \end{bmatrix} \cdot \begin{bmatrix} 1 & 0 & 0 & 0 \\ 0 & 1 & 0 & 0 \\ 0 & 0 & 1 & 0 \\ 0 & 0 & 0 & 1 \end{bmatrix} \right) \\ \text{tr}\left( \begin{bmatrix} 1 & 0 & 0 \\ 0 & 1 & 0 \\ 0 & 0 & 1 \\ 1 & 0 & 0 \end{bmatrix} \cdot \begin{bmatrix} 10 & 12 & 14 & 16 \\ 15 & 18 & 21 & 24 \\ 20 & 24 & 28 & 32 \end{bmatrix} \cdot \begin{bmatrix} 0 & 0 & 0 & 1 \\ 1 & 0 & 0 & 0 \\ 0 & 1 & 0 & 0 \\ 0 & 0 & 1 & 0 \end{bmatrix} \right) \\ \text{tr}\left( \begin{bmatrix} 1 & 0 & 0 \\ 0 & 1 & 0 \\ 0 & 0 & 1 \\ 1 & 0 & 0 \end{bmatrix} \cdot \begin{bmatrix} 10 & 12 & 14 & 16 \\ 15 & 18 & 21 & 24 \\ 20 & 24 & 28 & 32 \end{bmatrix} \cdot \begin{bmatrix} 0 & 0 & 1 & 0 \\ 0 & 0 & 0 & 1 \\ 1 & 0 & 0 & 0 \\ 0 & 1 & 0 & 0 \end{bmatrix} \right) \\ \text{tr}\left( \begin{bmatrix} 1 & 0 & 0 \\ 0 & 1 & 0 \\ 0 & 0 & 1 \\ 1 & 0 & 0 \end{bmatrix} \cdot \begin{bmatrix} 10 & 12 & 14 & 16 \\ 15 & 18 & 21 & 24 \\ 20 & 24 & 28 & 32 \end{bmatrix} \cdot \begin{bmatrix} 0 & 1 & 0 & 0 \\ 0 & 0 & 1 & 0 \\ 0 & 0 & 0 & 1 \\ 1 & 0 & 0 & 0 \end{bmatrix} \right) \end{bmatrix}$$

$$= \begin{bmatrix} \text{tr}\left( \begin{bmatrix} 1 & 0 & 0 \\ 0 & 1 & 0 \\ 0 & 0 & 1 \\ 1 & 0 & 0 \end{bmatrix} \cdot \begin{bmatrix} 10 & 12 & 14 & 16 \\ 15 & 18 & 21 & 24 \\ 20 & 24 & 28 & 32 \end{bmatrix} \right) \\ \text{tr}\left( \begin{bmatrix} 1 & 0 & 0 \\ 0 & 1 & 0 \\ 0 & 0 & 1 \\ 1 & 0 & 0 \end{bmatrix} \cdot \begin{bmatrix} 12 & 14 & 16 & 10 \\ 18 & 21 & 24 & 15 \\ 24 & 28 & 32 & 20 \end{bmatrix} \right) \\ \text{tr}\left( \begin{bmatrix} 1 & 0 & 0 \\ 0 & 1 & 0 \\ 0 & 0 & 1 \\ 1 & 0 & 0 \end{bmatrix} \cdot \begin{bmatrix} 14 & 16 & 10 & 12 \\ 21 & 24 & 15 & 18 \\ 28 & 32 & 20 & 24 \end{bmatrix} \right) \\ \text{tr}\left( \begin{bmatrix} 1 & 0 & 0 \\ 0 & 1 & 0 \\ 0 & 0 & 1 \\ 1 & 0 & 0 \end{bmatrix} \cdot \begin{bmatrix} 16 & 10 & 12 & 14 \\ 24 & 15 & 18 & 21 \\ 32 & 20 & 24 & 28 \end{bmatrix} \right) \end{bmatrix} = \begin{bmatrix} \text{tr}\left( \begin{bmatrix} 10 & 12 & 14 & 16 \\ 15 & 18 & 21 & 24 \\ 20 & 24 & 28 & 32 \\ 10 & 12 & 14 & 16 \end{bmatrix} \right) \\ \text{tr}\left( \begin{bmatrix} 12 & 14 & 16 & 10 \\ 18 & 21 & 24 & 15 \\ 24 & 28 & 32 & 20 \\ 12 & 14 & 16 & 10 \end{bmatrix} \right) \\ \text{tr}\left( \begin{bmatrix} 14 & 16 & 10 & 12 \\ 21 & 24 & 15 & 18 \\ 28 & 32 & 20 & 24 \\ 14 & 16 & 10 & 12 \end{bmatrix} \right) \\ \text{tr}\left( \begin{bmatrix} 16 & 10 & 12 & 14 \\ 24 & 15 & 18 & 21 \\ 32 & 20 & 24 & 28 \\ 16 & 10 & 12 & 14 \end{bmatrix} \right) \end{bmatrix} = \begin{bmatrix} 10 + 18 + 28 + 16 \\ 12 + 21 + 32 + 10 \\ 14 + 24 + 20 + 12 \\ 16 + 15 + 24 + 14 \end{bmatrix} = \begin{bmatrix} 72 \\ 75 \\ 70 \\ 69 \end{bmatrix}$$

To compute all N = 4 multiplied vectors requires obtaining the matrix $[P]_{L,N}$ only once, which requires N x L = 4 x 3 = 12 multiplication operations. Thus, in the present example, the ratio of the number of operations in the cyclical multiplication using the method of decomposition of the vector into a commutator and kernel to the number of operations for the same multiplication using a method without this decomposition is $C_+ = \frac{N \cdot (N-1)}{N \cdot (N-1)} = \frac{4 \cdot (4-1)}{4 \cdot (4-1)} = 1$ for addition and $C_* = \frac{N \cdot L}{N \cdot N} = \frac{4 \cdot 3}{4 \cdot 4} = \frac{3}{4} = 0.75$ for multiplication.

## 13. Definition of method and process for iterative scalar multiplication or iterative convolution of vectors.

Now we examine another fairly common case. Here the objective is sequential and continuous, which is to say iterative, multiplication of a known and constant vector by a series of vectors, each of which is formed from the preceding vector by a linear shift of its elements one position upward. In each succeeding vector of this series the lowest position of the vector is filled by a new element, and the uppermost element of the preceding vector is lost. We suppose that at each successive iteration the vector

$$[T]_N = \begin{bmatrix} t_1 \\ \dots \\ t_n \\ \dots \\ t_N \end{bmatrix} \qquad (47),$$

containing $L \leq N$ distinct nonzero elements and represented by the product of the commutator $[Z]_{N,L} = \begin{bmatrix} z_{1,1} & \dots & z_{1,L} \\ \vdots & z_{n,l} & \vdots \\ z_{N,1} & \dots & z_{N,L} \end{bmatrix}$ by the kernel $[U]_L = \begin{bmatrix} u_1 \\ \dots \\ u_l \\ \dots \\ u_L \end{bmatrix}$, is to be transposed and multiplied by the vector

$$[V_1]_N = \begin{bmatrix} v_1 \\ \dots \\ v_n \\ \dots \\ v_N \end{bmatrix}, \qquad (48)$$

having first obtained the matrix $[P_1]_{L,N}$ by multiplying the kernel $[U]_L$ of the vector $[T]_N$ by the transposed vector $[V_1]_N$:

$$[P_1]_{L,N} = [U]_L \cdot [V_1]_N^t = \begin{bmatrix} u_1 \\ \dots \\ u_l \\ \dots \\ u_L \end{bmatrix} \cdot [v_1 \dots v_n \dots v_N] = \begin{bmatrix} v_1 \cdot u_1 & \dots & v_N \cdot u_1 \\ \vdots & v_n \cdot u_l & \vdots \\ v_1 \cdot u_L & \dots & v_N \cdot u_L \end{bmatrix} = \begin{bmatrix} \begin{bmatrix} u_1 \\ \dots \\ u_l \\ \dots \\ u_L \end{bmatrix} \cdot v_1 & \begin{bmatrix} u_1 \\ \dots \\ u_l \\ \dots \\ u_L \end{bmatrix} \cdot v_2 & \dots & \begin{bmatrix} u_1 \\ \dots \\ u_l \\ \dots \\ u_L \end{bmatrix} \cdot v_n & \dots & \begin{bmatrix} u_1 \\ \dots \\ u_l \\ \dots \\ u_L \end{bmatrix} \cdot v_{N-1} & \begin{bmatrix} u_1 \\ \dots \\ u_l \\ \dots \\ u_L \end{bmatrix} \cdot v_N \end{bmatrix} \qquad (49)$$

Meanwhile the previous iteration consisted of the multiplication of the vector $[T]_N$ by the vector

$$[V_0]_N = \begin{bmatrix} v_0 \\ \dots \\ v_n \\ \dots \\ v_{N-1} \end{bmatrix} \qquad (50)$$

and consequently there exists a matrix $[P_0]_{L,N}$ formed from the multiplication of the kernel $[U]_L$ of the vector $[T]_N$ by the transposed vector $[V_0]_N$:

$$[P_0]_{L,N} = [U]_L \cdot [V_0]_N^t = \begin{bmatrix} u_1 \\ \dots \\ u_l \\ \dots \\ u_L \end{bmatrix} \cdot [v_0 \dots v_{n-1} \dots v_{N-1}] = \begin{bmatrix} v_0 \cdot u_1 & \dots & v_{N-1} \cdot u_1 \\ \vdots & v_{n-1} \cdot u_l & \vdots \\ v_0 \cdot u_L & \dots & v_{N-1} \cdot u_L \end{bmatrix} = \begin{bmatrix} \begin{bmatrix} u_1 \\ \dots \\ u_l \\ \dots \\ u_L \end{bmatrix} \cdot v_0 & \begin{bmatrix} u_1 \\ \dots \\ u_l \\ \dots \\ u_L \end{bmatrix} \cdot v_1 & \dots & \begin{bmatrix} u_1 \\ \dots \\ u_l \\ \dots \\ u_L \end{bmatrix} \cdot v_{n-1} & \dots & \begin{bmatrix} u_1 \\ \dots \\ u_l \\ \dots \\ u_L \end{bmatrix} \cdot v_{N-2} & \begin{bmatrix} u_1 \\ \dots \\ u_l \\ \dots \\ u_L \end{bmatrix} \cdot v_{N-1} \end{bmatrix} \qquad (51)$$

The matrix $[P_1]_{L,N}$ is equivalent to the matrix $[P_0]_{L,N}$, linearly shifted one position to the left, in which the rightmost column is the product of the kernel $[U]_L = \begin{bmatrix} u_1 \\ \ldots \\ u_l \\ \ldots \\ u_L \end{bmatrix}$ and the new element $v_N$:

$$[P_1]_{L,N} = \begin{bmatrix} v_1 \cdot u_1 & \cdots & v_N \cdot u_1 \\ \vdots & v_n \cdot u_l & \vdots \\ v_1 \cdot u_L & \cdots & v_N \cdot u_L \end{bmatrix} = \begin{bmatrix} v_1 \cdot u_1 & \cdots & 0 \\ \vdots & v_n \cdot u_l & \vdots \\ v_1 \cdot u_L & \cdots & 0 \end{bmatrix} + \begin{bmatrix} 0 & \cdots & u_1 \\ \vdots & 0 & \vdots \\ 0 & \cdots & u_L \end{bmatrix} \cdot v_N =$$

$$\begin{bmatrix} v_0 \cdot u_1 & \cdots & v_{N-1} \cdot u_1 \\ \vdots & v_{n-1} \cdot u_l & \vdots \\ v_0 \cdot u_L & \cdots & v_{N-1} \cdot u_L \end{bmatrix} \cdot \begin{bmatrix} 0\,0\,0\,0\ldots0\,0\,0\,0 \\ 1\,0\,0\,0\ldots0\,0\,0\,0 \\ 0\,1\,0\,0\ldots0\,0\,0\,0 \\ 0\,0\,1\,0\ldots0\,0\,0\,0 \\ \vdots \\ 0\,0\,0\,0\ldots0\,1\,0\,0 \\ 0\,0\,0\,0\ldots0\,0\,1\,0 \end{bmatrix} + \begin{bmatrix} u_1 \\ \ldots \\ u_l \\ \ldots \\ u_L \end{bmatrix} \cdot v_N \cdot [0 \ldots 0\,1]_N = [P_0]_{L,N} \cdot [\widetilde{C}]_{N,N} + [U]_L \cdot v_N \cdot [\widetilde{\widetilde{C}}]_N$$

(52)

Here, the matrix

$$[\widetilde{C}]_{N,N} = \begin{bmatrix} 0\,0\,0\,0\ldots0\,0\,0\,0 \\ 1\,0\,0\,0\ldots0\,0\,0\,0 \\ 0\,1\,0\,0\ldots0\,0\,0\,0 \\ 0\,0\,1\,0\ldots0\,0\,0\,0 \\ \vdots \\ 0\,0\,0\,0\ldots0\,1\,0\,0 \\ 0\,0\,0\,0\ldots0\,0\,1\,0 \end{bmatrix} \qquad (53)$$

and the vector

$$[\widetilde{\widetilde{C}}]_N = [0 \ldots 0\,1]_N \qquad (54)$$

are introduced in the iterative formation of the vector $[P]_{L,N}$.

Thus, the n$^{th}$ iteration takes the form:

$$\{[P_n]_{L,N} = [P_{n-1}]_{L,N} \cdot [\widetilde{C}]_{N,N} + [U]_L \cdot v_n \cdot [\widetilde{\widetilde{C}}]_N;\ r_n = tr([Z]_{N,L} \cdot [P_n]_{L,N}) | n \in [0, \infty[\} \qquad (55)$$

Since $[\widetilde{C}]_{N,N}$ and $[\widetilde{\widetilde{C}}]_N$ consist only of the elements 1 and 0, the iterative formation of $[P_n]_{L,N}$ requires the multiplication by the kernel not of all the components of the vector $[V_0]_N$, but only of the newest component $v_N$. Thus the ratio of the number of operations in the iterative multiplication using the method of decomposition of the vector into a commutator and kernel to the number of operations for the same multiplication using a method without this decomposition is $C_+ \leq \frac{(N-1)}{(N-1)} = 1$ for addition and $C_* \leq \frac{1 \cdot L}{N} = \frac{L}{N} \leq 1$ for multiplication.

## 14. Example of iterative scalar multiplication or iterative convolution of vectors.

The vector $[T]_N = \begin{bmatrix} 2 \\ 3 \\ 4 \\ 2 \end{bmatrix}$ of length $N = 4$ is to be transposed and iteratively multiplied by the vectors

$$\{[V_1]_N, [V_2]_N, [V_3]_N, [V_4]_N\} = \left\{ \begin{bmatrix} 0 \\ 0 \\ 0 \\ 5 \end{bmatrix}, \begin{bmatrix} 0 \\ 0 \\ 5 \\ 6 \end{bmatrix}, \begin{bmatrix} 0 \\ 5 \\ 6 \\ 7 \end{bmatrix}, \begin{bmatrix} 5 \\ 6 \\ 7 \\ 8 \end{bmatrix} \right\}.$$

The vector $[T]_N$, containing $L = 3$ distinct nonzero elements, is represented as the product of the commutator

$[Z]_{N,L} = \begin{bmatrix} 1 & 0 & 0 \\ 0 & 1 & 0 \\ 0 & 0 & 1 \\ 1 & 0 & 0 \end{bmatrix}$ and the kernel $[U]_L = \begin{bmatrix} 2 \\ 3 \\ 4 \end{bmatrix}$ :

$$[T]_N = [Z]_{N,L} \cdot [U]_L = \begin{bmatrix} z_{1,1} & \cdots & z_{1,L} \\ \vdots & z_{n,l} & \vdots \\ z_{N,1} & \cdots & z_{N,L} \end{bmatrix} \cdot \begin{bmatrix} u_1 \\ \cdots \\ u_l \\ \cdots \\ u_L \end{bmatrix} = \begin{bmatrix} 1 & 0 & 0 \\ 0 & 1 & 0 \\ 0 & 0 & 1 \\ 1 & 0 & 0 \end{bmatrix} \cdot \begin{bmatrix} 2 \\ 3 \\ 4 \end{bmatrix} = \begin{bmatrix} 2 \\ 3 \\ 4 \\ 2 \end{bmatrix}.$$

For N=4, the matrix $[\widetilde{C}]_{N,N} = \begin{bmatrix} 0 & 0 & 0 & 0 \\ 1 & 0 & 0 & 0 \\ 0 & 1 & 0 & 0 \\ 0 & 0 & 1 & 0 \end{bmatrix}$ and the vector $[\widetilde{\widetilde{C}}]_N = [0 \; 0 \; 0 \; 1]$.

The matrix $[P_0]_{L,N} = [\mathbf{0}]_{L,N} = \begin{bmatrix} 0 & 0 & 0 & 0 \\ 0 & 0 & 0 & 0 \\ 0 & 0 & 0 & 0 \end{bmatrix}$ since the condition $[V_n]_N = [\mathbf{0}]_N \;|\; \forall\, n < 1$ is fulfilled.

The process of iterative multiplication takes place as shown below.

Iteration 1:

$$[P_1]_{L,N} = [P_0]_{L,N} \cdot [\widetilde{C}]_{N,N} + [U]_L \cdot v_1 \cdot [\widetilde{\widetilde{C}}]_N = \begin{bmatrix} 0 & 0 & 0 & 0 \\ 0 & 0 & 0 & 0 \\ 0 & 0 & 0 & 0 \end{bmatrix} \cdot \begin{bmatrix} 0 & 0 & 0 & 0 \\ 1 & 0 & 0 & 0 \\ 0 & 1 & 0 & 0 \\ 0 & 0 & 1 & 0 \end{bmatrix} + \begin{bmatrix} 2 \\ 3 \\ 4 \end{bmatrix} \cdot 5 \cdot [0 \; 0 \; 0 \; 1] = \begin{bmatrix} 0 & 0 & 0 & 10 \\ 0 & 0 & 0 & 15 \\ 0 & 0 & 0 & 20 \end{bmatrix}$$

$$r_1 = \mathrm{tr}([Z]_{N,L} \cdot [P_1]_{L,N}) = \mathrm{tr}\left( \begin{bmatrix} 1 & 0 & 0 \\ 0 & 1 & 0 \\ 0 & 0 & 1 \\ 1 & 0 & 0 \end{bmatrix} \cdot \begin{bmatrix} 0 & 0 & 0 & 10 \\ 0 & 0 & 0 & 15 \\ 0 & 0 & 0 & 20 \end{bmatrix} \right) = \mathrm{tr}\left( \begin{bmatrix} 0 & 0 & 0 & 10 \\ 0 & 0 & 0 & 15 \\ 0 & 0 & 0 & 20 \\ 0 & 0 & 0 & 10 \end{bmatrix} \right) = 0 + 0 + 0 + 10 = 10$$

Iteration 2:

$$[P_2]_{L,N} = [P_1]_{L,N} \cdot [\widetilde{C}]_{N,N} + [U]_L \cdot v_2 \cdot [\widetilde{\widetilde{C}}]_N = \begin{bmatrix} 0 & 0 & 0 & 10 \\ 0 & 0 & 0 & 15 \\ 0 & 0 & 0 & 20 \end{bmatrix} \cdot \begin{bmatrix} 0 & 0 & 0 & 0 \\ 1 & 0 & 0 & 0 \\ 0 & 1 & 0 & 0 \\ 0 & 0 & 1 & 0 \end{bmatrix} + \begin{bmatrix} 2 \\ 3 \\ 4 \end{bmatrix} \cdot 6 \cdot [0 \; 0 \; 0 \; 1] = \begin{bmatrix} 0 & 0 & 10 & 12 \\ 0 & 0 & 15 & 18 \\ 0 & 0 & 20 & 24 \end{bmatrix}$$

$$r_2 = \mathrm{tr}([Z]_{N,L} \cdot [P_2]_{L,N}) = \mathrm{tr}\left( \begin{bmatrix} 1 & 0 & 0 \\ 0 & 1 & 0 \\ 0 & 0 & 1 \\ 1 & 0 & 0 \end{bmatrix} \cdot \begin{bmatrix} 0 & 0 & 10 & 12 \\ 0 & 0 & 15 & 18 \\ 0 & 0 & 20 & 24 \end{bmatrix} \right) = \mathrm{tr}\left( \begin{bmatrix} 0 & 0 & 10 & 12 \\ 0 & 0 & 15 & 18 \\ 0 & 0 & 20 & 24 \\ 0 & 0 & 10 & 12 \end{bmatrix} \right) = 0 + 0 + 20 + 12 = 32$$

Iteration 3:

$$[P_3]_{L,N} = [P_2]_{L,N} \cdot [\widetilde{C}]_{N,N} + [U]_L \cdot v_3 \cdot [\widetilde{\widetilde{C}}]_N = \begin{bmatrix} 0 & 0 & 10 & 12 \\ 0 & 0 & 15 & 18 \\ 0 & 0 & 20 & 24 \end{bmatrix} \cdot \begin{bmatrix} 0 & 0 & 0 & 0 \\ 1 & 0 & 0 & 0 \\ 0 & 1 & 0 & 0 \\ 0 & 0 & 1 & 0 \end{bmatrix} + \begin{bmatrix} 2 \\ 3 \\ 4 \end{bmatrix} \cdot 7 \cdot [0\ 0\ 0\ 1] = \begin{bmatrix} 0 & 10 & 12 & 14 \\ 0 & 15 & 18 & 21 \\ 0 & 20 & 24 & 28 \end{bmatrix}$$

$$r_3 = \mathrm{tr}([Z]_{N,L} \cdot [P_3]_{L,N}) = \mathrm{tr}\left(\begin{bmatrix} 1 & 0 & 0 \\ 0 & 1 & 0 \\ 0 & 0 & 1 \\ 1 & 0 & 0 \end{bmatrix} \cdot \begin{bmatrix} 0 & 10 & 12 & 14 \\ 0 & 15 & 18 & 21 \\ 0 & 20 & 24 & 28 \end{bmatrix}\right) = \mathrm{tr}\left(\begin{bmatrix} 0 & 10 & 12 & 14 \\ 0 & 15 & 18 & 21 \\ 0 & 20 & 24 & 28 \\ 0 & 10 & 12 & 14 \end{bmatrix}\right) = 0 + 15 + 24 + 14 = 53$$

Iteration 4:

$$[P_4]_{L,N} = [P_3]_{L,N} \cdot [\widetilde{C}]_{N,N} + [U]_L \cdot v_4 \cdot [\widetilde{\widetilde{C}}]_N = \begin{bmatrix} 0 & 10 & 12 & 14 \\ 0 & 15 & 18 & 21 \\ 0 & 20 & 24 & 28 \end{bmatrix} \cdot \begin{bmatrix} 0 & 0 & 0 & 0 \\ 1 & 0 & 0 & 0 \\ 0 & 1 & 0 & 0 \\ 0 & 0 & 1 & 0 \end{bmatrix} + \begin{bmatrix} 2 \\ 3 \\ 4 \end{bmatrix} \cdot 8 \cdot [0\ 0\ 0\ 1] = \begin{bmatrix} 10 & 12 & 14 & 16 \\ 15 & 18 & 21 & 24 \\ 20 & 24 & 28 & 32 \end{bmatrix}$$

$$r_4 = \mathrm{tr}([Z]_{N,L} \cdot [P_4]_{L,N}) = \mathrm{tr}\left(\begin{bmatrix} 1 & 0 & 0 \\ 0 & 1 & 0 \\ 0 & 0 & 1 \\ 1 & 0 & 0 \end{bmatrix} \cdot \begin{bmatrix} 10 & 12 & 14 & 16 \\ 15 & 18 & 21 & 24 \\ 20 & 24 & 28 & 32 \end{bmatrix}\right) = \mathrm{tr}\left(\begin{bmatrix} 10 & 12 & 14 & 16 \\ 15 & 18 & 21 & 24 \\ 20 & 24 & 28 & 32 \\ 10 & 12 & 14 & 16 \end{bmatrix}\right) = 10 + 18 + 28 + 16 = 72$$

the computation of all N = 4 vector convolutions requires, at each iteration, the multiplication of the kernel by only one new element т $v_n$, which necessitates $L = 3$ multiplication operations. Thus, in the given example, the ratio of the number of operations in the iterative multiplication using the method of decomposition of the vector into a commutator and kernel to the number of operations for the same multiplication using a method without this decomposition is $C_+ = \frac{(N-1)}{(N-1)} = \frac{(4-1)}{(4-1)} = 1$ for addition and $C_* = \frac{L}{N} = \frac{3}{4} = 0.75$ for multiplication.

## 15. Definition of method and process for the multiplication of a vector by a factored matrix.

The matrix

$$[T]_{M,N} = \begin{bmatrix} t_{1,1} & \cdots & t_{1,N} \\ \vdots & t_{m,n} & \vdots \\ t_{M,1} & \cdots & t_{M,N} \end{bmatrix} \tag{56}$$

of dimension M x $N$ containing $L \leq M \cdot N$ distinct nonzero elements, is to be multiplied by the vector $[V]_N = \begin{bmatrix} v_1 \\ \cdots \\ v_n \\ \cdots \\ v_N \end{bmatrix}$

. $\hspace{10cm}$ (57)

The matrix $[T]_{M,N}$ is represented by the product of the commutator $[Z]_{M,N,L} = \{ Z_{m,n,l} \mid m \in [1,M], n \in [1,N], l \in [1,L]\}$ and the kernel $[U]_L = \begin{bmatrix} u_1 \\ \dots \\ u_l \\ \dots \\ u_L \end{bmatrix}$:

$$[T]_{M,N} = [Z]_{M,N,L} \cdot [U]_L = \left\{ \sum_{l=1}^{l=L} z_{m,n,l} \cdot u_l \mid m \in [1,M], n \in [1,N] \right\} \qquad (58)$$

Then the product of the matrix $[T]_{M,N}$ by the vector $[V]_N$ may be written as:

$$[R]_N = [T]_{M,N} \cdot [V]_N = ([Z]_{M,N,L} \cdot [U]_L) \cdot [V]_N = \begin{bmatrix} \sum_{n=1}^{N} v_n \cdot \sum_{l=1}^{L} z_{1,n,l} \cdot u_l \\ \vdots \\ \sum_{n=1}^{N} v_n \cdot \sum_{l=1}^{L} z_{m,n,l} \cdot u_l \\ \vdots \\ \sum_{n=1}^{N} v_n \cdot \sum_{l=1}^{L} z_{M,n,l} \cdot u_l \end{bmatrix}_M = \begin{bmatrix} \sum_{n=1}^{N} \left(\sum_{l=1}^{L} z_{1,n,l} \cdot u_l\right) \cdot v_n \\ \vdots \\ \sum_{n=1}^{N} \left(\sum_{l=1}^{L} z_{m,n,l} \cdot u_l\right) \cdot v_n \\ \vdots \\ \sum_{n=1}^{N} \left(\sum_{l=1}^{L} z_{M,n,l} \cdot u_l\right) \cdot v_n \end{bmatrix}_M =$$

$$\begin{bmatrix} \sum_{n=1}^{N} \sum_{l=1}^{L} z_{1,n,l} \cdot u_l \cdot v_n \\ \vdots \\ \sum_{n=1}^{N} \sum_{l=1}^{L} z_{m,n,l} \cdot u_l \cdot v_n \\ \vdots \\ \sum_{n=1}^{N} \sum_{l=1}^{L} z_{M,n,l} \cdot u_l \cdot v_n \end{bmatrix}_M = \begin{bmatrix} \sum_{n=1}^{N} \sum_{l=1}^{L} z_{1,n,l} \cdot (u_l \cdot v_n) \\ \vdots \\ \sum_{n=1}^{N} \sum_{l=1}^{L} z_{m,n,l} \cdot (u_l \cdot v_n) \\ \vdots \\ \sum_{n=1}^{N} \sum_{l=1}^{L} z_{M,n,l} \cdot (u_l \cdot v_n) \end{bmatrix}_M \qquad (59)$$

In the above expression, under each nested sum the same coefficient $(u_l \cdot v_n)$ appears, which is an element of the matrix formed by multiplying the vector $[U]_L$ by the transposed vector $[V]_N$:

$$[P]_{L,N} = [U]_L \cdot [V]_N^t \qquad (60)$$

Taking this into account, the product of the matrix $[T]_{M,N}$ and the vector $[V]_N$ may be written as:

$$[R]_N = [T]_{M,N} \cdot [V]_N = \begin{bmatrix} \sum_{n=1}^{N} \sum_{l=1}^{L} z_{1,n,l} \cdot (u_l \cdot v_n) \\ \vdots \\ \sum_{n=1}^{N} \sum_{l=1}^{L} z_{m,n,l} \cdot (u_l \cdot v_n) \\ \vdots \\ \sum_{n=1}^{N} \sum_{l=1}^{L} z_{M,n,l} \cdot (u_l \cdot v_n) \end{bmatrix}_M = \begin{bmatrix} \sum_{n=1}^{N} \sum_{l=1}^{L} z_{1,n,l} \cdot p_{l,n} \\ \vdots \\ \sum_{n=1}^{N} \sum_{l=1}^{L} z_{m,n,l} \cdot p_{l,n} \\ \vdots \\ \sum_{n=1}^{N} \sum_{l=1}^{L} z_{M,n,l} \cdot p_{l,n} \end{bmatrix}_M =$$

$$\begin{bmatrix} \sum_{n=1}^{N}[z_{1,n,1},\dots,z_{1,n,l},\dots,z_{1,n,L}] \cdot \begin{bmatrix} p_{1,n} \\ \dots \\ p_{l,n} \\ \dots \\ p_{L,n} \end{bmatrix} \\ \vdots \\ \sum_{n=1}^{N}[z_{m,n,1},\dots,z_{m,n,l},\dots,z_{m,n,L}] \cdot \begin{bmatrix} p_{1,n} \\ \dots \\ p_{l,n} \\ \dots \\ p_{L,n} \end{bmatrix} \\ \vdots \\ \sum_{n=1}^{N}[z_{M,n,1},\dots,z_{M,n,l},\dots,z_{M,n,L}] \cdot \begin{bmatrix} p_{1,n} \\ \dots \\ p_{l,n} \\ \dots \\ p_{L,n} \end{bmatrix} \end{bmatrix}_M = \begin{bmatrix} \mathrm{tr}\left( \begin{bmatrix} z_{1,1,1} & \cdots & z_{1,1,L} \\ \vdots & z_{m,n,l} & \vdots \\ z_{1,N,1} & \cdots & z_{1,N,L} \end{bmatrix} \cdot [P]_{L,N} \right) \\ \vdots \\ \mathrm{tr}\left( \begin{bmatrix} z_{m,1,1} & \cdots & z_{m,1,L} \\ \vdots & z_{m,n,l} & \vdots \\ z_{m,N,1} & \cdots & z_{m,N,L} \end{bmatrix} \cdot [P]_{L,N} \right) \\ \vdots \\ \mathrm{tr}\left( \begin{bmatrix} z_{M,1,1} & \cdots & z_{M,1,L} \\ \vdots & z_{m,n,l} & \vdots \\ z_{M,N,1} & \cdots & z_{M,N,L} \end{bmatrix} \cdot [P]_{L,N} \right) \end{bmatrix}_M =$$

$$\begin{bmatrix} \mathrm{tr}\left( ([Z]_{M,N,L} \cdot [E_1]_M) \cdot [P]_{L,N} \right) \\ \vdots \\ \mathrm{tr}\left( ([Z]_{M,N,L} \cdot [E_m]_M) \cdot [P]_{L,N} \right) \\ \vdots \\ \mathrm{tr}\left( ([Z]_{M,N,L} \cdot [E_M]_M) \cdot [P]_{L,N} \right) \end{bmatrix}_M , \tag{61}$$

where

$$[E_m]_M = \begin{bmatrix} \begin{bmatrix} 0 \\ \vdots \\ 0 \end{bmatrix}_{m-1} \\ 1 \\ \begin{bmatrix} 0 \\ \vdots \\ 0 \end{bmatrix}_{M-m} \end{bmatrix}_M \tag{62}$$

is the selecting vector, necessarily containing exactly one unit element in the position 0<m<M+1 and all other elements of which must be 0. This vector has the property that the result of its multiplication by the commutator $[Z]_{M,N,L}$ is a matrix of dimension $(N, L)$ consisting of all the elements of the commutator with the first index equal to m:

$$[Z]_{M,N,L} \cdot [E_m]_M = \{z_{k,n,l} \mid k \in [1:M], n \in [1:N], l \in [1:L]\} \cdot \begin{bmatrix} \begin{bmatrix} 0 \\ \vdots \\ 0 \end{bmatrix}_{m-1} \\ 1 \\ \begin{bmatrix} 0 \\ \vdots \\ 0 \end{bmatrix}_{M-m} \end{bmatrix}_M = \{z_{k,n,l} \mid k = m, n \in [1:N], l \in [1:L] =$$

$$\begin{bmatrix} z_{m,1,1} & \cdots & z_{m,1,L} \\ \vdots & z_{m,n,l} & \vdots \\ z_{m,N,1} & \cdots & z_{m,N,L} \end{bmatrix} \} \tag{63}$$

Thus, the multiplication of a matrix by a vector of length N may be carried out in two stages. First, the matrix is obtained which contains the products of each element of the original vector with each element of the kernel of the original matrix. Second, Then each element of the resulting vector is computed as the trace of the matrix which is the product of the matrix obtained in the first step and the corresponding sub-tensor of the commutator, which is in turn obtained by the contraction of the commutator with an auxiliary vector containing a single unit element and L-1 zero elements. Thus all operations of multiplication take place in the first step, and their maximum number is equal not to the number M of rows of the original matrix multiplied by the number of columns N, as in the case of multiplication using a method that does not include the factorization of the matrix, but to the product of the length N of the original vector by the number L of distinct nonzero elements of the original matrix. All operations of addition are carried out in the second step, and their maximum number is $(M \cdot (N - 1))$. Thus the ratio of the number of operations using the method including the decomposition of the matrix into commutator and kernel to the number of operations in a method not including this decomposition is $Cm_+ \leq \frac{M \cdot (N-1)}{M \cdot (N-1)} = 1$ for addition and $Cm_* \leq \frac{N \cdot L}{M \cdot N} = \frac{L}{M}$ for multiplication.

## 16. Example of multiplication of a vector by a factored matrix.

The matrix $[T]_{M,N} = \begin{bmatrix} t_{1,1} & t_{1,2} & t_{1,3} \\ t_{2,1} & t_{2,2} & t_{2,3} \\ t_{3,1} & t_{3,2} & t_{3,3} \\ t_{4,1} & t_{4,2} & t_{4,3} \end{bmatrix} = \begin{bmatrix} 0 & 2 & 3 \\ 3 & 2 & 0 \\ 2 & 3 & 0 \\ 2 & 0 & 3 \end{bmatrix}$ of dimension M x $N = 4$ x 3 containing $L = 2$ distinct nonzero elements is to be multiplied by the vector

$[V]_N = \begin{bmatrix} v_1 \\ v_2 \\ v_3 \end{bmatrix} = \begin{bmatrix} 2 \\ 3 \\ 4 \end{bmatrix}$.

The matrix $[T]_{M,N}$ is represented by the tensor contraction of the commutator

$[Z]_{M,N,L} = \{ Z_{m,n,l} \mid m \in [1, M], n \in [1, N], l \in [1, L] \} = \begin{bmatrix} [0\ 0] & [1\ 0] & [0\ 1] \\ [0\ 1] & [1\ 0] & [0\ 0] \\ [1\ 0] & [0\ 1] & [0\ 0] \\ [1\ 0] & [0\ 0] & [0\ 1] \end{bmatrix}$

with the kernel $[U]_L = \begin{bmatrix} u_1 \\ u_2 \end{bmatrix} = \begin{bmatrix} 2 \\ 3 \end{bmatrix}$:

$[T]_{M,N} = [Z]_{M,N,L} \cdot [U]_L = \left\{ \sum_{l=1}^{l=L} z_{m,n,l} \cdot u_l \mid m \in [1, M], n \in [1, N] \right\} = \begin{bmatrix} [0\ 0] \cdot \begin{bmatrix} 2 \\ 3 \end{bmatrix} & [0\ 1] \cdot \begin{bmatrix} 2 \\ 3 \end{bmatrix} & [0\ 1] \cdot \begin{bmatrix} 2 \\ 3 \end{bmatrix} \\ [0\ 1] \cdot \begin{bmatrix} 2 \\ 3 \end{bmatrix} & [0\ 1] \cdot \begin{bmatrix} 2 \\ 3 \end{bmatrix} & [0\ 0] \cdot \begin{bmatrix} 2 \\ 3 \end{bmatrix} \\ [0\ 1] \cdot \begin{bmatrix} 2 \\ 3 \end{bmatrix} & [0\ 1] \cdot \begin{bmatrix} 2 \\ 3 \end{bmatrix} & [0\ 0] \cdot \begin{bmatrix} 2 \\ 3 \end{bmatrix} \\ [0\ 1] \cdot \begin{bmatrix} 2 \\ 3 \end{bmatrix} & [0\ 0] \cdot \begin{bmatrix} 2 \\ 3 \end{bmatrix} & [0\ 1] \cdot \begin{bmatrix} 2 \\ 3 \end{bmatrix} \end{bmatrix}$

$= \begin{bmatrix} [0\ 0] & [1\ 0] & [0\ 1] \\ [0\ 1] & [1\ 0] & [0\ 0] \\ [1\ 0] & [0\ 1] & [0\ 0] \\ [1\ 0] & [0\ 0] & [0\ 1] \end{bmatrix} \cdot \begin{bmatrix} 2 \\ 3 \end{bmatrix} = \begin{bmatrix} 0 & 3 & 2 \\ 3 & 2 & 0 \\ 2 & 3 & 0 \\ 2 & 0 & 3 \end{bmatrix}$

The selecting vectors are: $[E_1]_M = \begin{bmatrix} 1 \\ 0 \\ 0 \\ 0 \end{bmatrix}$, $[E_2]_M = \begin{bmatrix} 0 \\ 1 \\ 0 \\ 0 \end{bmatrix}$, $[E_3]_M = \begin{bmatrix} 0 \\ 0 \\ 1 \\ 0 \end{bmatrix}$ и $[E_4]_M = \begin{bmatrix} 0 \\ 0 \\ 0 \\ 1 \end{bmatrix}$.

The corresponding tensor contraction of the commutator $[Z]_{M,N,L}$ are:

$$[Z]_{M,N,L} \cdot [E_1]_M = \begin{bmatrix} ([0\ 0]\ [1\ 0]\ [0\ 1]) \\ ([0\ 1]\ [1\ 0]\ [0\ 0]) \\ ([1\ 0]\ [0\ 1]\ [0\ 0]) \\ ([1\ 0]\ [0\ 0]\ [0\ 1]) \end{bmatrix}^t \cdot \begin{bmatrix} 1 \\ 0 \\ 0 \\ 0 \end{bmatrix} = \begin{bmatrix} [0\ 0] \\ [1\ 0] \\ [0\ 1] \end{bmatrix} = \begin{bmatrix} 0\ 0 \\ 1\ 0 \\ 0\ 1 \end{bmatrix}$$

$$[Z]_{M,N,L} \cdot [E_2]_M = \begin{bmatrix} ([0\ 0]\ [1\ 0]\ [0\ 1]) \\ ([0\ 1]\ [1\ 0]\ [0\ 0]) \\ ([1\ 0]\ [0\ 1]\ [0\ 0]) \\ ([1\ 0]\ [0\ 0]\ [0\ 1]) \end{bmatrix}^t \cdot \begin{bmatrix} 0 \\ 1 \\ 0 \\ 0 \end{bmatrix} = \begin{bmatrix} [0\ 1] \\ [1\ 0] \\ [0\ 0] \end{bmatrix} = \begin{bmatrix} 0\ 1 \\ 1\ 0 \\ 0\ 0 \end{bmatrix}$$

$$[Z]_{M,N,L} \cdot [E_3]_M = \begin{bmatrix} ([0\ 0]\ [1\ 0]\ [0\ 1]) \\ ([0\ 1]\ [1\ 0]\ [0\ 0]) \\ ([1\ 0]\ [0\ 1]\ [0\ 0]) \\ ([1\ 0]\ [0\ 0]\ [0\ 1]) \end{bmatrix}^t \cdot \begin{bmatrix} 0 \\ 0 \\ 1 \\ 0 \end{bmatrix} = \begin{bmatrix} [1\ 0] \\ [0\ 1] \\ [0\ 0] \end{bmatrix} = \begin{bmatrix} 1\ 0 \\ 0\ 1 \\ 0\ 0 \end{bmatrix}$$

$$[Z]_{M,N,L} \cdot [E_4]_M = \begin{bmatrix} ([0\ 0]\ [1\ 0]\ [0\ 1]) \\ ([0\ 1]\ [1\ 0]\ [0\ 0]) \\ ([1\ 0]\ [0\ 1]\ [0\ 0]) \\ ([1\ 0]\ [0\ 0]\ [0\ 1]) \end{bmatrix}^t \cdot \begin{bmatrix} 0 \\ 0 \\ 0 \\ 1 \end{bmatrix} = \begin{bmatrix} [1\ 0] \\ [0\ 0] \\ [0\ 1] \end{bmatrix} = \begin{bmatrix} 1\ 0 \\ 0\ 0 \\ 0\ 1 \end{bmatrix}$$

The product of the matrix $[T]_{M,N}$ by the vector $[V]_N$ is the matrix $[P]_{L,N}$, equal to:

$$[P]_{L,N} = [U]_L \cdot [V]_N^t = \begin{bmatrix} 2 \\ 3 \end{bmatrix} \cdot [2\ 3\ 4] = \begin{bmatrix} 2\cdot 2 & 2\cdot 3 & 2\cdot 4 \\ 3\cdot 2 & 3\cdot 3 & 3\cdot 4 \end{bmatrix} = \begin{bmatrix} 4 & 6 & 8 \\ 6 & 9 & 12 \end{bmatrix}$$

Thus we can write the product of the matrix $[T]_{M,N}$ by the vector $[V]_N$:

$$[R]_N = [T]_{M,N} \cdot [V]_N = \begin{bmatrix} \mathrm{tr}\big(([Z]_{M,N,L} \cdot [E_1]_M) \cdot [P]_{L,N}\big) \\ \mathrm{tr}\big(([Z]_{M,N,L} \cdot [E_2]_M) \cdot [P]_{L,N}\big) \\ \mathrm{tr}\big(([Z]_{M,N,L} \cdot [E_3]_M) \cdot [P]_{L,N}\big) \\ \mathrm{tr}\big(([Z]_{M,N,L} \cdot [E_4]_M) \cdot [P]_{L,N}\big) \end{bmatrix} = \begin{bmatrix} \mathrm{tr}\left(\begin{bmatrix}0\ 0\\1\ 0\\0\ 1\end{bmatrix}\cdot\begin{bmatrix}4&6&8\\6&9&12\end{bmatrix}\right) \\ \mathrm{tr}\left(\begin{bmatrix}0\ 1\\1\ 0\\0\ 0\end{bmatrix}\cdot\begin{bmatrix}4&6&8\\6&9&12\end{bmatrix}\right) \\ \mathrm{tr}\left(\begin{bmatrix}1\ 0\\0\ 1\\0\ 0\end{bmatrix}\cdot\begin{bmatrix}4&6&8\\6&9&12\end{bmatrix}\right) \\ \mathrm{tr}\left(\begin{bmatrix}1\ 0\\0\ 0\\0\ 1\end{bmatrix}\cdot\begin{bmatrix}4&6&8\\6&9&12\end{bmatrix}\right) \end{bmatrix} = \begin{bmatrix} \mathrm{tr}\left(\begin{bmatrix}0&0&0\\4&6&8\\6&9&12\end{bmatrix}\right) \\ \mathrm{tr}\left(\begin{bmatrix}6&9&12\\4&6&8\\0&0&0\end{bmatrix}\right) \\ \mathrm{tr}\left(\begin{bmatrix}4&6&8\\6&9&12\\0&0&0\end{bmatrix}\right) \\ \mathrm{tr}\left(\begin{bmatrix}4&6&8\\0&0&0\\6&9&12\end{bmatrix}\right) \end{bmatrix}$$

$$= \begin{bmatrix} 0+6+12 \\ 6+6+0 \\ 4+9+0 \\ 4+0+12 \end{bmatrix} = \begin{bmatrix} 18 \\ 12 \\ 13 \\ 16 \end{bmatrix}$$

In this example the multiplication of a matrix with dimensions M=4 x N=3 by a vector of length N=3 required L x N =2 x 3 = 6 multiplication operations. Thus, in the given example the ratio of the number of operations using a method including decomposition into commutator and kernel to the number of operations using a method that does not include this decomposition is $Cm_+ = \frac{M \cdot (N-1)}{M \cdot (N-1)} = \frac{4 \cdot (3-1)}{4 \cdot (3-1)} = 1$ for addition and $Cm_* = \frac{L}{M} = \frac{2}{4} = 0.5$ for multiplication.

## 17. Definition of method and process for recursive multiplication of a vector by a factored matrix.

The matrix

$$[T]_{M,N} = \begin{bmatrix} t_{1,1} & \cdots & t_{1,N} \\ \vdots & t_{m,n} & \vdots \\ t_{M,1} & \cdots & t_{M,N} \end{bmatrix} \tag{64}$$

of dimensions M x $N$ containing $L \leq M \cdot N$ distinct nonzero elements is to be multiplied by the vector

$$[V]_N = [V_0]_N = \begin{bmatrix} v_1 \\ \cdots \\ v_n \\ \cdots \\ v_N \end{bmatrix} \tag{65}$$

and all its circularly-shifted variants:

$$\{[V_1]_N, [V_2]_N, \ldots, [V_{N-1}]_N\} = \left\{ \begin{bmatrix} v_2 \\ \cdots \\ \cdots \\ v_N \\ v_1 \end{bmatrix}, \begin{bmatrix} v_3 \\ \cdots \\ \cdots \\ v_1 \\ v_2 \end{bmatrix}, \ldots, \begin{bmatrix} v_N \\ v_1 \\ \cdots \\ \cdots \\ v_{N-1} \end{bmatrix} \right\}. \tag{66}$$

The matrix $[T]_{M,N}$ is represented as the product of the commutator $[Z]_{M,N,L} = \{ Z_{m,n,l} \mid m \in [1,M], n \in [1,N], l \in [1,L]\}$ by the kernel $[U]_L = \begin{bmatrix} u_1 \\ \cdots \\ u_l \\ \cdots \\ u_L \end{bmatrix}$:

$$[T]_{M,N} = [Z]_{M,N,L} \cdot [U]_L = \left\{ \sum_{l=1}^{l=L} z_{m,n,l} \cdot u_l \mid m \in [1,M], n \in [1,N] \right\} \tag{67}$$

First it is necessary to compute the product of multiplication of the matrix $[T]_{M,N}$ by the vector $[V]_N$. This multiplication may be written as:

$$[R]_N = [T]_{M,N} \cdot [V]_N = \begin{bmatrix} \text{tr}\left(([Z]_{M,N,L} \cdot [E_1]_M) \cdot [P]_{L,N}\right) \\ \vdots \\ \text{tr}\left(([Z]_{M,N,L} \cdot [E_m]_M) \cdot [P]_{L,N}\right) \\ \vdots \\ \text{tr}\left(([Z]_{M,N,L} \cdot [E_M]_M) \cdot [P]_{L,N}\right) \end{bmatrix}_M \tag{68}$$

, where

$$[E_m]_M = \left\{ \begin{bmatrix} \begin{bmatrix} 0 \\ \vdots \\ 0 \end{bmatrix}_{m-1} \\ 1 \\ \begin{bmatrix} 0 \\ \vdots \\ 0 \end{bmatrix}_{M-m} \end{bmatrix}_M \right\} \tag{69}$$

are selecting vectors, and the matrix $[P]_{L,N}$ is obtained from the multiplication of the kernel $[U]_L$ by the transposed vector $[V]_N$:

$$[P]_{L,N} = [U]_L \cdot [V]_N^t = \begin{bmatrix} u_1 \\ \ldots \\ u_l \\ \ldots \\ u_L \end{bmatrix} \cdot [v_1 \ldots v_n \ldots v_N] = \begin{bmatrix} v_1 \cdot u_1 & \cdots & v_N \cdot u_1 \\ \vdots & v_n \cdot u_l & \vdots \\ v_1 \cdot u_L & \cdots & v_N \cdot u_L \end{bmatrix} \tag{70}$$

To obtain the subsequent value, the result of the multiplication of the matrix $[T]_{M,N}$ by the first shifted variant of the vector $[V]_N$, which is the vector

$$[V_1]_N = \begin{bmatrix} v_2 \\ \ldots \\ \ldots \\ v_N \\ v_1 \end{bmatrix}, \tag{71}$$

it is necessary to obtain the new matrix $[P_1]_{L,N}$:

$$[P_1]_{L,N} = [U]_L \cdot [V_1]_N^t = \begin{bmatrix} u_1 \\ \ldots \\ u_l \\ \ldots \\ u_L \end{bmatrix} \cdot [v_2 \ldots v_{n+1} \ldots v_N \ v_1] = \begin{bmatrix} v_2 \cdot u_1 & \cdots & v_N \cdot u_1 & v_1 \cdot u_1 \\ \vdots & & \vdots & \vdots \\ v_1 \cdot u_L & \cdots & v_N \cdot u_L & v_1 \cdot u_L \end{bmatrix}. \tag{72}$$

Clearly the matrix $[P_1]_{L,N}$ is equivalent to the matrix $[P]_{L,N}$ cyclically shifted to the left by one position. Thus, the matrix $[P_1]_{L,N}$ is obtained from the matrix $[P]_{L,N}$ via multiplication by the displacement matrix

$$[C_1]_{N,N} = \begin{bmatrix} 0 & 0 & 0 & \cdots & 0 & 0 & 1 \\ 1 & 0 & 0 & \cdots & 0 & 0 & 0 \\ 0 & 1 & 0 & \cdots & 0 & 0 & 0 \\ & & & \vdots & & & \\ 0 & 0 & 0 & \cdots & 0 & 1 & 0 \end{bmatrix} \tag{73}$$

of dimension N x N , which is equivalent to a diagonal matrix circularly shifted to the left by one position.

$$[P_1]_{L,N} = [P]_{L,N} \cdot [C_1]_{N,N} = \begin{bmatrix} v_1 \cdot u_1 & \cdots & v_N \cdot u_1 \\ \vdots & v_n \cdot u_l & \vdots \\ v_1 \cdot u_L & \cdots & v_N \cdot u_L \end{bmatrix} \cdot \begin{bmatrix} 0 & 0 & 0 & \cdots & 0 & 0 & 1 \\ 1 & 0 & 0 & \cdots & 0 & 0 & 0 \\ 0 & 1 & 0 & \cdots & 0 & 0 & 0 \\ & & & \vdots & & & \\ 0 & 0 & 0 & \cdots & 0 & 1 & 0 \end{bmatrix} \tag{74}$$

To obtain the remaining results of multiplying the transposed vector $[T]_N$ by the second and subsequent shifted variants of the vector $[V]_N$, which are the vectors $[V_2]_N, [V_3]_N, \ldots, [V_{N-1}]_N$, it is necessary to obtain the matrices $[P_2]_{L,N}, [P_3]_{L,N}, \ldots, [P_{N-1}]_{L,N}$. Each successive matrix $[P_n]_{L,N}$ is obtained by cyclically shifting the matrix $[P_{n-1}]_{L,N}$ by one position to the left, or equivalently by multiplying the matrix $[P_{n-1}]_{L,N}$ by the displacement matrix $[C_1]_{N,N}$, or by multiplying the matrix $[P]_{L,N}$ by the displacement matrix $[C_n]_{N,N}$ which is the n$^{th}$ power of the matrix $[C_n]_{N,N}$:

$$[P_n]_{L,N} = [U]_L \cdot [V_n]_N^t = [P_{n-1}]_{L,N} \cdot [C_1]_{N,N} = [P_{n-k}]_{L,N} \cdot [C_1]_{N,N}^k = [P]_{L,N} \cdot [C_1]_{N,N}^n = [P]_{L,N} \cdot [C_n]_{N,N}, \tag{75}$$

where $n, k \in [0, N-1]$. For the sake of generality we take

$$[V_0]_N = [V]_N \tag{76}$$

and

$$[C_1]_{N,N}^n = [C]_{N,N}^n. \tag{77}$$

Thus, the result of the cyclic multiplication may be represented by the matrix

$$[R]_{M,N} = \begin{bmatrix} r_{1,1} & \cdots & r_{1,N} \\ \vdots & r_{m,n} & \vdots \\ r_{M,1} & \cdots & r_{M,N} \end{bmatrix} = \begin{bmatrix} [T]_{M,N} \cdot [V_0]_N \\ \cdots \\ [T]_{M,N} \cdot [V_{n-1}]_N \\ \cdots \\ [T]_{M,N} \cdot [V_{N-1}]_N \end{bmatrix} =$$

$$\begin{bmatrix} \mathrm{tr}\left(([Z]_{M,N,L} \cdot [E_1]_M) \cdot [P_0]_{L,N}\right) & \cdots & \mathrm{tr}\left(([Z]_{M,N,L} \cdot [E_1]_M) \cdot [P_{N-1}]_{L,N}\right) \\ \vdots & \mathrm{tr}\left(([Z]_{M,N,L} \cdot [E_m]_M) \cdot [P_{n-1}]_{L,N}\right) & \vdots \\ \mathrm{tr}\left(([Z]_{M,N,L} \cdot [E_M]_M) \cdot [P_0]_{L,N}\right) & \cdots & \mathrm{tr}\left(([Z]_{M,N,L} \cdot [E_M]_M) \cdot [P_{N-1}]_{L,N}\right) \end{bmatrix} =$$

$$\begin{bmatrix} \mathrm{tr}\left(([Z]_{M,N,L} \cdot [E_1]_M) \cdot [P]_{L,N} \cdot [C]_{N,N}^0\right) & \cdots & \mathrm{tr}\left(([Z]_{M,N,L} \cdot [E_1]_M) \cdot [P]_{L,N} \cdot [C]_{N,N}^{N-1}\right) \\ \vdots & \mathrm{tr}\left(([Z]_{M,N,L} \cdot [E_m]_M) \cdot [P]_{L,N} \cdot [C]_{N,N}^{n-1}\right) & \vdots \\ \mathrm{tr}\left(([Z]_{M,N,L} \cdot [E_M]_M) \cdot [P]_{L,N} \cdot [C]_{N,N}^0\right) & \cdots & \mathrm{tr}\left(([Z]_{M,N,L} \cdot [E_M]_M) \cdot [P]_{L,N} \cdot [C]_{N,N}^{N-1}\right) \end{bmatrix}$$

$$\tag{78}$$

The recursive multiplication of a matrix of length N may be carried out in two steps. First the matrix is obtained which contains the products of each element of the original vector with each element of the kernel of the original matrix. Then each element of the resulting matrix is computed as the trace of the matrix which is the product of the matrix obtained in the first step and the corresponding sub-matrix of the commutator. Thus, all operations of multiplication are carried out in the first step and have a maximum number is the product of the length N of the original vector and the number L of distinct nonzero elements of the original matrix, rather than the product of the number M of rows of the original matrix and the square $N^2$ of the number of columns, as in the case of multiplication without factorization of the matrix. All operations of addition are carried out in the second step, and their maximal number is $(M \cdot (N-1) \cdot N)$. Thus the ratio of the number of operations required by the method including the decomposition of the matrix into commutator and kernel to the number of operations required by a method that does not include this decomposition is $Cm_+ \leq \frac{M \cdot (N-1) \cdot N}{M \cdot (N-1) \cdot N} = 1$ for addition and $Cm_* \leq \frac{N \cdot L}{M \cdot N^2} = \frac{L}{M \cdot N}$ for multiplication.

## 18. Example of recursive multiplication of a vector by a factored matrix.

The matrix $[T]_{M,N} = \begin{bmatrix} t_{1,1} & t_{1,2} & t_{1,3} \\ t_{2,1} & t_{2,2} & t_{2,3} \\ t_{3,1} & t_{3,2} & t_{3,3} \\ t_{4,1} & t_{4,2} & t_{4,3} \end{bmatrix} = \begin{bmatrix} 0 & 2 & 3 \\ 3 & 2 & 0 \\ 2 & 3 & 0 \\ 2 & 0 & 3 \end{bmatrix}$ of dimension M x $N$ = 4 x 3 containing $L$ = 2 distinct nonzero elements is to be recursively multiplied by the vector

$[V]_N = [V_0]_N = \begin{bmatrix} v_1 \\ v_2 \\ v_3 \end{bmatrix} = \begin{bmatrix} 2 \\ 3 \\ 4 \end{bmatrix}$ and all its circularly-shifted variants: $\{[V_1]_N, [V_2]_N\} = \left\{ \begin{bmatrix} 3 \\ 4 \\ 2 \end{bmatrix}, \begin{bmatrix} 4 \\ 2 \\ 3 \end{bmatrix} \right\}$.

The matrix $[T]_{M,N}$ is represented by the tensor contraction of the commutator

$[Z]_{M,N,L} = \{ Z_{m,n,l} \mid m \in [1, M], n \in [1, N], l \in [1, L]\} = \begin{bmatrix} [0\ 0] & [1\ 0] & [0\ 1] \\ [0\ 1] & [1\ 0] & [0\ 0] \\ [1\ 0] & [0\ 1] & [0\ 0] \\ [1\ 0] & [0\ 0] & [0\ 1] \end{bmatrix}$

with the kernel $[U]_L = \begin{bmatrix} u_1 \\ u_2 \end{bmatrix} = \begin{bmatrix} 2 \\ 3 \end{bmatrix}$:

$[T]_{M,N} = [Z]_{M,N,L} \cdot [U]_L = \left\{ \sum_{l=1}^{l=L} z_{m,n,l} \cdot u_l \mid m \in [1, M], n \in [1, N] \right\} = \begin{bmatrix} [0\ 0]\cdot\begin{bmatrix}2\\3\end{bmatrix} & [0\ 1]\cdot\begin{bmatrix}2\\3\end{bmatrix} & [0\ 1]\cdot\begin{bmatrix}2\\3\end{bmatrix} \\ [0\ 1]\cdot\begin{bmatrix}2\\3\end{bmatrix} & [0\ 1]\cdot\begin{bmatrix}2\\3\end{bmatrix} & [0\ 0]\cdot\begin{bmatrix}2\\3\end{bmatrix} \\ [0\ 1]\cdot\begin{bmatrix}2\\3\end{bmatrix} & [0\ 1]\cdot\begin{bmatrix}2\\3\end{bmatrix} & [0\ 0]\cdot\begin{bmatrix}2\\3\end{bmatrix} \\ [0\ 1]\cdot\begin{bmatrix}2\\3\end{bmatrix} & [0\ 0]\cdot\begin{bmatrix}2\\3\end{bmatrix} & [0\ 1]\cdot\begin{bmatrix}2\\3\end{bmatrix} \end{bmatrix}$

$= \begin{bmatrix} [0\ 0] & [1\ 0] & [0\ 1] \\ [0\ 1] & [1\ 0] & [0\ 0] \\ [1\ 0] & [0\ 1] & [0\ 0] \\ [1\ 0] & [0\ 0] & [0\ 1] \end{bmatrix} \cdot \begin{bmatrix} 2 \\ 3 \end{bmatrix} = \begin{bmatrix} 0 & 3 & 2 \\ 3 & 2 & 0 \\ 2 & 3 & 0 \\ 2 & 0 & 3 \end{bmatrix}$

The selecting vectors are: $[E_1]_M = \begin{bmatrix} 1 \\ 0 \\ 0 \\ 0 \end{bmatrix}, [E_2]_M = \begin{bmatrix} 0 \\ 1 \\ 0 \\ 0 \end{bmatrix}, [E_3]_M = \begin{bmatrix} 0 \\ 0 \\ 1 \\ 0 \end{bmatrix}$ и $[E_4]_M = \begin{bmatrix} 0 \\ 0 \\ 0 \\ 1 \end{bmatrix}$.

The corresponding tensor contraction of the commutator $[Z]_{M,N,L}$ equal:

$[Z]_{M,N,L} \cdot [E_1]_M = \begin{bmatrix} ([0\ 0]\ [1\ 0]\ [0\ 1]) \\ ([0\ 1]\ [1\ 0]\ [0\ 0]) \\ ([1\ 0]\ [0\ 1]\ [0\ 0]) \\ ([1\ 0]\ [0\ 0]\ [0\ 1]) \end{bmatrix}^t \cdot \begin{bmatrix} 1 \\ 0 \\ 0 \\ 0 \end{bmatrix} = \begin{bmatrix} [0\ 0] \\ [1\ 0] \\ [0\ 1] \end{bmatrix} = \begin{bmatrix} 0 & 0 \\ 1 & 0 \\ 0 & 1 \end{bmatrix}$

$[Z]_{M,N,L} \cdot [E_2]_M = \begin{bmatrix} ([0\ 0]\ [1\ 0]\ [0\ 1]) \\ ([0\ 1]\ [1\ 0]\ [0\ 0]) \\ ([1\ 0]\ [0\ 1]\ [0\ 0]) \\ ([1\ 0]\ [0\ 0]\ [0\ 1]) \end{bmatrix}^t \cdot \begin{bmatrix} 0 \\ 1 \\ 0 \\ 0 \end{bmatrix} = \begin{bmatrix} [0\ 1] \\ [1\ 0] \\ [0\ 0] \end{bmatrix} = \begin{bmatrix} 0 & 1 \\ 1 & 0 \\ 0 & 0 \end{bmatrix}$

$$[Z]_{M,N,L} \cdot [E_3]_M = \begin{bmatrix} ([0\ 0]\ [1\ 0]\ [0\ 1]) \\ ([0\ 1]\ [1\ 0]\ [0\ 0]) \\ ([1\ 0]\ [0\ 1]\ [0\ 0]) \\ ([1\ 0]\ [0\ 0]\ [0\ 1]) \end{bmatrix}^t \cdot \begin{bmatrix} 0 \\ 0 \\ 1 \\ 0 \end{bmatrix} = \begin{bmatrix} [1\ 0] \\ [0\ 1] \\ [0\ 0] \end{bmatrix} = \begin{bmatrix} 1\ 0 \\ 0\ 1 \\ 0\ 0 \end{bmatrix}$$

$$[Z]_{M,N,L} \cdot [E_4]_M = \begin{bmatrix} ([0\ 0]\ [1\ 0]\ [0\ 1]) \\ ([0\ 1]\ [1\ 0]\ [0\ 0]) \\ ([1\ 0]\ [0\ 1]\ [0\ 0]) \\ ([1\ 0]\ [0\ 0]\ [0\ 1]) \end{bmatrix}^t \cdot \begin{bmatrix} 0 \\ 0 \\ 0 \\ 1 \end{bmatrix} = \begin{bmatrix} [1\ 0] \\ [0\ 0] \\ [0\ 1] \end{bmatrix} = \begin{bmatrix} 1\ 0 \\ 0\ 0 \\ 0\ 1 \end{bmatrix}$$

The product of the matrix $[U]_L [T]_{M,N}$ and the vector $[V]_N$ is the matrix $[P]_{L,N}$, equal to:

$$[P]_{L,N} = [U]_L \cdot [V]_N^t = \begin{bmatrix} 2 \\ 3 \end{bmatrix} \cdot [2\ 3\ 4] = \begin{bmatrix} 2\cdot 2 & 2\cdot 3 & 2\cdot 4 \\ 3\cdot 2 & 3\cdot 3 & 3\cdot 4 \end{bmatrix} = \begin{bmatrix} 4 & 6 & 8 \\ 6 & 9 & 12 \end{bmatrix}$$

The displacement matrix $[C_1]_{N,N} = [C]_{N,N} = \begin{bmatrix} 0 & 0 & 1 \\ 1 & 0 & 0 \\ 0 & 1 & 0 \end{bmatrix}$ of dimensions $N \times N = 3 \times 3$ is obtained, and from it are obtained the other N-1=2 displacement matrices:

$$[C]^0_{N,N} = \begin{bmatrix} 0 & 0 & 1 \\ 1 & 0 & 0 \\ 0 & 1 & 0 \end{bmatrix}^0 = \begin{bmatrix} 1 & 0 & 0 \\ 0 & 1 & 0 \\ 0 & 0 & 1 \end{bmatrix}, [C]^2_{N,N} = \begin{bmatrix} 0 & 0 & 1 \\ 1 & 0 & 0 \\ 0 & 1 & 0 \end{bmatrix}^2 = \begin{bmatrix} 0 & 1 & 0 \\ 0 & 0 & 1 \\ 1 & 0 & 0 \end{bmatrix}.$$

Now it is possible to write the product of the recursive multiplication of the matrix $[T]_{M,N}$ by the vector $[V]_N$:

$$[R]_{M,N}$$
$$= \begin{bmatrix} \text{tr}\left(([Z]_{M,N,L} \cdot [E_1]_M) \cdot [P]_{L,N} \cdot [C]^0_{N,N}\right) & \text{tr}\left(([Z]_{M,N,L} \cdot [E_1]_M) \cdot [P]_{L,N} \cdot [C]_{N,N}\right) & \text{tr}\left(([Z]_{M,N,L} \cdot [E_1]_M) \cdot [P]_{L,N} \cdot [C]^2_{N,N}\right) \\ \text{tr}\left(([Z]_{M,N,L} \cdot [E_2]_M) \cdot [P]_{L,N} \cdot [C]^0_{N,N}\right) & \text{tr}\left(([Z]_{M,N,L} \cdot [E_2]_M) \cdot [P]_{L,N} \cdot [C]_{N,N}\right) & \text{tr}\left(([Z]_{M,N,L} \cdot [E_2]_M) \cdot [P]_{L,N} \cdot [C]^2_{N,N}\right) \\ \text{tr}\left(([Z]_{M,N,L} \cdot [E_3]_M) \cdot [P]_{L,N} \cdot [C]^0_{N,N}\right) & \text{tr}\left(([Z]_{M,N,L} \cdot [E_3]_M) \cdot [P]_{L,N} \cdot [C]_{N,N}\right) & \text{tr}\left(([Z]_{M,N,L} \cdot [E_3]_M) \cdot [P]_{L,N} \cdot [C]^2_{N,N}\right) \\ \text{tr}\left(([Z]_{M,N,L} \cdot [E_4]_M) \cdot [P]_{L,N} \cdot [C]^0_{N,N}\right) & \text{tr}\left(([Z]_{M,N,L} \cdot [E_4]_M) \cdot [P]_{L,N} \cdot [C]_{N,N}\right) & \text{tr}\left(([Z]_{M,N,L} \cdot [E_4]_M) \cdot [P]_{L,N} \cdot [C]^2_{N,N}\right) \end{bmatrix}$$

$$= \begin{bmatrix} \text{tr}\left(\begin{bmatrix} 0 & 0 & 0 \\ 4 & 6 & 8 \\ 6 & 9 & 12 \end{bmatrix}\right) & \text{tr}\left(\begin{bmatrix} 0 & 0 & 0 \\ 6 & 8 & 4 \\ 9 & 12 & 6 \end{bmatrix}\right) & \text{tr}\left(\begin{bmatrix} 0 & 0 & 0 \\ 8 & 4 & 6 \\ 12 & 6 & 9 \end{bmatrix}\right) \\ \text{tr}\left(\begin{bmatrix} 6 & 9 & 12 \\ 4 & 6 & 8 \\ 0 & 0 & 0 \end{bmatrix}\right) & \text{tr}\left(\begin{bmatrix} 9 & 12 & 6 \\ 6 & 8 & 4 \\ 0 & 0 & 0 \end{bmatrix}\right) & \text{tr}\left(\begin{bmatrix} 12 & 6 & 9 \\ 8 & 4 & 6 \\ 0 & 0 & 0 \end{bmatrix}\right) \\ \text{tr}\left(\begin{bmatrix} 4 & 6 & 8 \\ 6 & 9 & 12 \\ 0 & 0 & 0 \end{bmatrix}\right) & \text{tr}\left(\begin{bmatrix} 6 & 8 & 4 \\ 9 & 12 & 6 \\ 0 & 0 & 0 \end{bmatrix}\right) & \text{tr}\left(\begin{bmatrix} 8 & 4 & 6 \\ 12 & 6 & 9 \\ 0 & 0 & 0 \end{bmatrix}\right) \\ \text{tr}\left(\begin{bmatrix} 4 & 6 & 8 \\ 0 & 0 & 0 \\ 6 & 9 & 12 \end{bmatrix}\right) & \text{tr}\left(\begin{bmatrix} 6 & 8 & 4 \\ 0 & 0 & 0 \\ 9 & 12 & 6 \end{bmatrix}\right) & \text{tr}\left(\begin{bmatrix} 8 & 4 & 6 \\ 0 & 0 & 0 \\ 12 & 6 & 9 \end{bmatrix}\right) \end{bmatrix} = \begin{bmatrix} 0+6+12 & 0+8+6 & 0+4+9 \\ 6+6+0 & 9+8+0 & 12+4+0 \\ 4+9+0 & 6+12+0 & 8+6+0 \\ 4+0+12 & 6+0+6 & 8+0+9 \end{bmatrix} = \begin{bmatrix} 18 & 14 & 13 \\ 12 & 17 & 16 \\ 13 & 18 & 14 \\ 16 & 12 & 17 \end{bmatrix}$$

In this example the recursive multiplication of a matrix of dimensions M=4 x N=3 by a vector of length N=3 requires L x N =2 x 3 = 6 operations of multiplication. Thus, in the given example the ratio of the number of operations required by the method including the decomposition of the matrix into commutator and kernel to the number of operations required by a method that does not include this decomposition is $Cm_+ = \frac{M \cdot (N-1) \cdot N}{M \cdot (N-1) \cdot N} = \frac{4 \cdot (3-1) \cdot 4}{4 \cdot (3-1) \cdot 4} = 1$ for addition and $Cm_* = \frac{L}{M \cdot N} = \frac{2}{4 \cdot 3} = \frac{1}{6}$ for multiplication.

## 19. Definition of method and process for iterative multiplication of a vector by a factored matrix.

Now we examine the problem of sequential and continuous, that is iterative, computation of the product of a known and constant matrix 
$$[T]_{M,N} = \begin{bmatrix} t_{1,1} & \cdots & t_{1,N} \\ \vdots & t_{m,n} & \vdots \\ t_{M,1} & \cdots & t_{M,N} \end{bmatrix} \tag{79}$$

of dimensions M x N, containing $L \leq M \cdot N$ distinct nonzero elements, and a series of vectors each of which is formed from the preceding vector via a linear shift of all its elements upward by one position. At each iteration the lowest position of this vector is filled with a new element, and the element in the uppermost position is lost. At each iteration the matrix $[T]_{M,N}$ is multiplied by the vector

$$[V_1]_N = \begin{bmatrix} v_1 \\ \cdots \\ v_n \\ \cdots \\ v_N \end{bmatrix}, \tag{80}$$

having first obtained the matrix $[P_1]_{L,N}$ from the product of the kernel $[U]_L$ of this matrix by the transposed vector $[V_1]_N$:

$$[P_1]_{L,N} = [U]_L \cdot [V_1]_N^t = \begin{bmatrix} u_1 \\ \cdots \\ u_l \\ \cdots \\ u_L \end{bmatrix} \cdot [v_1 \ \ldots \ v_n \ \ldots \ v_N] = \begin{bmatrix} v_1 \cdot u_1 & \cdots & v_N \cdot u_1 \\ \vdots & v_n \cdot u_l & \vdots \\ v_1 \cdot u_L & \cdots & v_N \cdot u_L \end{bmatrix} = = \begin{bmatrix} \begin{bmatrix} u_1 \\ \cdots \\ u_l \\ \cdots \\ u_L \end{bmatrix} \cdot v_1 \ \begin{bmatrix} u_1 \\ \cdots \\ u_l \\ \cdots \\ u_L \end{bmatrix} \cdot v_2 \ \ldots \ \begin{bmatrix} u_1 \\ \cdots \\ u_l \\ \cdots \\ u_L \end{bmatrix} \cdot v_n \ \ldots \ \begin{bmatrix} u_1 \\ \cdots \\ u_l \\ \cdots \\ u_L \end{bmatrix} \cdot v_{N-1} \ \begin{bmatrix} u_1 \\ \cdots \\ u_l \\ \cdots \\ u_L \end{bmatrix} \cdot v_N \end{bmatrix} \tag{81}$$

In its turn the matrix $[T]_{M,N}$ is represented as the product of the commutator $[Z]_{M,N,L} = \{ z_{m,n,l} \mid m \in [1,M], n \in [1,N], l \in [1,L]\}$ and the kernel $[U]_L = \begin{bmatrix} u_1 \\ \cdots \\ u_l \\ \cdots \\ u_L \end{bmatrix}$:

$$[T]_{M,N} = [Z]_{M,N,L} \cdot [U]_L = \{ \sum_{l=1}^{l=L} z_{m,n,l} \cdot u_l \mid m \in [1,M], n \in [1,N]\} \tag{82}$$

At the previous iteration the matrix $[T]_{M,N}$ was multiplied by the vector

$$[V_0]_N = \begin{bmatrix} v_0 \\ \vdots \\ v_n \\ \vdots \\ v_{N-1} \end{bmatrix}, \tag{83}$$

and therefore there exists a matrix $[P_0]_{L,N}$ formed from the product of the kernel $[U]_L$ of the matrix $[T]_{M,N}$ and the transposed matrix $[V_0]_N$:

$$[P_0]_{L,N} = [U]_L \cdot [V_0]_N^t = \begin{bmatrix} u_1 \\ \vdots \\ u_l \\ \vdots \\ u_L \end{bmatrix} \cdot [v_0 \ \ldots \ v_{n-1} \ \ldots \ v_{N-1}] = \begin{bmatrix} v_0 \cdot u_1 & \cdots & v_{N-1} \cdot u_1 \\ \vdots & v_{n-1} \cdot u_l & \vdots \\ v_0 \cdot u_L & \cdots & v_{N-1} \cdot u_L \end{bmatrix} = \begin{bmatrix} u_1 \\ \vdots \\ u_l \\ \vdots \\ u_L \end{bmatrix} \cdot v_0 \ \begin{bmatrix} u_1 \\ \vdots \\ u_l \\ \vdots \\ u_L \end{bmatrix} \cdot$$

$$v_1 \ \ldots \ \begin{bmatrix} u_1 \\ \vdots \\ u_l \\ \vdots \\ u_L \end{bmatrix} \cdot v_{n-1} \ \ldots \ \begin{bmatrix} u_1 \\ \vdots \\ u_l \\ \vdots \\ u_L \end{bmatrix} \cdot v_{N-2} \ \begin{bmatrix} u_1 \\ \vdots \\ u_l \\ \vdots \\ u_L \end{bmatrix} \cdot v_{N-1} \tag{84}$$

The matrix $[P_1]_{L,N}$ is equivalent to a linear shift one position to the left of the matrix $[P_0]_{L,N}$, in which the rightmost column is the product of the kernel

$$[U]_L = \begin{bmatrix} u_1 \\ \vdots \\ u_l \\ \vdots \\ u_L \end{bmatrix} \tag{85}$$

and the new value $v_N$:

$$[P_1]_{L,N} = \begin{bmatrix} v_1 \cdot u_1 & \cdots & v_N \cdot u_1 \\ \vdots & v_n \cdot u_l & \vdots \\ v_1 \cdot u_L & \cdots & v_N \cdot u_L \end{bmatrix} = \begin{bmatrix} v_1 \cdot u_1 & \cdots & 0 \\ \vdots & v_n \cdot u_l & \vdots \\ v_1 \cdot u_L & \cdots & 0 \end{bmatrix} + \begin{bmatrix} 0 & \cdots & u_1 \\ \vdots & 0 & \vdots \\ 0 & \cdots & u_L \end{bmatrix} \cdot v_N =$$

$$\begin{bmatrix} v_0 \cdot u_1 & \cdots & v_{N-1} \cdot u_1 \\ \vdots & v_{n-1} \cdot u_l & \vdots \\ v_0 \cdot u_L & \cdots & v_{N-1} \cdot u_L \end{bmatrix} \cdot \begin{bmatrix} 0\,0\,0\,0\,\ldots\,0\,0\,0\,0 \\ 1\,0\,0\,0\,\ldots\,0\,0\,0\,0 \\ 0\,1\,0\,0\,\ldots\,0\,0\,0\,0 \\ 0\,0\,1\,0\,\ldots\,0\,0\,0\,0 \\ \vdots \\ 0\,0\,0\,0\,\ldots\,0\,1\,0\,0 \\ 0\,0\,0\,0\,\ldots\,0\,0\,1\,0 \end{bmatrix} + \begin{bmatrix} u_1 \\ \vdots \\ u_l \\ \vdots \\ u_L \end{bmatrix} \cdot v_N \cdot [0 \ \ldots \ 0 \ 1]_N = [P_0]_{L,N} \cdot [\widetilde{C}]_{N,N} + [U]_L \cdot v_N \cdot [\widetilde{\widetilde{C}}]_N$$

$$\tag{86}$$

The matrix

$$[\widetilde{C}]_{N,N} = \begin{bmatrix} 0\,0\,0\,0\,\ldots\,0\,0\,0\,0 \\ 1\,0\,0\,0\,\ldots\,0\,0\,0\,0 \\ 0\,1\,0\,0\,\ldots\,0\,0\,0\,0 \\ 0\,0\,1\,0\,\ldots\,0\,0\,0\,0 \\ \vdots \\ 0\,0\,0\,0\,\ldots\,0\,1\,0\,0 \\ 0\,0\,0\,0\,\ldots\,0\,0\,1\,0 \end{bmatrix} \tag{87}$$

and vector

$$[\widetilde{\widetilde{C}}]_N = [0 \ldots 0\ 1]_N \tag{88}$$

introduced here participate in the iterative formation of the matrix $[P]_{L,N}$.

Thus, the n$^{th}$ iteration takes the form:

$$\begin{cases} [P_n]_{L,N} = [P_{n-1}]_{L,N} \cdot [\widetilde{C}]_{N,N} + [U]_L \cdot v_n \cdot [\widetilde{\widetilde{C}}]_N, \\ [R_n]_M = \begin{bmatrix} \operatorname{tr}\left(([Z]_{M,N,L} \cdot [E_1]_M) \cdot [P_n]_{L,N}\right) \\ \vdots \\ \operatorname{tr}\left(([Z]_{M,N,L} \cdot [E_m]_M) \cdot [P_n]_{L,N}\right) \\ \vdots \\ \operatorname{tr}\left(([Z]_{M,N,L} \cdot [E_M]_M) \cdot [P_n]_{L,N}\right) \end{bmatrix}_M \\ n \in [0, \infty[ \end{cases} \tag{89}$$

where

$$\left\{ [E_m]_M = \begin{bmatrix} \begin{bmatrix} 0 \\ \vdots \\ 0 \end{bmatrix}_{m-1} \\ 1 \\ \begin{bmatrix} 0 \\ \vdots \\ 0 \end{bmatrix}_{M-m} \end{bmatrix}_M \right\} \tag{90}$$

are selecting vectors.

Since $[\widetilde{C}]_{N,N}$ and $[\widetilde{\widetilde{C}}]_N$ consist only of zero and unit elements, the iterative formation of $[P_n]_{L,N}$ requires multiplying the elements of the kernel by only one component of the vector $[V_0]_N$ - the newest component $v_N$. Thus the ratio of the number of operations for one iteration of iterative multiplication with a method using the decomposition of the vector into a kernel and a commutator to the number of operations required with a method that does not include such a decomposition is $C_+ \leq \frac{(N-1) \cdot M}{(N-1) \cdot M} = 1$ for addition and $C_* \leq \frac{1 \cdot L}{N \cdot M} = \frac{L}{N \cdot M} \leq 1$ for multiplication.

## 20. Example of iterative multiplication of a vector by a factored matrix.

The matrix $[T]_{M,N} = \begin{bmatrix} t_{1,1} & t_{1,2} & t_{1,3} \\ t_{2,1} & t_{2,2} & t_{2,3} \\ t_{3,1} & t_{3,2} & t_{3,3} \\ t_{4,1} & t_{4,2} & t_{4,3} \end{bmatrix} = \begin{bmatrix} 0 & 2 & 3 \\ 3 & 2 & 0 \\ 2 & 3 & 0 \\ 2 & 0 & 3 \end{bmatrix}$ of dimensions M x $N$ = 4 x 3, containing $L = 2$ distinct nonzero elements, is to be iteratively multiplied by the vectors

$$\{[V_1]_N, [V_2]_N, [V_3]_N\} = \left\{ \begin{bmatrix} 0 \\ 0 \\ 2 \end{bmatrix}, \begin{bmatrix} 0 \\ 2 \\ 3 \end{bmatrix}, \begin{bmatrix} 2 \\ 3 \\ 4 \end{bmatrix} \right\}.$$

The matrix $[T]_{M,N}$ is written as the tensor contraction of the commutator

$$[Z]_{M,N,L} = \{ Z_{m,n,l} \mid m \in [1,M], n \in [1,N], l \in [1,L]\} = \begin{bmatrix} [0\ 0] & [1\ 0] & [0\ 1] \\ [0\ 1] & [1\ 0] & [0\ 0] \\ [1\ 0] & [0\ 1] & [0\ 0] \\ [1\ 0] & [0\ 0] & [0\ 1] \end{bmatrix}$$

with the kernel $[U]_L = \begin{bmatrix} u_1 \\ u_2 \end{bmatrix} = \begin{bmatrix} 2 \\ 3 \end{bmatrix}$:

$$[T]_{M,N} = [Z]_{M,N,L} \cdot [U]_L = \left\{ \sum_{l=1}^{l=L} z_{m,n,l} \cdot u_l \mid m \in [1,M], n \in [1,N] \right\} = \begin{bmatrix} [0\ 0]\cdot\begin{bmatrix}2\\3\end{bmatrix} & [0\ 1]\cdot\begin{bmatrix}2\\3\end{bmatrix} & [0\ 1]\cdot\begin{bmatrix}2\\3\end{bmatrix} \\ [0\ 1]\cdot\begin{bmatrix}2\\3\end{bmatrix} & [0\ 1]\cdot\begin{bmatrix}2\\3\end{bmatrix} & [0\ 0]\cdot\begin{bmatrix}2\\3\end{bmatrix} \\ [0\ 1]\cdot\begin{bmatrix}2\\3\end{bmatrix} & [0\ 1]\cdot\begin{bmatrix}2\\3\end{bmatrix} & [0\ 0]\cdot\begin{bmatrix}2\\3\end{bmatrix} \\ [0\ 1]\cdot\begin{bmatrix}2\\3\end{bmatrix} & [0\ 0]\cdot\begin{bmatrix}2\\3\end{bmatrix} & [0\ 1]\cdot\begin{bmatrix}2\\3\end{bmatrix} \end{bmatrix}$$

$$= \begin{bmatrix} [0\ 0] & [1\ 0] & [0\ 1] \\ [0\ 1] & [1\ 0] & [0\ 0] \\ [1\ 0] & [0\ 1] & [0\ 0] \\ [1\ 0] & [0\ 0] & [0\ 1] \end{bmatrix} \cdot \begin{bmatrix} 2 \\ 3 \end{bmatrix} = \begin{bmatrix} 0 & 3 & 2 \\ 3 & 2 & 0 \\ 2 & 3 & 0 \\ 2 & 0 & 3 \end{bmatrix}$$

The selecting vectors are: $[E_1]_M = \begin{bmatrix}1\\0\\0\\0\end{bmatrix}, [E_2]_M = \begin{bmatrix}0\\1\\0\\0\end{bmatrix}, [E_3]_M = \begin{bmatrix}0\\0\\1\\0\end{bmatrix}$ и $[E_4]_M = \begin{bmatrix}0\\0\\0\\1\end{bmatrix}$.

The corresponding tensor contraction of the commutator $[Z]_{M,N,L}$ are:

$$[Z]_{M,N,L} \cdot [E_1]_M = \begin{bmatrix} ([0\ 0]\ [1\ 0]\ [0\ 1]) \\ ([0\ 1]\ [1\ 0]\ [0\ 0]) \\ ([1\ 0]\ [0\ 1]\ [0\ 0]) \\ ([1\ 0]\ [0\ 0]\ [0\ 1]) \end{bmatrix}^t \cdot \begin{bmatrix}1\\0\\0\\0\end{bmatrix} = \begin{bmatrix}[0\ 0]\\[1\ 0]\\[0\ 1]\end{bmatrix} = \begin{bmatrix}0&0\\1&0\\0&1\end{bmatrix}$$

$$[Z]_{M,N,L} \cdot [E_2]_M = \begin{bmatrix} ([0\ 0]\ [1\ 0]\ [0\ 1]) \\ ([0\ 1]\ [1\ 0]\ [0\ 0]) \\ ([1\ 0]\ [0\ 1]\ [0\ 0]) \\ ([1\ 0]\ [0\ 0]\ [0\ 1]) \end{bmatrix}^t \cdot \begin{bmatrix}0\\1\\0\\0\end{bmatrix} = \begin{bmatrix}[0\ 1]\\[1\ 0]\\[0\ 0]\end{bmatrix} = \begin{bmatrix}0&1\\1&0\\0&0\end{bmatrix}$$

$$[Z]_{M,N,L} \cdot [E_3]_M = \begin{bmatrix} ([0\ 0]\ [1\ 0]\ [0\ 1]) \\ ([0\ 1]\ [1\ 0]\ [0\ 0]) \\ ([1\ 0]\ [0\ 1]\ [0\ 0]) \\ ([1\ 0]\ [0\ 0]\ [0\ 1]) \end{bmatrix}^t \cdot \begin{bmatrix}0\\0\\1\\0\end{bmatrix} = \begin{bmatrix}[1\ 0]\\[0\ 1]\\[0\ 0]\end{bmatrix} = \begin{bmatrix}1&0\\0&1\\0&0\end{bmatrix}$$

$$[Z]_{M,N,L} \cdot [E_4]_M = \begin{bmatrix} ([0\ 0]\ [1\ 0]\ [0\ 1]) \\ ([0\ 1]\ [1\ 0]\ [0\ 0]) \\ ([1\ 0]\ [0\ 1]\ [0\ 0]) \\ ([1\ 0]\ [0\ 0]\ [0\ 1]) \end{bmatrix}^t \cdot \begin{bmatrix}0\\0\\0\\1\end{bmatrix} = \begin{bmatrix}[1\ 0]\\[0\ 0]\\[0\ 1]\end{bmatrix} = \begin{bmatrix}1&0\\0&0\\0&1\end{bmatrix}$$

The product of the matrix $[T]_{M,N}$ and the vector $[V]_N$ is the matrix $[P]_{L,N}$, equal to:

$$[P]_{L,N} = [U]_L \cdot [V]_N^t = \begin{bmatrix} 2 \\ 3 \end{bmatrix} \cdot [2\ 3\ 4] = \begin{bmatrix} 2 \cdot 2 & 2 \cdot 3 & 2 \cdot 4 \\ 3 \cdot 2 & 3 \cdot 3 & 3 \cdot 4 \end{bmatrix} = \begin{bmatrix} 4 & 6 & 8 \\ 6 & 9 & 12 \end{bmatrix}$$

For N=3, the matrix $[\widetilde{C}]_{N,N} = \begin{bmatrix} 0 & 0 & 0 \\ 1 & 0 & 0 \\ 0 & 1 & 0 \end{bmatrix}$ and the vector $[\widetilde{\widetilde{C}}]_N = [0\ 0\ 1]$.

The matrix $[P_0]_{L,N} = [\mathbf{0}]_{L,N} = \begin{bmatrix} 0 & 0 & 0 \\ 0 & 0 & 0 \end{bmatrix}$, since the condition $[V_n]_N = [\mathbf{0}]_N\ |\ \forall\ n < 1$ is fulfilled.

The process of iterative multiplication is carried out as follows.

Iteration 1:

$$[P_1]_{L,N} = [P_0]_{L,N} \cdot [\widetilde{C}]_{N,N} + [U]_L \cdot v_1 \cdot [\widetilde{\widetilde{C}}]_N = \begin{bmatrix} 0 & 0 & 0 \\ 0 & 0 & 0 \end{bmatrix} \cdot \begin{bmatrix} 0 & 0 & 0 \\ 1 & 0 & 0 \\ 0 & 1 & 0 \end{bmatrix} + \begin{bmatrix} 2 \\ 3 \end{bmatrix} \cdot 2 \cdot [0\ 0\ 1] = \begin{bmatrix} 0 & 0 & 4 \\ 0 & 0 & 6 \end{bmatrix}$$

$$[R_1]_M = \begin{bmatrix} \mathrm{tr}\left(([Z]_{M,N,L} \cdot [E_1]_M) \cdot [P_1]_{L,N}\right) \\ \mathrm{tr}\left(([Z]_{M,N,L} \cdot [E_2]_M) \cdot [P_1]_{L,N}\right) \\ \mathrm{tr}\left(([Z]_{M,N,L} \cdot [E_3]_M) \cdot [P_1]_{L,N}\right) \\ \mathrm{tr}\left(([Z]_{M,N,L} \cdot [E_4]_M) \cdot [P_1]_{L,N}\right) \end{bmatrix} = \begin{bmatrix} \mathrm{tr}\left(\begin{bmatrix} 0 & 0 \\ 1 & 0 \\ 0 & 1 \end{bmatrix} \cdot \begin{bmatrix} 0 & 0 & 4 \\ 0 & 0 & 6 \end{bmatrix}\right) \\ \mathrm{tr}\left(\begin{bmatrix} 0 & 1 \\ 1 & 0 \\ 0 & 0 \end{bmatrix} \cdot \begin{bmatrix} 0 & 0 & 4 \\ 0 & 0 & 6 \end{bmatrix}\right) \\ \mathrm{tr}\left(\begin{bmatrix} 1 & 0 \\ 0 & 1 \\ 0 & 0 \end{bmatrix} \cdot \begin{bmatrix} 0 & 0 & 4 \\ 0 & 0 & 6 \end{bmatrix}\right) \\ \mathrm{tr}\left(\begin{bmatrix} 1 & 0 \\ 0 & 0 \\ 0 & 1 \end{bmatrix} \cdot \begin{bmatrix} 0 & 0 & 4 \\ 0 & 0 & 6 \end{bmatrix}\right) \end{bmatrix} = \begin{bmatrix} \mathrm{tr}\left(\begin{bmatrix} 0 & 0 & 0 \\ 0 & 0 & 4 \\ 0 & 0 & 6 \end{bmatrix}\right) \\ \mathrm{tr}\left(\begin{bmatrix} 0 & 0 & 6 \\ 0 & 0 & 4 \\ 0 & 0 & 0 \end{bmatrix}\right) \\ \mathrm{tr}\left(\begin{bmatrix} 0 & 0 & 4 \\ 0 & 0 & 6 \\ 0 & 0 & 0 \end{bmatrix}\right) \\ \mathrm{tr}\left(\begin{bmatrix} 0 & 0 & 4 \\ 0 & 0 & 0 \\ 0 & 0 & 6 \end{bmatrix}\right) \end{bmatrix} = \begin{bmatrix} 6 \\ 0 \\ 0 \\ 6 \end{bmatrix}$$

Iteration 2:

$$[P_2]_{L,N} = [P_1]_{L,N} \cdot [\widetilde{C}]_{N,N} + [U]_L \cdot v_2 \cdot [\widetilde{\widetilde{C}}]_N = \begin{bmatrix} 0 & 0 & 4 \\ 0 & 0 & 6 \end{bmatrix} \cdot \begin{bmatrix} 0 & 0 & 0 \\ 1 & 0 & 0 \\ 0 & 1 & 0 \end{bmatrix} + \begin{bmatrix} 2 \\ 3 \end{bmatrix} \cdot 3 \cdot [0\ 0\ 1] = \begin{bmatrix} 0 & 4 & 6 \\ 0 & 6 & 9 \end{bmatrix}$$

$$[R_2]_M = \begin{bmatrix} \mathrm{tr}\left(([Z]_{M,N,L} \cdot [E_1]_M) \cdot [P_2]_{L,N}\right) \\ \mathrm{tr}\left(([Z]_{M,N,L} \cdot [E_2]_M) \cdot [P_2]_{L,N}\right) \\ \mathrm{tr}\left(([Z]_{M,N,L} \cdot [E_3]_M) \cdot [P_2]_{L,N}\right) \\ \mathrm{tr}\left(([Z]_{M,N,L} \cdot [E_4]_M) \cdot [P_2]_{L,N}\right) \end{bmatrix} = \begin{bmatrix} \mathrm{tr}\left(\begin{bmatrix} 0 & 0 \\ 1 & 0 \\ 0 & 1 \end{bmatrix} \cdot \begin{bmatrix} 0 & 4 & 6 \\ 0 & 6 & 9 \end{bmatrix}\right) \\ \mathrm{tr}\left(\begin{bmatrix} 0 & 1 \\ 1 & 0 \\ 0 & 0 \end{bmatrix} \cdot \begin{bmatrix} 0 & 4 & 6 \\ 0 & 6 & 9 \end{bmatrix}\right) \\ \mathrm{tr}\left(\begin{bmatrix} 1 & 0 \\ 0 & 1 \\ 0 & 0 \end{bmatrix} \cdot \begin{bmatrix} 0 & 4 & 6 \\ 0 & 6 & 9 \end{bmatrix}\right) \\ \mathrm{tr}\left(\begin{bmatrix} 1 & 0 \\ 0 & 0 \\ 0 & 1 \end{bmatrix} \cdot \begin{bmatrix} 0 & 4 & 6 \\ 0 & 6 & 9 \end{bmatrix}\right) \end{bmatrix} = \begin{bmatrix} \mathrm{tr}\left(\begin{bmatrix} 0 & 0 & 0 \\ 0 & 4 & 6 \\ 0 & 6 & 9 \end{bmatrix}\right) \\ \mathrm{tr}\left(\begin{bmatrix} 0 & 6 & 9 \\ 0 & 4 & 6 \\ 0 & 0 & 0 \end{bmatrix}\right) \\ \mathrm{tr}\left(\begin{bmatrix} 0 & 4 & 6 \\ 0 & 6 & 9 \\ 0 & 0 & 0 \end{bmatrix}\right) \\ \mathrm{tr}\left(\begin{bmatrix} 0 & 4 & 6 \\ 0 & 0 & 0 \\ 0 & 6 & 9 \end{bmatrix}\right) \end{bmatrix} = \begin{bmatrix} 13 \\ 4 \\ 6 \\ 9 \end{bmatrix}$$

Iteration 3:

$$[P_3]_{L,N} = [P_2]_{L,N} \cdot [\widetilde{C}]_{N,N} + [U]_L \cdot v_3 \cdot [\widetilde{\widetilde{C}}]_N = \begin{bmatrix} 0 & 4 & 6 \\ 0 & 6 & 9 \end{bmatrix} \cdot \begin{bmatrix} 0 & 0 & 0 \\ 1 & 0 & 0 \\ 0 & 1 & 0 \end{bmatrix} + \begin{bmatrix} 2 \\ 3 \end{bmatrix} \cdot 4 \cdot [0\ 0\ 1] = \begin{bmatrix} 4 & 6 & 8 \\ 6 & 9 & 12 \end{bmatrix}$$

$$[R_3]_M = \begin{bmatrix} \operatorname{tr}\left(([Z]_{M,N,L} \cdot [E_1]_M) \cdot [P_3]_{L,N}\right) \\ \operatorname{tr}\left(([Z]_{M,N,L} \cdot [E_2]_M) \cdot [P_3]_{L,N}\right) \\ \operatorname{tr}\left(([Z]_{M,N,L} \cdot [E_3]_M) \cdot [P_3]_{L,N}\right) \\ \operatorname{tr}\left(([Z]_{M,N,L} \cdot [E_4]_M) \cdot [P_3]_{L,N}\right) \end{bmatrix} = \begin{bmatrix} \operatorname{tr}\left(\begin{bmatrix} 0 & 0 \\ 1 & 0 \\ 0 & 1 \end{bmatrix} \cdot \begin{bmatrix} 4 & 6 & 8 \\ 6 & 9 & 12 \end{bmatrix}\right) \\ \operatorname{tr}\left(\begin{bmatrix} 0 & 1 \\ 1 & 0 \\ 0 & 0 \end{bmatrix} \cdot \begin{bmatrix} 4 & 6 & 8 \\ 6 & 9 & 12 \end{bmatrix}\right) \\ \operatorname{tr}\left(\begin{bmatrix} 1 & 0 \\ 0 & 1 \\ 0 & 0 \end{bmatrix} \cdot \begin{bmatrix} 4 & 6 & 8 \\ 6 & 9 & 12 \end{bmatrix}\right) \\ \operatorname{tr}\left(\begin{bmatrix} 1 & 0 \\ 0 & 0 \\ 0 & 1 \end{bmatrix} \cdot \begin{bmatrix} 4 & 6 & 8 \\ 6 & 9 & 12 \end{bmatrix}\right) \end{bmatrix} = \begin{bmatrix} \operatorname{tr}\left(\begin{bmatrix} 0 & 0 & 0 \\ 4 & 6 & 8 \\ 6 & 9 & 12 \end{bmatrix}\right) \\ \operatorname{tr}\left(\begin{bmatrix} 6 & 9 & 12 \\ 4 & 6 & 8 \\ 0 & 0 & 0 \end{bmatrix}\right) \\ \operatorname{tr}\left(\begin{bmatrix} 4 & 6 & 8 \\ 6 & 9 & 12 \\ 0 & 0 & 0 \end{bmatrix}\right) \\ \operatorname{tr}\left(\begin{bmatrix} 4 & 6 & 8 \\ 0 & 0 & 0 \\ 6 & 9 & 12 \end{bmatrix}\right) \end{bmatrix} = \begin{bmatrix} 18 \\ 12 \\ 13 \\ 16 \end{bmatrix}$$

To compute all N = 3 operations of multiplication of a matrix by a vector, it is necessary at each iteration to multiply the kernel by only one new element $v_n$, which uses $L = 2$ multiplication operations. Thus, in the given example, the ratio of the number of operations with a method using the decomposition of the vector into a kernel and a commutator to the number of operations required with a method that does not include such a decomposition is $C_+ = \frac{M \cdot (N-1)}{M \cdot (N-1)} = \frac{4 \cdot (3-1)}{4 \cdot (3-1)} = 1$ for addition and $C_* = \frac{L}{M \cdot N} = \frac{2}{4 \cdot 3} = \frac{1}{6}$ for multiplication.

## 21. Definition of method and process for the multiplication of a vector by a factored tensor.

The tensor

$$[T]_{N_1, N_2, \ldots, N_m, \ldots, N_M} = \{ t_{n_1, n_2, \ldots, n_m, \ldots, n_M} \mid n_m \in [1, N_m], m \in [1, M] \} \tag{91}$$

of rank, or dimension, $(N_1, N_2, \ldots, N_m, \ldots, N_M)$ containing

$$L \leq \prod_{k=1}^{M} N_k \tag{92}$$

distinct nonzero elements, is to be multiplied by the vector

$$[V]_{N_m} = \begin{bmatrix} v_1 \\ \ldots \\ v_n \\ \ldots \\ v_{N_m} \end{bmatrix}. \tag{93}$$

The tensor $[T]_{N_1, N_2, \ldots, N_m, \ldots, N_M}$ is written as the product of the commutator

$$[Z]_{N_1, N_2, \ldots, N_m, \ldots, N_M, L} = \{ z_{n_1, n_2, \ldots, n_m, \ldots, n_M, l} \mid n_m \in [1, N_m], m \in [1, M], l \in [1, L] \} \tag{94}$$

and the kernel

$$[U]_L = \begin{bmatrix} u_1 \\ \ldots \\ u_l \\ \ldots \\ u_L \end{bmatrix}: \tag{95}$$

$$[T]_{N_1,N_2,\dots,N_m,\dots,N_M} = [Z]_{N_1,N_2,\dots,N_m,\dots,N_M,L} \cdot [U]_L = \left\{ \sum_{l=1}^{l=L} z_{n_1,n_2,\dots,n_m,\dots,n_M,l} \cdot u_l \mid n_m \in [1, N_m], m \in [1, M] \right\} \quad (96)$$

Then the product of the tensor $[T]_{N_1,N_2,\dots,N_m,\dots,N_M}$ and the vector $[V]_{N_m}$ may be written as:

$$[R]_{N_1,N_2,\dots,N_{m-1},N_{m+1},\dots,N_M} = [T]_{N_1,N_2,\dots,N_m,\dots,N_M} \cdot [V]_{N_m} = \left([Z]_{N_1,N_2,\dots,N_m,\dots,N_M,L} \cdot [U]_L\right) \cdot [V]_{N_m} = \left\{ \sum_{n=1}^{N_m} v_n \cdot \sum_{l=1}^{L} z_{n_1,n_2,\dots,n_{m-1},n,n_{m+1},\dots,n_M,l} \cdot u_l \mid n_k \in [1, N_k], k \in \{[1, m-1], [m+1, M]\} \right\} =$$
$$\left\{ \sum_{n=1}^{N_m} \left( \sum_{l=1}^{L} z_{n_1,n_2,\dots,n_{m-1},n,n_{m+1},\dots,n_M,l} \cdot u_l \right) \cdot v_n \mid n_k \in [1, N_k], k \in \{[1, m-1], [m+1, M]\} \right\} =$$
$$\left\{ \sum_{n=1}^{N_m} \sum_{l=1}^{L} z_{n_1,n_2,\dots,n_{m-1},n,n_{m+1},\dots,n_M,l} \cdot u_l \cdot v_n \mid n_k \in [1, N_k], k \in \{[1, m-1], [m+1, M]\} \right\} =$$
$$\left\{ \sum_{n=1}^{N_m} \sum_{l=1}^{L} z_{n_1,n_2,\dots,n_{m-1},n,n_{m+1},\dots,n_M,l} \cdot (u_l \cdot v_n) \mid n_k \in [1, N_k], k \in \{[1, m-1], [m+1, M]\} \right\} \quad (97)$$

In this expression each nested sum contains the same coefficient $(u_l \cdot v_n)$ which is an element of the matrix which is the product of the vector $[U]_L$ and the transposed vector $[V]_{N_m}$:

$$[P]_{L,N} = [U]_L \cdot [V]_{N_m}^t \quad (98)$$

Thus the product of the matrix $[T]_{M,N}$ and the vector $[V]_N$ may be written as:

$$[R]_{N_1,N_2,\dots,N_{m-1},N_{m+1},\dots,N_M} = [T]_{N_1,N_2,\dots,N_m,\dots,N_M} \cdot [V]_{N_m} == \left\{ \sum_{n=1}^{N_m} \sum_{l=1}^{L} z_{n_1,n_2,\dots,n_{m-1},n,n_{m+1},\dots,n_M,l} \cdot (u_l \cdot v_n) \mid n_k \in [1, N_k], k \in \{[1, m-1], [m+1, M]\} \right\} = \left\{ \sum_{n=1}^{N_m} \sum_{l=1}^{L} z_{n_1,n_2,\dots,n_{m-1},n,n_{m+1},\dots,n_M,l} \cdot p_{l,n} \mid n_k \in [1, N_k], k \in \{[1, m-1], [m+1, M]\} \right\} \quad (99)$$

Thus the multiplication of a tensor by a vector of length $N_m$ may be carried out in two steps. First, the matrix is obtained which contains the product of each element of the original vector and each element of the kernel $[T]_{N_1,N_2,\dots,N_m,\dots,N_M}$ of the initial tensor. Then each element of the resulting tensor $[R]_{N_1,N_2,\dots,N_{m-1},N_{m+1},\dots,N_M}$ is calculated as the tensor contraction of the commutator with the matrix obtained in the first step. This sequence means that all multiplication operations are carried out in the first step, and their maximum number is equal to the product of the length $N_m$ of the original vector and the number L of distinct nonzero elements of the original tensor $[T]_{N_1,N_2,\dots,N_m,\dots,N_M}$, rather than the number of elements of the original tensor $[T]_{N_1,N_2,\dots,N_m,\dots,N_M}$, which is equal to $\prod_{k=1}^{M} N_k$, as in the case of multiplication without factorization of the tensor. All addition operations are carried out in the second step, and their maximal number is $\frac{N_m-1}{N_m} \cdot \prod_{k=1}^{M} N_k$. Thus the ratio of the number of operations with a method using the decomposition of the vector into a kernel and a commutator to the number of operations required with a method that does not include such a decomposition is $Cm_+ \leq \frac{\frac{N_m-1}{N_m} \prod_{k=1}^{M} N_k}{\frac{N_m-1}{N_m} \prod_{k=1}^{M} N_k} = 1$ for addition and $Cm_* \leq \frac{N_m \cdot L}{\prod_{k=1}^{M} N_k} = \frac{L}{(\prod_{k=1}^{m-1} N_k) \cdot (\prod_{k=m+1}^{M} N_k)}$ for multiplication.

## 22. Definition of method and process for the recursive multiplication of a vector by a factored tensor.

The tensor

$$[T]_{N_1,N_2,\dots,N_m,\dots,N_M} = \left\{ t_{n_1,n_2,\dots,n_m,\dots,n_M} \mid n_m \in [1, N_m], m \in [1, M] \right\} \quad (100)$$

containing

$$L \leq \prod_{k=1}^{M} N_k \tag{101}$$

distinct nonzero elements is to be multiplied by the vector

$$[V]_{N_m} = [V_0]_{N_m} = \begin{bmatrix} v_1 \\ \dots \\ v_n \\ \dots \\ v_{N_m} \end{bmatrix} \tag{102}$$

and all its circularly-shifted variants:

$$\left\{[V_1]_{N_m}, [V_2]_{N_m}, \dots, [V_{N_m-1}]_{N_m}\right\} = \left\{ \begin{bmatrix} v_2 \\ \dots \\ \dots \\ v_{N_m} \\ v_1 \end{bmatrix}, \begin{bmatrix} v_3 \\ \dots \\ \dots \\ v_1 \\ v_2 \end{bmatrix}, \dots, \begin{bmatrix} v_{N_m} \\ v_1 \\ \dots \\ \dots \\ v_{N_m-1} \end{bmatrix} \right\}. \tag{103}$$

The tensor $[T]_{N_1, N_2, \dots, N_m, \dots, N_M}$ is written as the product of the commutator

$$[Z]_{N_1, N_2, \dots, N_m, \dots, N_M, L} = \{ z_{n_1, n_2, \dots, n_m, \dots, n_M, l} \mid n_m \in [1, N_m], m \in [1, M], l \in [1, L] \} \tag{104}$$

and the kernel

$$[U]_L = \begin{bmatrix} u_1 \\ \dots \\ u_l \\ \dots \\ u_L \end{bmatrix}: \tag{105}$$

$$[T]_{N_1, N_2, \dots, N_m, \dots, N_M} = [Z]_{N_1, N_2, \dots, N_m, \dots, N_M, L} \cdot [U]_L = \left\{ \sum_{l=1}^{l=L} z_{n_1, n_2, \dots, n_m, \dots, n_M, l} \cdot u_l \mid n_m \in [1, N_m], m \in [1, M] \right\} \tag{106}$$

First the product of the tensor $[T]_{N_1, N_2, \dots, N_m, \dots, N_M}$ and the vector $[V]_{N_m}$ is obtained. This product may be written as:

$$[R]_{N_1, N_2, \dots, N_{m-1}, N_{m+1}, \dots, N_M} = [T]_{N_1, N_2, \dots, N_m, \dots, N_M} \cdot [V]_{N_m} == \left\{ \sum_{n=1}^{N_m} \sum_{l=1}^{L} z_{n_1, n_2, \dots, n_{m-1}, n, n_{m+1}, \dots, n_M, l} \cdot p_{l,n} \mid n_k \in [1, N_k], k \in \{[1, m-1], [m+1, M]\} \right\} \tag{107}$$

, where $p_{l,n}$ are the elements of the matrix $[P]_{L, N_m}$ obtained from the multiplication of the kernel $[U]_L$ by the transposed vector $[V]_{N_m}$:

$$[P]_{L, N_m} = [U]_L \cdot [V]_{N_m}^t = \begin{bmatrix} u_1 \\ \dots \\ u_l \\ \dots \\ u_L \end{bmatrix} \cdot [v_1 \dots v_n \dots v_{N_m}] = \begin{bmatrix} v_1 \cdot u_1 & \dots & v_{N_m} \cdot u_1 \\ \vdots & v_n \cdot u_l & \vdots \\ v_1 \cdot u_L & \dots & v_{N_m N} \cdot u_L \end{bmatrix} \tag{108}$$

To obtain the succeeding value, the product of the tensor $[T]_{N_1, N_2, \dots, N_m, \dots, N_M}$ and the first circularly-shifted variant of the vector $[V]_{N_m}$, which is the vector

$$[V_1]_{N_m} = \begin{bmatrix} v_2 \\ \vdots \\ \vdots \\ v_{N_m} \\ v_1 \end{bmatrix}, \tag{109}$$

the new matrix $[P_1]_{L,N_m}$ is obtained:

$$[P_1]_{L,N_m} = [U]_L \cdot [V_1]_{N_m}^t = \begin{bmatrix} u_1 \\ \vdots \\ u_l \\ \vdots \\ u_L \end{bmatrix} \cdot [v_2 \ \ldots \ v_{n+1} \ \ldots \ v_{N_m} \ v_1] = \begin{bmatrix} v_2 \cdot u_1 & \cdots & v_{N_m} \cdot u_1 & v_1 \cdot u_1 \\ \vdots & \vdots & \vdots & \vdots \\ v_2 \cdot u_L & \cdots & v_{N_m} \cdot u_L & v_1 \cdot u_L \end{bmatrix} \tag{110}$$

Clearly, the matrix $[P_1]_{L,N_m}$ is equivalent to the matrix $[P]_{L,N_m}$ cyclically shifted one position to the left. Each element $p1_{l,n}$ of the matrix $[P_1]_{L,N_m}$ is a copy of the element $p_{l,1+(n-2)\mod(N_m)}$ of the matrix $[P]_{L,N_m}$, the element $p2_{l,n}$ of the matrix $[P_2]_{L,N_m}$ is a copy of the element $p1_{l,1+(n-2)\mod(N_m)}$ of the matrix $[P_1]_{L,N_m}$ and also a copy of the element $p_{l,1+(n-3)\mod(N_m)}$ of the matrix $[P]_{L,N_m}$. The general rule of representing an element of any matrix $[P_k]_{L,N_m}, k \in [0, N_m - 1]$ in terms of elements of the matrix $[P]_{L,N_m}$ may be written as:

$$pk_{l,1+(n-1-k)\mod(N_m)} = p_{l,n} \tag{111}$$

$$pk_{l,n} = p_{l,1+(n-1+k)\mod(N_m)} \tag{112}$$

All elements $pk_{l,n}$ may be included in a tensor $[P]_{N_m,L,N_m}$ of rank 3, and thus the result of cyclical multiplication of a tensor by a vector may be written as:

$$[R]_{N_m,N_1,N_2,\ldots,N_{m-1},N_{m+1},\ldots,N_M} = \{[T]_{N_1,N_2,\ldots,N_m,\ldots,N_M} \cdot [V_k]_{N_m} | k \in [0, N_m - 1]\} = [Z]_{N_1,N_2,\ldots,N_m,\ldots,N_M,L} \cdot [P]_{N_m,L,N_m}$$

$$= \left\{ \sum_{n=1}^{N_m} \sum_{l=1}^{L} z_{n_1,n_2,\ldots,n_{m-1},n,n_{m+1},\ldots,n_M,l} \cdot pk_{l,n} \middle| n_i \in [1, N_i], i \in \{[1, m-1], [m+1, M]\}, k \in [0, N_m - 1] \right\}$$

$$= \{\sum_{n=1}^{N_m} \sum_{l=1}^{L} z_{n_1,n_2,\ldots,n_{m-1},n,n_{m+1},\ldots,n_M,l} \cdot p_{l,1+(n-1+k)\mod(N_m)} | n_i \in [1, N_i], i \in \{[1, m-1], [m+1, M]\}, k \in [0, N_m - 1]\} \tag{113}$$

The recursive multiplication of a tensor by a vector of length $N_m$ may be carried out in two steps. First the tensor $[P]_{N_m,L,N_m}$ is obtained, consisting of all $N_m$ cyclically shifted variants of the matrix containing the product of each element of the initial vector and each element of the kernel of the initial tensor $[T]_{N_1,N_2,\ldots,N_m,\ldots,N_M}$. Then each element of the resulting tensor $[R]_{N_m,N_1,N_2,\ldots,N_{m-1},N_{m+1},\ldots,N_M}$ is obtained as the tensor contraction of the commutator with the tensor $[P]_{N_m,L,N_m}$ obtained in the first steo. Thus all multiplication operations take place during the first step, and their maximal number is equal to the product of the length $N_m$ of the original vector and the number L of distinct nonzero elements of the initial tensor $[T]_{N_1,N_2,\ldots,N_m,\ldots,N_M}$, not the product of the length $N_m$ of the original vector and the total number of elements in the original tensor $[T]_{N_1,N_2,\ldots,N_m,\ldots,N_M}$, which is $\prod_{k=1}^{M} N_k$, as in the case of multiplication without factorization of the tensor. All addition operations take place during the second step, and their maximal number is $\frac{N_m}{N_m} \cdot \frac{N_m - 1}{N_m} \cdot \prod_{k=1}^{M} N_k$. Thus the ratio of the number of operations with a method using the decomposition of the vector into a kernel and a commutator to the number of operations required with a method that does not include such a decomposition is $Cm_+ \leq \frac{\frac{N_m-1}{N_m} \cdot \prod_{k=1}^{M} N_k}{\frac{N_m-1}{N_m} \cdot \prod_{k=1}^{M} N_k} = 1$ for addition and $Cm_* \leq \frac{N_m \cdot L}{N_m \cdot \prod_{k=1}^{M} N_k} = \frac{L}{\prod_{k=1}^{M} N_k}$ for multiplication.

## 23. Definition of method and process for the iterative multiplication of a vector by a factored tensor.

Here the objective is sequential and continuous, which is to say iterative multiplication of a known and constant tensor

$$[T]_{N_1, N_2, \ldots, N_m, \ldots, N_M} = \{ t_{n_1, n_2, \ldots, n_m, \ldots, n_M} \mid n_m \in [1, N_m], m \in [1, M] \} \tag{114}$$

containing

$$L \leq \prod_{k=1}^{M} N_k \tag{115}$$

distinct nonzero elements, by a series of vectors, each of which is obtained from the preceding vector by a linear shift of each of its elements one position upward. At each successive iteration the lowest position of the vector is filled by a new element, and the uppermost element is lost. At each iteration the tensor $[T]_{N_1, N_2, \ldots, N_m, \ldots, N_M}$ is multiplied by the vector

$$[V_1]_{N_m} = \begin{bmatrix} v_1 \\ \ldots \\ v_n \\ \ldots \\ v_{N_m} \end{bmatrix}, \tag{116}$$

after obtaining the matrix $[P_1]_{L, N_m}$, which is the product of the kernel $[U]_L$ of tensor $[T]_{N_1, N_2, \ldots, N_m, \ldots, N_M}$ and the transposed vector $[V_1]_{N_m}$:

$$[P_1]_{L, N_m} = [U]_L \cdot [V_1]_{N_m}^t = \begin{bmatrix} u_1 \\ \ldots \\ u_l \\ \ldots \\ u_L \end{bmatrix} \cdot [v_1 \ldots v_n \ldots v_{N_m}] = \begin{bmatrix} v_1 \cdot u_1 & \cdots & v_{N_m} \cdot u_1 \\ \vdots & v_n \cdot u_l & \vdots \\ v_1 \cdot u_L & \cdots & v_{N_m} \cdot u_L \end{bmatrix} == \begin{bmatrix} \begin{bmatrix} u_1 \\ \ldots \\ u_l \\ \ldots \\ u_L \end{bmatrix} \cdot v_1 & \begin{bmatrix} u_1 \\ \ldots \\ u_l \\ \ldots \\ u_L \end{bmatrix} \cdot \end{bmatrix}$$

$$v_2 \ldots \begin{bmatrix} u_1 \\ \ldots \\ u_l \\ \ldots \\ u_L \end{bmatrix} \cdot v_n \ldots \begin{bmatrix} u_1 \\ \ldots \\ u_l \\ \ldots \\ u_L \end{bmatrix} \cdot v_{N_m - 1} \begin{bmatrix} u_1 \\ \ldots \\ u_l \\ \ldots \\ u_L \end{bmatrix} \cdot v_{N_m} \Bigg] \tag{117}$$

In its turn the tensor $[T]_{N_1, N_2, \ldots, N_m, \ldots, N_M}$ is represented as the product of the commutator

$$[Z]_{N_1, N_2, \ldots, N_m, \ldots, N_M, L} = \{ z_{n_1, n_2, \ldots, n_m, \ldots, n_M, l} \mid n_m \in [1, N_m], m \in [1, M], l \in [1, L] \} \tag{118}$$

and the kernel

$$[U]_L = \begin{bmatrix} u_1 \\ \ldots \\ u_l \\ \ldots \\ u_L \end{bmatrix}; \tag{119}$$

$$[T]_{N_1,N_2,\ldots,N_m,\ldots,N_M} = [Z]_{N_1,N_2,\ldots,N_m,\ldots,N_M,L} \cdot [U]_L = \left\{ \sum_{l=1}^{l=L} z_{n_1,n_2,\ldots,n_m,\ldots,n_M,l} \cdot u_l \mid n_m \in [1,N_m], m \in [1,M] \right\} \tag{120}$$

Obviously, at the previous iteration the tensor $[T]_{N_1,N_2,\ldots,N_m,\ldots,N_M}$ was multiplied by the vector

$$[V_0]_{N_m} = \begin{bmatrix} v_0 \\ \ldots \\ v_n \\ \ldots \\ v_{N_m-1} \end{bmatrix}, \tag{121}$$

and therefore there exists a matrix $[P_0]_{L,N_m}$ which is obtained by the multiplication of the kernel $[U]_L$ of the tensor $[T]_{N_1,N_2,\ldots,N_m,\ldots,N_M}$ by the transposed vector $[V_0]_{N_m}$:

$$[P_0]_{L,N_m} = [U]_L \cdot [V_0]_{N_m}^t = \begin{bmatrix} u_1 \\ \ldots \\ u_l \\ \ldots \\ u_L \end{bmatrix} \cdot [v_0 \ldots v_{n-1} \ldots v_{N_m-1}] = \begin{bmatrix} v_0 \cdot u_1 & \cdots & v_{N_m-1} \cdot u_1 \\ \vdots & v_{n-1} \cdot u_l & \vdots \\ v_0 \cdot u_L & \cdots & v_{N_m-1} \cdot u_L \end{bmatrix} = \begin{bmatrix} \begin{bmatrix} u_1 \\ \ldots \\ u_l \\ \ldots \\ u_L \end{bmatrix} \cdot v_0 & \begin{bmatrix} u_1 \\ \ldots \\ u_l \\ \ldots \\ u_L \end{bmatrix} \cdot v_{n-1} & \ldots & \begin{bmatrix} u_1 \\ \ldots \\ u_l \\ \ldots \\ u_L \end{bmatrix} \cdot v_{N_m-2} & \begin{bmatrix} u_1 \\ \ldots \\ u_l \\ \ldots \\ u_L \end{bmatrix} \cdot v_{N_m-1} \end{bmatrix} \tag{122}$$

The matrix $[P_1]_{L,N_m}$ is equivalent to the matrix $[P_0]_{L,N_m}$ linearly shifted to the left, where the rightmost column is the product of the kernel

$$[U]_L = \begin{bmatrix} u_1 \\ \ldots \\ u_l \\ \ldots \\ u_L \end{bmatrix} \tag{123}$$

and the new value $v_{N_m}$.

Each element $\{p1_{l,n} | l \in [1,L], n \in [1, N_m - 1]\}$ of the matrix $[P_1]_{L,N_m}$ is a copy of the element $\{p1_{l,n+1} | l \in [1,L], n \in [1, N_m - 1]\}$ of the matrix $[P]_{L,N_m}$ obtained in the previous iteration, and may be used in the current iteration, thereby obviating the need to use a multiplication operation to obtain them. Each element $\{p1_{l,N_m} | l \in [1,L]\}$ - which is an element of the rightmost column of the matrix $[P]_{L,N_m}$ is formed from the multiplication of each element of the kernel and the new value of $v_{N_m}$ of the new input vector. A general rule for the formation of the elements of the matrix $[P_i]_{L,N_m}$ from the elements of the matrix $[P_{i-1}]_{L,N_m}$ may be written as:

$$p_{i_{l,n}} = \begin{cases} p_{i-1_{l,n+1}}, & |n \in [1, N_m - 1] \\ u_l \cdot v_{N_m}, & |n = N_m \end{cases}, l \in [1,L], i \in [1, \infty[ \tag{124}$$

Thus, iteration $i \in [1, \infty[$ is written as:

$$\left\{ \begin{array}{l} p_{i_{l,n}} = \begin{cases} p_{i-1_{l,n+1}}, & |n \in [1, N_m - 1] \\ u_l \cdot v_{N_m}, & |n = N_m \end{cases}, l \in [1, L] \\ [R_i]_{N_1, N_2, \ldots, N_{m-1}, N_{m+1}, \ldots, N_M} = \left\{ \sum_{n=1}^{N_m} \sum_{l=1}^{L} z_{n_1, n_2, \ldots, n_{m-1}, n, n_{m+1}, \ldots, n_M, l} \cdot p_{i_{l,n}} \,|\, n_k \in [1, N_k], k \in \{[1, m-1], [m+1, M]\} \right\} \end{array} \right\}$$
(125)

Every such iteration consists of two steps – the first step contains all operations of multiplication and the formation of the matrix $[P_i]_{L, N_m}$, and in the second step the result $[R_i]_{N_1, N_2, \ldots, N_{m-1}, N_{m+1}, \ldots, N_M}$ is obtained via tensor contraction of the commutator and the new matrix $[P_i]_{L, N_m}$. Since the iterative formation of $[P_i]_{L, N}$ requires the multiplication of only the newest component $v_{N_m}$ of the vector $[V]_{N_m}$ by the kernel, the maximum number of operations in a single iteration is the number L of distinct nonzero elements of the original tensor $[T]_{N_1, N_2, \ldots, N_m, \ldots, N_M}$ rather than the total number of elements in the original tensor $[T]_{N_1, N_2, \ldots, N_m, \ldots, N_M}$, which is $\prod_{k=1}^{M} N_k$. The maximum number of addition operations is $\frac{N_m - 1}{N_m} \cdot \prod_{k=1}^{M} N_k$. Thus the ratio of the number of operations with a method using the decomposition of the vector into a kernel and a commutator to the number of operations required with a method that does not include such a decomposition is $Cm_+ \leq \frac{\frac{N_m - 1}{N_m} \prod_{k=1}^{M} N_k}{\frac{N_m - 1}{N_m} \prod_{k=1}^{M} N_k} = 1$ for addition and $Cm_* \leq \frac{L}{\prod_{k=1}^{M} N_k}$ for multiplication.

## 24. Definition of method and process for the multiplication of factored tensors.

The tensor

$$[T]_{N_1, N_2, \ldots, N_{\tilde{n}}, M_1, M_2, \ldots, M_{\widetilde{m}}} = \left\{ t_{n_1, n_2, \ldots, n_{\tilde{n}}, m_1, m_2, \ldots, m_{\widetilde{m}}} \,|\, n_i \in [1, N_i], i \in [1, \tilde{n}], m_j \in [1, M_j], j \in [1, \widetilde{m}] \right\}$$

(126)

of dimensions $(N_1, N_2, \ldots, N_{\tilde{n}}, M_1, M_2, \ldots, M_{\widetilde{m}})$ containing

$$L_t \leq \prod_{i=1}^{\tilde{n}} N_i \cdot \prod_{j=1}^{\widetilde{m}} M_j \tag{127}$$

distinct nonzero elements is to be multiplied by the tensor

$$[W]_{N_1, N_2, \ldots, N_{\tilde{n}}, K_1, K_2, \ldots, K_{\tilde{k}}} = \left\{ w_{n_1, n_2, \ldots, n_{\tilde{n}}, k_1, k_2, \ldots, k_{\tilde{k}}} \,\middle|\, n_i \in [1, N_i], i \in [1, \tilde{n}], k_j \in [1, K_j], j \in [1, \tilde{k}] \right\} \tag{128}$$

of dimensions $(N_1, N_2, \ldots, N_{\tilde{n}}, K_1, K_2, \ldots, K_{\tilde{k}})$ containing

$$L_w \leq \prod_{i=1}^{\tilde{n}} N_i \cdot \prod_{j=1}^{\tilde{k}} K_j \tag{129}$$

distinct nonzero elements.

The identical dimensions present in both tensors are contracted, and the others appear in the result, a tensor of dimensions $(M_1, M_2, \ldots, M_{\widetilde{m}}, K_1, K_2, \ldots, K_{\tilde{k}})$:

$$[R]_{M_1,M_2,...,M_{\widetilde{m}},K_1,K_2,...,K_{\widetilde{k}}} = [T]_{N_1,N_2,...,N_{\widetilde{n}},M_1,M_2,...,M_{\widetilde{m}}} \cdot [W]_{N_1,N_2,...,N_{\widetilde{n}},K_1,K_2,...,K_{\widetilde{k}}} =$$

$$\left\{ \sum_{n_1=1}^{N_1} \sum_{n_2=1}^{N_2} \cdots \sum_{n_{\widetilde{n}}=1}^{N_{\widetilde{n}}} t_{n_1,n_2,...,n_{\widetilde{n}},m_1,m_2,...,m_{\widetilde{m}}} \cdot w_{n_1,n_2,...,n_{\widetilde{n}},k_1,k_2,...,k_{\widetilde{k}}} \Big| m_i \in [1,M_i], i \in [1,\widetilde{m}], k_j \in [1,K_j], j \in [1,\widetilde{k}] \right\} =$$

$$= \left\{ \sum_{\substack{1 \leq n_l \leq N_l \\ 1 \leq l \leq \widetilde{n}}} t_{n_1,n_2,...,n_{\widetilde{n}},m_1,m_2,...,m_{\widetilde{m}}} \cdot w_{n_1,n_2,...,n_{\widetilde{n}},k_1,k_2,...,k_{\widetilde{k}}} \Big| m_i \in [1,M_i], i \in [1,\widetilde{m}], k_j \in [1,K_j], j \in [1,\widetilde{k}] \right\} =$$

$$\left\{ r_{m_1,m_2,...,m_{\widetilde{m}},k_1,k_2,...,k_{\widetilde{k}}} \Big| m_i \in [1,M_i], i \in [1,\widetilde{m}], k_j \in [1,K_j], j \in [1,\widetilde{k}] \right\} \tag{130}$$

The tensor $[T]_{N_1,N_2,...,N_{\widetilde{n}},M_1,M_2,...,M_{\widetilde{m}}}$ is written as the product of the commutator

$$[Z_t]_{N_1,N_2,...,N_{\widetilde{n}},M_1,M_2,...,M_{\widetilde{m}},L_t} =$$

$$\left\{ z_{t_{n_1,n_2,...,n_{\widetilde{n}},m_1,m_2,...,m_{\widetilde{m}},l_t}} \Big| n_i \in [1,N_i], i \in [1,\widetilde{n}], m_j \in [1,M_j], j \in [1,\widetilde{m}], l_t \in [1,L_t] \right\} \tag{131}$$

and the kernel

$$[U_t]_{L_t} = \left\{ u_{t_{l_t}} \Big| l_t \in [1,L_t] \right\}: \tag{132}$$

$$[T]_{N_1,N_2,...,N_{\widetilde{n}},M_1,M_2,...,M_{\widetilde{m}}} = [Z_t]_{N_1,N_2,...,N_{\widetilde{n}},M_1,M_2,...,M_{\widetilde{m}},L_t} \cdot [U_t]_{L_t} =$$

$$\left\{ \sum_{l=1}^{L_t} z_{t_{n_1,n_2,...,n_{\widetilde{n}},m_1,m_2,...,m_{\widetilde{m}},l}} \cdot u_{t_l} \Big| n_i \in [1,N_i], i \in [1,\widetilde{n}], m_j \in [1,M_j], j \in [1,\widetilde{m}] \right\} \tag{133}$$

The tensor $[W]_{N_1,N_2,...,N_{\widetilde{n}},K_1,K_2,...,K_{\widetilde{k}}}$ is written as the product of the commutator

$$[Z_w]_{N_1,N_2,...,N_{\widetilde{n}},K_1,K_2,...,K_{\widetilde{k}},L_w} =$$

$$\left\{ z_{w_{n_1,n_2,...,n_{\widetilde{n}},k_1,k_2,...,k_{\widetilde{k}},l_w}} \Big| n_i \in [1,N_i], i \in [1,\widetilde{n}], k_j \in [1,K_j], j \in [1,\widetilde{k}], l_w \in [1,L_w] \right\} \tag{134}$$

and the kernel

$$[U_w]_{L_w} = \left\{ u_{w_{l_w}} \Big| l_w \in [1,L_w] \right\}: \tag{135}$$

$$[W]_{N_1,N_2,...,N_{\widetilde{n}},K_1,K_2,...,K_{\widetilde{k}}} = [Z_w]_{N_1,N_2,...,N_{\widetilde{n}},K_1,K_2,...,K_{\widetilde{k}},L_w} \cdot [U_w]_{L_w} =$$

$$\left\{ \sum_{l=1}^{L_w} z_{w_{n_1,n_2,...,n_{\widetilde{n}},k_1,k_2,...,k_{\widetilde{k}},l}} \cdot u_{w_l} \Big| n_i \in [1,N_i], i \in [1,\widetilde{n}], k_j \in [1,K_j], j \in [1,\widetilde{k}] \right\} \tag{136}$$

Then the product of the tensor $[T]_{N_1,N_2,...,N_{\widetilde{n}},M_1,M_2,...,M_{\widetilde{m}}}$ and the tensor $[W]_{N_1,N_2,...,N_{\widetilde{n}},K_1,K_2,...,K_{\widetilde{k}}}$ may be written as:

$$[R]_{M_1,M_2,...,M_{\widetilde{m}},K_1,K_2,...,K_{\widetilde{k}}} = [T]_{N_1,N_2,...,N_{\widetilde{n}},M_1,M_2,...,M_{\widetilde{m}}} \cdot [W]_{N_1,N_2,...,N_{\widetilde{n}},K_1,K_2,...,K_{\widetilde{k}}} =$$

$$\left\{ \sum_{\substack{1 \leq n_l \leq N_l \\ 1 \leq l \leq \tilde{n}}} \left( \sum_{l_t=1}^{L_t} z_{t_{n_1,n_2,\ldots,n_{\tilde{n}},m_1,m_2,\ldots,m_{\widetilde{m}},l_t}} \cdot u_{t_{l_t}} \right) \cdot \left( \sum_{l_w=1}^{L_w} z_{w_{n_1,n_2,\ldots,n_{\tilde{n}},k_1,k_2,\ldots,k_{\widetilde{k}},l_w}} \cdot u_{w_{l_w}} \right) \middle| m_i \in [1, M_i], i \in [1, \widetilde{m}], k_j \in [1, K_j], j \in [1, \tilde{k}] \right\} =$$

$$\left\{ \sum_{\substack{1 \leq n_l \leq N_l \\ 1 \leq l \leq \tilde{n}}} \sum_{l_t=1}^{L_t} \sum_{l_w=1}^{L_w} z_{t_{n_1,n_2,\ldots,n_{\tilde{n}},m_1,m_2,\ldots,m_{\widetilde{m}},l_t}} \cdot u_{t_{l_t}} \cdot z_{w_{n_1,n_2,\ldots,n_{\tilde{n}},k_1,k_2,\ldots,k_{\widetilde{k}},l_w}} \cdot u_{w_{l_w}} \middle| m_i \in [1, M_i], i \in [1, \widetilde{m}], k_j \in [1, K_j], j \in [1, \tilde{k}] \right\} =$$

$$\left\{ \sum_{\substack{1 \leq n_l \leq N_l \\ 1 \leq l \leq \tilde{n} \\ 1 \leq l_t \leq L_t \\ 1 \leq l_w \leq L_w}} \left( z_{t_{n_1,n_2,\ldots,n_{\tilde{n}},m_1,m_2,\ldots,m_{\widetilde{m}},l_t}} \cdot u_{t_{l_t}} \right) \cdot \left( z_{w_{n_1,n_2,\ldots,n_{\tilde{n}},k_1,k_2,\ldots,k_{\widetilde{k}},l_w}} \cdot u_{w_{l_w}} \right) \middle| m_i \in [1, M_i], i \in [1, \widetilde{m}], k_j \in [1, K_j], j \in [1, \tilde{k}] \right\} =$$

$$\left\{ \sum_{\substack{1 \leq n_l \leq N_l \\ 1 \leq l \leq \tilde{n} \\ 1 \leq l_t \leq L_t \\ 1 \leq l_w \leq L_w}} \left( z_{t_{n_1,n_2,\ldots,n_{\tilde{n}},m_1,m_2,\ldots,m_{\widetilde{m}},l_t}} \cdot z_{w_{n_1,n_2,\ldots,n_{\tilde{n}},k_1,k_2,\ldots,k_{\widetilde{k}},l_w}} \right) \cdot \left( u_{t_{l_t}} \cdot u_{w_{l_w}} \right) \middle| m_i \in [1, M_i], i \in [1, \widetilde{m}], k_j \in [1, K_j], j \in [1, \tilde{k}] \right\} =$$

$$\left\{\sum_{\substack{1\leq l_t\leq L_t \\ 1\leq l_w\leq L_w}} \left(u_{t_{l_t}} \cdot u_{w_{l_w}} \cdot \sum_{\substack{1\leq n_l\leq N_l \\ 1\leq l\leq \widetilde{n}}} z_{t_{n_1,n_2,\ldots,n_{\widetilde{n}},m_1,m_2,\ldots,m_{\widetilde{m}},l_t}} \cdot z_{w_{n_1,n_2,\ldots,n_{\widetilde{n}},k_1,k_2,\ldots,k_{\widetilde{k}},l_w}}\right) \Bigg| m_i \in [1, M_i], i \in [1, \widetilde{m}], k_j \in [1, K_j], j \in [1, \widetilde{k}]\right\} =$$

$$\left\{\sum_{\substack{1\leq l_t\leq L_t \\ 1\leq l_w\leq L_w}} u_{l_t,l_w} \cdot z_{m_1,m_2,\ldots,m_{\widetilde{m}},l_t,k_1,k_2,\ldots,k_{\widetilde{k}},l_w} \Bigg| m_i \in [1, M_i], i \in [1, \widetilde{m}], k_j \in [1, K_j], j \in [1, \widetilde{k}]\right\} \qquad (137)$$

Here

$u_{l_t,l_w}$ is an element of the tensor $[U]_{L_t,L_w}$ of rank $(L_t, L_w)$ which is the result of multiplication of the kernels of the original tensors:

$$[U]_{L_t,L_w} = [U_t]_{L_t} \cdot [U_w]_{L_w} = \left\{u_{l_t,l_w} = u_{t_{l_t}} \cdot u_{w_{l_w}} \big| l_t \in [1, L_t], l_w \in [1, L_w]\right\}, \qquad (138)$$

and $z_{m_1,m_2,\ldots,m_{\widetilde{m}},l_t,k_1,k_2,\ldots,k_{\widetilde{k}},l_w}$ is an element of the tensor $[Z]_{M_1,M_2,\ldots,M_{\widetilde{m}},L_w,K_1,K_2,\ldots,K_{\widetilde{k}},L_t}$ of dimensions $(M_1, M_2, \ldots, M_{\widetilde{m}}, L_w, K_1, K_2, \ldots, K_{\widetilde{k}}, L_t)$ which is the result of multiplication of the commutators of the original tensors:

$$[Z]_{M_1,M_2,\ldots,M_{\widetilde{m}},L_t,K_1,K_2,\ldots,K_{\widetilde{k}},L_w} = [Z_t]_{N_1,N_2,\ldots,N_{\widetilde{n}},M_1,M_2,\ldots,M_{\widetilde{m}},L_t} \cdot [Z_w]_{N_1,N_2,\ldots,N_{\widetilde{n}},K_1,K_2,\ldots,K_{\widetilde{k}},L_w} =$$

$$\left\{z_{m_1,m_2,\ldots,m_{\widetilde{m}},l_t,k_1,k_2,\ldots,k_{\widetilde{k}},l_w} \Big| m_i \in [1, M_i], i \in [1, \widetilde{m}], k_j \in [1, K_j], j \in [1, \widetilde{k}], l_t \in [1, L_t], l_w \in [1, L_w]\right\} =$$

$$\left\{\sum_{\substack{1\leq n_l\leq N_l \\ 1\leq l\leq \widetilde{n}}} z_{n_1,n_2,\ldots,n_{\widetilde{n}},m_1,m_2,\ldots,m_{\widetilde{m}},l_t} \cdot w_{n_1,n_2,\ldots,n_{\widetilde{n}},k_1,k_2,\ldots,k_{\widetilde{k}},l_w} \Big| m_i \in [1, M_i], i \in [1, \widetilde{m}], k_j \in [1, K_j], j \in [1, \widetilde{k}], l_t \in [1, L_t], l_w \in [1, L_w]\right\} \qquad (139)$$

Thus, we may write:

$$[R]_{M_1,M_2,\ldots,M_{\widetilde{m}},K_1,K_2,\ldots,K_{\widetilde{k}}} = \left\{r_{m_1,m_2,\ldots,m_{\widetilde{m}},k_1,k_2,\ldots,k_{\widetilde{k}}} \Big| m_i \in [1, M_i], i \in [1, \widetilde{m}], k_j \in [1, K_j], j \in [1, \widetilde{k}]\right\} =$$

$$\left\{\sum_{\substack{1\leq l_t\leq L_t \\ 1\leq l_w\leq L_w}} u_{l_t,l_w} \cdot z_{m_1,m_2,\ldots,m_{\widetilde{m}},l_t,k_1,k_2,\ldots,k_{\widetilde{k}},l_w} \Bigg| m_i \in [1, M_i], i \in [1, \widetilde{m}], k_j \in [1, K_j], j \in [1, \widetilde{k}]\right\} =$$

$$[Z]_{M_1,M_2,\ldots,M_{\widetilde{m}},L_t,K_1,K_2,\ldots,K_{\widetilde{k}},L_w} \cdot [U]_{L_t,L_w} \qquad (140)$$

Thus, the multiplication of factored tensors requires the multiplication of all the elements of their kernels, which uses a maximum of $L_t \cdot L_w$ operations of scalar multiplications, and not $\prod_{i=1}^{\widetilde{n}} N_i \cdot \prod_{j=1}^{\widetilde{m}} M_j \cdot \prod_{j=1}^{\widetilde{k}} K_j$ as required for multiplication of two tensors without factorization.

## 25. Construction of a computational scheme to perform the iterative multiplication of a vector, matrix, or tensor by a factored vector, matrix or tensor employing a minimal number of addition operations.

It is shown above that the number of operations required for multiplication of a vector, matrix, or tensor by a vector, matrix, or tensor depends on the size of the kernel rather than the size of the original vector, matrix, or tensor, and may be substantially smaller than the number of operations required without preliminary factorization. Now we turn our attention to the fact that in many cases it is possible to substantially reduce the number of addition operations required at each iteration, and hence in the entire process of multiplication of a vector, matrix, or tensor by a vector, matrix, or tensor. In any of the possible cases such an iteration is carried out on the factored vector, matrix, or tensor and an element of a vector, matrix, or tensor. Henceforth any vector, matrix, or tensor will be represented in tensor form, where vectors and matrices are specific cases of tensors.

The process of iterative multiplication of a factored vector, matrix, or tensor by an arbitrary vector, matrix, or tensor is examined in linear-algebraic terms. The process of factorization of a tensor and specifically the process of describing a computational scheme for performing one iteration of multiplying a vector, matrix, or tensor by a factored tensor will require a system conducive to the description of algorithms or sequential processes. The symbol "$\Leftarrow$" will indicate the operation of assignment; for example the expression "$a \Leftarrow b$" will indicate that the value $b$ is assigned to the symbol $a$. The result of any logical operation of comparison will be either true or false, represented respectively as 1 or 0.

The entire process of factorization of a tensor and subsequent construction of a description of the computational scheme for performing one iteration of multiplication by a factored tensor consists of the operations described below.

In the general case, suppose that there exists a tensor

$$[\tilde{T}]_{N_1,N_2,\ldots,N_m,\ldots,N_M} = \{ \tilde{t}_{n_1,n_2,\ldots,n_m,\ldots,n_M} | n_m \in [1, N_m], m \in [1, M] \} \tag{141}$$

$[\tilde{T}]$ may be a matrix if $M = 2$, or a vector if $M = 1$.

If necessary, the elements of the tensor $[\tilde{T}]_{N_1,N_2,\ldots,N_m,\ldots,N_M}$ are rounded to a given precision $\varepsilon$:

$$[T]_{N_1,N_2,\ldots,N_m,\ldots,N_M} = \{ t_{n_1,n_2,\ldots,n_m,\ldots,n_M} | n_m \in [1, N_m], m \in [1, M] \} \Leftarrow$$
$$\left\{ \varepsilon \cdot round\left(\frac{\tilde{t}_{n_1,n_2,\ldots,n_m,\ldots,n_M}}{\varepsilon}\right) \middle| n_m \in [1, N_m], m \in [1, M] \right\} \tag{142}$$

To obtain the kernel and commutator the tensor $[T]_{N_1,N_2,\ldots,N_m,\ldots,N_M}$ is factored according to the algorithm described below. The initial conditions are as follows.

The length of the kernel is set to 0:

$$L \Leftarrow 0; \tag{143}$$

Initially the kernel is an empty vector of length zero:

$$[U]_L \Leftarrow [\,]; \tag{144}$$

The commutator is the tensor $[Y]_{N_1,N_2,...,N_m,...,N_M}$ of dimensions equal to the dimensions of the tensor $[T]_{N_1,N_2,...,N_m,...,N_M}$, all of whose elements are initially set equal to 0:

$$[Y]_{N_1,N_2,...,N_m,...,N_M} \Leftarrow \{0_{n_1,n_2,...,n_m,...,n_M} | n_m \in [1, N_m], m \in [1, M]\} \qquad (145)$$

The indices $n_1, n_2, ..., n_m, ..., n_M$ are initially set to 1:

$$n_1 \Leftarrow 1; n_2 \Leftarrow 1; ...; n_m \Leftarrow 1; ...; n_M \Leftarrow 1; \qquad (146)$$

$n_1, n_2, ..., n_m, ..., n_M n_m \in [1, N_m], m \in [1, M]$ This is the end of the initialization phase, and the beginning of the iterative phase of the computation.

Then for each set of indices $n_1, n_2, ..., n_m, ..., n_M$, where $n_m \in [1, N_m], m \in [1, M]$, the following operations are carried out:

Step 1:

If the element $t_{n_1,n_2,...,n_m,...,n_M}$ of the tensor $[T]_{N_1,N_2,...,N_m,...,N_M}$ is equal to 0, skip to step 3. Otherwise, go to step 2.

Step 2:

The length of the kernel is increased by 1:

$$L \Leftarrow L + 1; \qquad (147)$$

The element $t_{n_1,n_2,...,n_m,...,n_M}$ of the tensor $[T]_{N_1,N_2,...,N_m,...,N_M}$ is added to the kernel:

$$[U]_L \Leftarrow \begin{bmatrix} [U]_{L-1} \\ t_{n_1,n_2,...,n_m,...,n_M} \end{bmatrix} = \begin{bmatrix} [U]_{L-1} \\ u_L \end{bmatrix}; \qquad (148)$$

The intermediate tensor $[P]_{N_1,N_2,...,N_m,...,N_M}$ is formed, containing values of 0 in those positions where elements of the tensor $[T]_{N_1,N_2,...,N_m,...,N_M}$ are not equal to the last obtained element of the kernel $u_L$, and in all other positions values of $u_L$:

$$[P]_{N_1,N_2,...,N_m,...,N_M} = \{p_{n_1,n_2,...,n_m,...,n_M} | n_m \in [1, N_m], m \in [1, M]\} \Leftarrow u_L \cdot 0^{|[T]_{N_1,N_2,...,N_m,...,N_M} - u_L|} =$$
$$\{u_L \cdot 0^{|t_{\eta_1,\eta_2,...,\eta_m,...,\eta_M} - u_L|} | n_m \in [1, N_m], m \in [1, M]\} \qquad (149)$$

All elements of the tensor $[T]_{N_1,N_2,...,N_m,...,N_M}$ equal to the newly obtained element of the kernel are set equal to 0:

$$[T]_{N_1,N_2,...,N_m,...,N_M} \Leftarrow [T]_{N_1,N_2,...,N_m,...,N_M} - [P]_{N_1,N_2,...,N_m,...,N_M}; \qquad (150)$$

To the representation of the commutator, the tensor $[Y]_{N_1,N_2,...,N_m,...,N_M}$, the tensor $[P]_{N_1,N_2,...,N_m,...,N_M}$ is added:

$$[Y]_{N_1,N_2,...,N_m,...,N_M} \Leftarrow [Y]_{N_1,N_2,...,N_m,...,N_M} + [P]_{N_1,N_2,...,N_m,...,N_M} = \{y_{n_1,n_2,...,n_m,...,n_M} + p_{n_1,n_2,...,n_m,...,n_M} | n_m \in [1, N_m], m \in [1, M]\}; \qquad (151)$$

Next go to step 3.

Step 3:

The index $m$ is set equal to M:

$$m \Leftarrow M; \qquad (152)$$

Next go to step 4.

Step 4:

The index $n_m$ is increased by 1:

$$n_m \Leftarrow n_m + 1; \qquad (153)$$

If $n_m \leq N_m$, go to step 1. Otherwise, go to step 5.

Step 5:

The index $n_m$ is set equal to 1:

$$n_m \Leftarrow 1; \qquad (154)$$

The index $m$ is reduced by 1:

$$m \Leftarrow m - 1; \qquad (155)$$

If $m \geq 1$, go to step 4. Otherwise the process is terminated.

The results of the process described herein for the factorization of the tensor $[T]_{N_1,N_2,\ldots,N_m,\ldots,N_M}$ are the kernel $[U]_L$ and the tensor representing the commutator $[Y]_{N_1,N_2,\ldots,N_m,\ldots,N_M}$, which is the tensor contraction of the commutator $[Z]_{N_1,N_2,\ldots,N_m,\ldots,N_M,L}$ with the auxiliary vector

$$[Y]_L = \begin{bmatrix} 1 \\ 2 \\ \ldots \\ l \\ \ldots \\ L-1 \\ L \end{bmatrix}; \qquad (156)$$

$$[Y]_{N_1,N_2,\ldots,N_m,\ldots,N_M} = \left\{ \sum_{l=1}^{L} z_{n_1,n_2,\ldots,n_m,\ldots,n_M,l} \cdot l \,\middle|\, n_m \in [1, N_m], m \in [1, M] \right\} \qquad (157)$$

These two objects - $[U]_L$ and $[Y]_{N_1,N_2,\ldots,N_m,\ldots,N_M}$ - and the two parameters of operational delay and number of channels are the initial inputs to the process of constructing a computational structure to perform one iteration of multiplication by a factored tensor. An operational delay of $\delta$ indicates the number of system clock cycles required to perform the addition of two arguments in the computational scheme for which a computational system is described. The number of channels $\sigma$ determines the number of distinct independent vectors that compose the vector that is multiplied by the factored tensor. Then for N elements, the elements $\{M | M \in [1, \infty]\}$ of channel K, where $1 \leq K \leq N$, are resent in the resultant vector as elements $\{K + (M - 1) \cdot N | K \in [1, N], M \in [0, \infty]\}$.

The process of constructing a description of the computational system for performing one iteration of multiplication by a factored tensor contains the steps described below.

For a given kernel $[U]_L$, commutator tensor $[Y]_{N_1,N_2,\ldots,N_m,\ldots,N_M}$, operational delay $\delta$ and number of channels $\sigma$, the initialization of this process consists of the following steps.

The empty matrix

$$[Q]_{0,4} \Leftarrow [\,];  \tag{158}$$

is initialized, to which the combinations

$$[P]_4 = [p_1\ p_2\ p_3\ p_4]  \tag{159}$$

are to be added. These combinations are represented by vectors of length 4. In every such vector the first element $p_1$ is the identifier or index of the combination. These numbers are an extension of the numeration of elements of the kernel. Thus the index of the first combination is $L + 1$, and each successive combination has an index one more than the preceding combination:

$$q_{1,1} = L + 1, q_{n,1} = q_{n-1,1} + 1, n > 1  \tag{160}$$

The second element $p_2$ of each combination is an element of the subset

$$\{[Y]_{n_1,N_2,\ldots,N_m,\ldots,N_M} | n_1 \in [1, N_1 - p_4 - 1], p_4 \in [1, N_1 - 1]\}  \tag{161}$$

of elements of the commutator tensor $[Y]_{N_1,N_2,\ldots,N_m,\ldots,N_M}$ as shown below.

The third element $p_3$ of the combination represents an element of the subset

$$\{[Y]_{n_1,N_2,\ldots,N_m,\ldots,N_M} | n_1 \in [p_4, N_1], p_4 \in [1, N_1 - 1]\}  \tag{162}$$

of elements of the commutator tensor $[Y]_{N_1,N_2,\ldots,N_m,\ldots,N_M}$ as shown below.

The fourth element $p_4 \in [1, N_1 - 1]$ of the combination represents the distance along the dimension $N_1$ between the elements equal to $p_2$ and $p_3$ in the commutator tensor $[Y]_{N_1,N_2,\ldots,N_m,\ldots,N_M}$.

The index of the first element of the combination is set equal to the dimension of the kernel:

$$p_1 \Leftarrow L;  \tag{163}$$

Here ends the initialization and begins the iterative section of the process of constructing a description of the computational structure.

Step 1:

The variable containing the number of occurrences of the most frequent combination is set equal to 0:

$$\alpha \Leftarrow 0;  \tag{164}$$

Go to step 2.

Step 2:

The index of the second element is set equal to 1:

$$p_2 \Leftarrow 1;  \tag{165}$$

Go to step 3.

Step 3:

The index of the third element of the combination is set equal to 1:

$$p_3 \Leftarrow 1; \qquad (166)$$

Go to step 4.

Step 4:

The index of the fourth element is set equal to 1:

$$p_4 \Leftarrow 1; \qquad (167)$$

Go to step 5.

Step 5:

The variable containing the number of occurrences of the combination is set equal to 0:

$$\beta \Leftarrow 0; \qquad (168)$$

The indices $n_1, n_2, \ldots, n_m, \ldots, n_M$ are set equal to 1:

$$n_1 \Leftarrow 1; n_2 \Leftarrow 1; \ldots; n_m \Leftarrow 1; \ldots; n_M \Leftarrow 1; \qquad (169)$$

Go to step 6.

Step 6:

The elements of the commutator tensor $[Y]_{N_1,N_2,\ldots,N_m,\ldots,N_M}$ form the vector

$$[\Theta]_{N_M} = \{\theta_\eta | \eta \in [1, N_M]\} \Leftarrow \{y_{n_1,n_2,\ldots,n_m,\ldots,\eta} | \eta \in [1, N_M]\} \qquad (170)$$

Go to step 7.

Step 7:

If $\theta_{n_M} \neq p_2$ or $\theta_{n_M+p_4} \neq p_3$, skip to step 9. Otherwise, go to step 8.

Step 8:

The variable containing the number of occurrences of the combination is increased by 1:

$$\beta \Leftarrow \beta + 1; \qquad (171)$$

The elements $\theta_{n_M}$ and $\theta_{n_M+p_4}$ of the vector $[\Theta]_{N_M}$ are set equal to 0:

$$\theta_{n_M} \Leftarrow 0; \qquad (172)$$

$$\theta_{n_M+p_4} \Leftarrow 0; \qquad (173)$$

If $\beta \leq \alpha$, skip to step 10. Otherwise, go to step 9.

Step 9:

The variable containing the number of occurrences of the most frequently occurring combination is set equal to the number of occurrences of the combination:

$$\alpha \Leftarrow \beta; \tag{174}$$

The most frequently occurring combination is recorded:

$$[P]_4 \Leftarrow [p_1 + 1 \; p_2 \; p_3 \; p_4]; \tag{175}$$

Go to step 10.

Step 10:

The index $m$ is set equal to M:

$$m \Leftarrow M; \tag{176}$$

Go to step 11.

Step 11:

The index $n_m$ is increased by 1:

$$n_m \Leftarrow n_m + 1; \tag{177}$$

If $n_m \leq N_m$, then if $m = M$, go to step 7, and if $m < M$, go to step 6. If $n_m > N_m$, go to step 12.

Step 12:

The index $n_m$ is set equal to 1:

$$n_m \Leftarrow 1; \tag{178}$$

The index $m$ is decreased by 1:

$$m \Leftarrow m - 1; \tag{179}$$

If $m \geq 1$, go to step 11. Otherwise, go to step 13.

Step 13:

The index of the fourth element of the combination is increased by 1:

$$p_4 \Leftarrow p_4 + 1; \tag{180}$$

If $p_4 < N_M$, go to step 4. Otherwise go to step 14.

Step 14:

The index of the third element of the combination is increased by 1:

$$p_3 \Leftarrow p_3 + 1; \tag{181}$$

If $p_3 \leq p_1$, go to step 3. Otherwise, go to step 15.

Step 15:

The index of the second element of the combination is increased by 1:

$$p_2 \Leftarrow p_2 + 1; \tag{182}$$

If $p_2 \leq p_1$, go to step 2. Otherwise, go to step 16.

Step 16:

If $\alpha > 0$, go to step 17. Otherwise, skip to step 18.

Step 17:

The index of the first element is increased by 1:

$$p_1 \Leftarrow p_1 + 1; \tag{183}$$

To the matrix of combinations the most frequently occurring combination is added:

$$[Q]_{p_1-L,4} \Leftarrow \begin{bmatrix} [Q]_{p_1-L-1,4} \\ [P]_4 \end{bmatrix}; \tag{184}$$

Go to step 18.

Step 18:

The indices $n_1, n_2, \ldots, n_m, \ldots, n_M$ are set equal to 1:

$$n_1 \Leftarrow 1; n_2 \Leftarrow 1; \ldots; n_m \Leftarrow 1; \ldots; n_M \Leftarrow 1; \tag{185}$$

Go to step 19.

Step 19:

If $y_{n_1,n_2,\ldots,n_m,\ldots,n_M} \neq p_2$ or $y_{n_1,n_2,\ldots,n_m,\ldots,n_M+p_4} \neq p_3$, skip to step 21. Otherwise, go to step 20.

Step 20:

The element $y_{n_1,n_2,\ldots,n_m,\ldots,n_M}$ of the commutator tensor $[Y]_{N_1,N_2,\ldots,N_m,\ldots,N_M}$ is set equal to 0:

$$y_{n_1,n_2,\ldots,n_m,\ldots,n_M} \Leftarrow 0; \tag{186}$$

The element $y_{n_1,n_2,\ldots,n_m,\ldots,n_M+p_4}$ of the commutator tensor $[Y]_{N_1,N_2,\ldots,N_m,\ldots,N_M}$ is set equal to the current value of the index of the first element of the combination:

$$y_{n_1,n_2,\ldots,n_m,\ldots,n_M} \Leftarrow p_1; \tag{187}$$

Go to step 21.

Step 21:

The index $m$ is set equal to M:

$$m \Leftarrow M; \qquad (188)$$

Go to step 22.

Step 22:

The index $n_m$ is increased by 1:

$$n_m \Leftarrow n_m + 1; \qquad (189)$$

If $m < M$ and $n_m \leq N_m$ or $m = M$ and $n_m \leq N_m - p_4$, then go to step 19. Otherwise, go to step 23.

Step 23:

The index $n_m$ is set equal to 1:

$$n_m \Leftarrow 1; \qquad (190)$$

The index $m$ is decreased by 1:

$$m \Leftarrow m - 1; \qquad (191)$$

If $m \geq 1$, go to step 22. Otherwise, go to step 24.

Step 24:

At the end of each row of the matrix of combinations, append a zero element:

$$[Q]_{p_1-L,5} \Leftarrow \left[ [Q]_{p_1-L,4} \begin{bmatrix} 0 \\ \ldots \\ 0 \\ \ldots \\ 0 \end{bmatrix}_{p_1-L} \right]; \qquad (192)$$

Go to step 25.

Step 25:

The variable $\Omega$ is set equal to the number $p_1 - L$ of rows in the resulting matrix of combinations $[Q]_{p_1-L,5}$:

$$\Omega \Leftarrow p_1 - L; \qquad (193)$$

Go to step 26.

Step 26:

The index $\mu$ is set equal to 1:

$$\mu \Leftarrow 1; \qquad (194)$$

Go to step 27.

Step 27:

The index $\xi$ is set equal to one more than the index $\mu$:

$$\xi \Leftarrow \mu + 1; \tag{195}$$

Go to step 28.

Step 28:

If $p_{\mu,1} \neq q_{\xi,2}$, skip to step 30. Otherwise, go to step 29.

Step 29:

The element $q_{\xi,4}$ of the matrix of combinations is decreased by the value of the operational delay $\delta$:

$$q_{\xi,4} \Leftarrow q_{\xi,4} - \delta; \tag{196}$$

Go to step 30.

Step 20:

If $p_{\mu,1} \neq q_{\xi,3}$, skip to step 32. Otherwise, go to step 31.

Step 31:

The element $q_{\xi,5}$ of the matrix of combinations is decreased by the value of the operational delay $\delta$:

$$q_{\xi,5} \Leftarrow q_{\xi,5} - \delta; \tag{197}$$

Go to step 32.

Step 32:

The index $\xi$ is increased by 1:

$$\xi \Leftarrow \xi + 1; \tag{198}$$

If $\xi \leq \Omega$, go to step 28. Otherwise go to step 33.

Step 33:

The index $\mu$ is increased by 1:

$$\mu \Leftarrow \mu + 1; \tag{199}$$

If $\mu < \Omega$, go to step 27. Otherwise go to step 34.

Step 34:

The cumulative operational delay of the computational scheme is set equal to 0:

$$\Delta \Leftarrow 0; \tag{200}$$

The index $\mu$ is set equal to 1:

$$\mu \Leftarrow 1; \tag{201}$$

Go to step 35.

Step 35:

The index $\xi$ is set equal to 4:

$$\xi \Leftarrow 4; \tag{202}$$

Go to step 36.

Step 36:

If $\Delta > q_{\mu,\xi}$, skip to step 38. Otherwise, go to step 37.

Step 37:

The value of the cumulative operational delay of the computational scheme is set equal to the value of $q_{\mu,\xi}$:

$$\Delta \Leftarrow q_{\mu,\xi}; \tag{203}$$

Go to step 38.

Step 38:

The index $n$ is increased by 1:

$$\xi \Leftarrow \xi + 1; \tag{204}$$

If $\xi \leq 5$, go to step 36. Otherwise, go to step 39.

Step 39:

The index $\mu$ is increased by 1:

$$\mu \Leftarrow \mu + 1; \tag{205}$$

If $\mu < \Omega$, go to step 35. Otherwise, go to step 40.

Step 40:

To each element of the two rightmost columns of the matrix of combinations, add the calculated value of the cumulative operational delay of the computational scheme:

$$\{q_{\mu,\xi} \Leftarrow q_{\mu,\xi} + \Delta | \mu \in [1, \Omega], \xi \in [4,5]\}; \tag{206}$$

Go to step 41.

Step 41:

After step 24, any subset $\{y_{n_1,n_2,\ldots,n_m,\gamma}|m \in [1, M-1], \gamma \in [1, N_M]\}$ of elements of the commutator tensor $[Y]_{N_1,N_2,\ldots,N_m,\ldots,N_M}$ contains no more than one nonzero element. These elements contain the result of the constructed computational scheme represented by the matrix of combinations $[Q]_{\Omega,5}$. Moreover, the position of each such element along the dimension $n_M$ determines the delay in calculating each of the elements relative to the input and each other.

The tensor $[D]_{N_1,N_2,\ldots,N_m,\ldots,N_{M-1}}$ of dimension $(N_1, N_2, \ldots, N_m, \ldots, N_{M-1})$, containing the delay in calculating each corresponding element of the resultant may be found using the following operation:

$$[D]_{N_1,N_2,\ldots,N_m,\ldots,N_{M-1}} = \{d_{n_1,n_2,\ldots,n_m,n_{M-1}}|m \in [1, M-1], n_m \in [1, N_M]\} \Leftarrow$$
$$\left\{\sum_{\gamma=1}^{n_{M-1}} \gamma \cdot \left(1 - 0^{|y_{n_1,n_2,\ldots,n_m,\gamma}|}\right) \bigg| m \in [1, M-1], n_m \in [1, N_M]\right\} \quad (207)$$

The indices of the combinations comprising the resultant tensor $[R]_{N_1,N_2,\ldots,N_m,\ldots,N_{M-1}}$ of dimensions $(N_1, N_2, \ldots, N_m, \ldots, N_{M-1})$ may be determined using the following operation:

$$[R]_{N_1,N_2,\ldots,N_m,\ldots,N_{M-1}} = \{r_{n_1,n_2,\ldots,n_m,n_{M-1}}|m \in [1, M-1], n_m \in [1, N_M]\} \Leftarrow$$
$$\left\{y_{n_1,n_2,\ldots,n_m,n_{M-1},d_{n_1,n_2,\ldots,n_m,n_{M-1}}} \bigg| m \in [1, M-1], n_m \in [1, N_M]\right\} \quad (208)$$

Go to step 42.

Step 42:

Each of the elements of the two rightmost columns of the matrix of combinations is multiplied by the number of channels $\sigma$:

$$\{q_{\mu,\xi} \Leftarrow \sigma \cdot q_{\mu,\xi} | \mu \in [1, \Omega], \xi \in [4,5]\}; \quad (209)$$

The construction of the computational scheme is concluded. The results of this process are:

- The cumulative value of the operational delay $\Delta$;

- The matrix of combinations $[Q]_{\Omega,5}$;

- The tensor of indices $[R]_{N_1,N_2,\ldots,N_m,\ldots,N_{M-1}}$;

- The tensor of delays $[D]_{N_1,N_2,\ldots,N_m,\ldots,N_{M-1}}$.

## 26. Construction of an algorithm for the iterative multiplication of a tensor by a vector using the constructed computational scheme.

The construction in the previous section of a representation of the computational scheme contains sufficient information for the synthesis of a manual or automated system or program. Any such program consists of two parts – the initialization and the iterative section.

The initialization consists of allocating memory within the computational system for the storage of copies of all components with the corresponding time delays. The iterative section is contained within the waiting loop or is activated by an interrupt caused by the arrival of a new element of the input tensor. It results in the movement through the memory of the components that have already been calculated, the performance of operations represented by the rows of the matrix of combinations $[Q]_{\Omega,5}$ and the computation of the result. The following is a more detailed discussion of one of the many possible examples of such a process.

For a given initial vector of length $N_M$, number $\sigma$ of channels, cumulative operational delay $\Delta$, matrix $[Q]_{\Omega,5}$ of combinations, kernel vector $[U]_{\omega_{1,1}-1}$, tensor $[R]_{N_1,N_2,...,N_m,...,N_{M-1}}$ of indices and tensor $[D]_{N_1,N_2,...,N_m,...,N_{M-1}}$ of delays, the steps given below constitute a process for iterative multiplication.

Step 1 (initialization):

A two-dimensional array is allocated and initialized, represented here by the matrix $[\Phi]_{\omega_{\Omega,1},\sigma\cdot(N_M+\Delta)}$ of dimension $\omega_{\Omega,1}, \sigma \cdot (N_M + \Delta)$:

$$[\Phi]_{\omega_{\Omega,1},\sigma\cdot(N_M+\Delta)} = \{\varphi_{\mu,\eta} \Leftarrow 0 | \mu \in [1, \omega_{\Omega,1}], \eta \in [1, \sigma \cdot (N_M + \Delta)]\}; \tag{210}$$

The variable $\xi$, serving as the indicator of the current column of the matrix $[\Phi]_{\omega_{\Omega,1},\sigma\cdot(N_M+\Delta)}$, is initialized:

$$\xi \Leftarrow \sigma \cdot (N_M + \Delta); \tag{211}$$

Go to step 2.

Step 2:

Obtain the value of the next element of the input vector and record it in variable $\chi$.

The indicator $\xi$ of the current column of the matrix $[\Phi]_{\omega_{\Omega,1},\sigma\cdot(N_M+\Delta)}$ is cyclically shifted to the right:

$$\xi \Leftarrow 1 + (\xi) mod(\sigma \cdot (N_M + \Delta)); \tag{212}$$

The product of the variable $\chi$ by the elements of the kernel $[U]_{\omega_{1,1}-1}$ are obtained and recorded in the corresponding positions of the matrix $[\Phi]_{\omega_{\Omega,1},\sigma\cdot(N_M+\Delta)}$:

$$\{\varphi_{\mu,\xi} \Leftarrow \chi \cdot u_\mu | \mu \in [1, \omega_{1,1} - 1]\}; \tag{213}$$

The variable $\mu$, serving as an indicator of the current row of the matrix of combinations $[Q]_{\Omega,5}$ is initialized:

$$\mu \Leftarrow 1; \tag{214}$$

Go to step 3.

Step 3:

Find the new value of combination $\mu$ and assign it to the element $\varphi_{\mu+\omega_{1,1}-1,\xi}$ of the matrix $[\Phi]_{\omega_{\Omega,1},\sigma\cdot(N_M+\Delta)}$:

$$\varphi_{\mu+\omega_{1,1}-1,\xi} \Leftarrow \sum_{\tau=0}^{1} \varphi_{q_{\mu,2+\tau},1+(\xi-1-q_{\mu,4+\tau})mod(\sigma\cdot(N_M+\Delta))}; \tag{215}$$

The variable $\mu$ is increased by 1:

$$\mu \Leftarrow \mu + 1; \tag{216}$$

Go to step 4.

Step 4:

If $\mu \leq \Omega$, go to step 3. Otherwise, go to step 5.

Step 5:

The elements of the tensor $[P]_{N_1,N_2,\ldots,N_m,\ldots,N_{M-1}}$, containing the result, are determined:

$$[P]_{N_1,N_2,\ldots,N_m,\ldots,N_{M-1}} = \left\{ \rho_{n_1,n_2,\ldots,n_m,n_{M-1}} \Leftarrow \varphi_{r_{n_1,n_2,\ldots,n_m,n_{M-1}},1+\left(\xi-1-d_{n_1,n_2,\ldots,n_m,n_{M-1}}\right)\mathrm{mod}(\sigma\cdot(N_M+\Delta))} \middle| m \in [1, M-1], n_m \in [1, N_M] \right\}; \tag{217}$$

If all elements of the input vector have been processed, the process is concluded and the tensor $[P]_{N_1,N_2,\ldots,N_m,\ldots,N_{M-1}}$ is the product of the multiplication. Otherwise, go to step 2.

## 27. Construction of a block diagram for a system to perform iterative multiplication of a tensor by a vector according to the constructed computational scheme.

It was remarked in the previous section that the obtained computational scheme may be used to construct an algorithm for a corresponding program or a block diagram for a corresponding calculation system. One possible method of constructing such an algorithm was discussed in detail in the preceding section.

Discussed in this section will be one of the many possible ways of synthesizing a block diagram for a computational system defined by the computational scheme.

Any equipment system is composed of basic components connected in a specific way. The computational scheme is constructed for a given set of components and their properties. There are a total of three elements. For a synchronous digital system these are: a time delay element of one system count, a two-input summator with an operational delay of $\delta$ system counts, and a scalar multiplication operator. For an asynchronous analog system or an impulse system, these are a line of time delay between successive elements of the input vector, a two-input summator with a time delay of $\delta$ element counts, and a scalar multiplication component in the form of an amplifier or attenuator.

Thus, for an input vector of length $N_M$, number of channels $\sigma$, matrix $[Q]_{\Omega,5}$ of combinations, kernel vector $[U]_{\omega_{1,1}-1}$, tensor $[R]_{N_1,N_2,\ldots,N_m,\ldots,N_{M-1}}$ of indices and tensor $[D]_{N_1,N_2,\ldots,N_m,\ldots,N_{M-1}}$ of time delays, the steps shown below describe the process of formation of a schematics description for a system for the iterative multiplication of a vector by a tensor. For convenience in representing the process of synthesis, the following convention is introduced: any variable enclosed in triangular brackets, for example $\langle \xi \rangle$, represents the alphanumeric value currently assigned to that variable. This value in tern may be part of a value identifying a node or component of the block diagram. Alphanumeric strings will be enclosed in double quotes. Step 1:

The initially empty block diagram of the system is generated, and within it the node "N_0" which is the entry port for the elements of the input vector.

The variable $\xi$ is initialized, serving as the indicator of the current element of the kernel $[U]_{\omega_{1,1}-1}$:

$$\xi \Leftarrow 1; \tag{218}$$

Go to step 2.

Step 2:

To the block diagram of the apparatus add the node "N_⟨$\xi$⟩_0" and the multiplier "M_⟨$\xi$⟩" the entry to which is connected to the node "N_0" , and the outlet to the node "N_⟨$\xi$⟩_0".

The value of the indicator $\xi$ of the current element of the kernel $[U]_{\omega_{1,1}-1}$ is increased by 1:

$$\xi \Leftarrow \xi + 1; \tag{219}$$

Go to step 3.

Step 3:

If $\xi \geq \omega_{1,1}$ , go to step 2. Otherwise, go to step 4.

Step 4:

The variable $\mu$ is initialized, serving as an indicator of the current row of the matrix of combinations $[Q]_{\Omega,5}$:

$$\mu \Leftarrow 1; \tag{220}$$

Go to step 5.

Step 5:

To the block diagram of the system add the node "N_⟨q_{$\mu,1$}⟩_0" and the summator "A_⟨q_{$\mu,1$}⟩" the exit of which is connected to the node "N_⟨q_{$\mu,1$}⟩_0".

The variable $\xi$ is initialized, serving as an indicator of the number of the entry to the summator "A_⟨q_{$\mu,1$}⟩":

$$\xi \Leftarrow 1; \tag{221}$$

Go to step 6.

Step 6:

The variable $\gamma$ is initialized, storing the delay component index offset:

$$\gamma \Leftarrow 0; \tag{222}$$

Go to step 7.

Step 7:

If the node $N\_\langle q_{\mu,\xi+1}\rangle\_\langle q_{\mu,\xi+3} - \gamma\rangle$ has already been initialized, skip to step 12. Otherwise, go to step 8.

Step 8:

To the block diagram of the system add the node $N\_\langle q_{\mu,\xi+1}\rangle\_\langle q_{\mu,\xi+3} - \gamma\rangle$ and a unit delay $Z\_\langle q_{\mu,\xi+1}\rangle\_\langle q_{\mu,\xi+3} - \gamma\rangle$, the exit of which is connected to the node $N\_\langle q_{\mu,\xi+1}\rangle\_\langle q_{\mu,\xi+3} - \gamma\rangle$.

If $\gamma > 0$, go to step 10. Otherwise, go to step 9.

Step 9:

Entry number $\xi$ of the summator "$A\_\langle q_{\mu,1}\rangle$" is connected to the node $N\_\langle q_{\mu,\xi+1}\rangle\_\langle q_{\mu,\xi+3}\rangle$.

Go to step 11

Step 10:

The entry to the element of one count delay $Z\_\langle q_{\mu,\xi+1}\rangle\_\langle q_{\mu,\xi+3} - \gamma\rangle$ is connected to the node $N\_\langle q_{\mu,\xi+1}\rangle\_\langle q_{\mu,\xi+3} - \gamma + 1\rangle$.

Go to step 11.

Step 11:

The delay component index offset is increased by 1:

$$\gamma \Leftarrow \gamma + 1; \tag{223}$$

If $\gamma < 2$, go to step 7. Otherwise, go to step 12.

Step 12:

The indicator $\mu$ of the current row of the matrix of combinations $[Q]_{\Omega,5}$ is increased by 1:

$$\mu \Leftarrow \mu + 1; \tag{224}$$

If $\mu \leq \Omega$, go to step 5. Otherwise, go to step 13.

Step 13:

From each element of the delay tensor $[D]_{N_1,N_2,\ldots,N_m,\ldots,N_{M-1}}$ subtract the value of the least element of that matrix:

$$[D]_{N_1,N_2,\ldots,N_m,\ldots,N_{M-1}} \Leftarrow [D]_{N_1,N_2,\ldots,N_m,\ldots,N_{M-1}} - min(d_{n_1,n_2,\ldots,n_m,\ldots,n_{M-1}}|m \in [1, M-1], n_m \in [1, N_m]); \tag{225}$$

The indices $n_1, n_2, \ldots, n_m, \ldots, n_{M-1}$ are set equal to 1:

$$n_1 \Leftarrow 1; n_2 \Leftarrow 1; \ldots; n_m \Leftarrow 1; \ldots; n_M \Leftarrow 1; \tag{226}$$

Go to step 14.

Step 14:

To the block diagram of the system add the node N_⟨n₁⟩_⟨n₂⟩_..._⟨nₘ⟩_..._⟨n_{M-1}⟩ at the exit of the element $n_1, n_2, \ldots, n_m, \ldots, n_{M-1}$ of the result of multiplying the tensor by the vector.

Go to step 15.

Step 15:

The variable $\gamma$ is initialized, storing the delay component index offset :

$$\gamma \Leftarrow 0; \tag{227}$$

Go to step 16.

Step 16:

If the node $N\_\langle r_{n_1,n_2,\ldots,n_m,\ldots,n_{M-1}}\rangle\_\langle d_{n_1,n_2,\ldots,n_m,\ldots,n_{M-1}} - \gamma \rangle$ has already been initialized, skip to step 21. Otherwise, go to step 16.

Step 17:

To the block diagram of the system introduce the node $N\_\langle r_{n_1,n_2,\ldots,n_m,\ldots,n_{M-1}}\rangle\_\langle d_{n_1,n_2,\ldots,n_m,\ldots,n_{M-1}} - \gamma \rangle$ and the unit delay $Z\_\langle r_{n_1,n_2,\ldots,n_m,\ldots,n_{M-1}}\rangle\_\langle d_{n_1,n_2,\ldots,n_m,\ldots,n_{M-1}} - \gamma \rangle$.

If $\gamma > 0$, Go to step 18. Otherwise skip to step 19.

Step 18:

The exit of the delay element $Z\_\langle r_{n_1,n_2,\ldots,n_m,\ldots,n_{M-1}}\rangle\_\langle d_{n_1,n_2,\ldots,n_m,\ldots,n_{M-1}} - \gamma \rangle$ is connected to the node N_⟨n₁⟩_⟨n₂⟩_..._⟨nₘ⟩_..._⟨n_{M-1}⟩.

Go to step 19.

Step 19:

The exit of the delay element $Z\_\langle r_{n_1,n_2,\ldots,n_m,\ldots,n_{M-1}}\rangle\_\langle d_{n_1,n_2,\ldots,n_m,\ldots,n_{M-1}} - \gamma \rangle$ is connected to the node $N\_\langle r_{n_1,n_2,\ldots,n_m,\ldots,n_{M-1}}\rangle\_\langle d_{n_1,n_2,\ldots,n_m,\ldots,n_{M-1}} - \gamma + 1 \rangle$.

Go to step 20.

Step 20:

The delay component index offset is increased by 1:

$$\gamma \Leftarrow \gamma + 1; \tag{228}$$

Go to step 16.

Step 21:

If $\gamma > 0$, skip to step 23. Otherwise, go to step 22.

Step 22:

The node $N\_\langle r_{n_1,n_2,...,n_m,...,n_{M-1}}\rangle\_\langle d_{n_1,n_2,...,n_m,...,n_{M-1}} - \gamma\rangle$ is connected to the node $N\_\langle n_1\rangle\_\langle n_2\rangle\_..._\langle n_m\rangle\_..._\langle n_{M-1}\rangle$.

Go to step 23.

Step 23:

The index $m$ is set equal to M:

$$m \Leftarrow M; \tag{229}$$

Go to step 24.

Step 24:

The index $n_m$ is increased by 1:

$$n_m \Leftarrow n_m + 1; \tag{230}$$

If $m < M$ and $n_m \leq N_m$ then go to step 14. Otherwise, go to step 25.

Step 25:

The index $n_m$ is set equal to 1:

$$n_m \Leftarrow 1; \tag{231}$$

The index $m$ is decreased by 1:

$$m \Leftarrow m - 1; \tag{232}$$

If $m \geq 1$, go to step 24. Otherwise, the process is concluded.

## 28. Architecture of a parametrically configurable system for the iterative multiplication of a vector by a tensor.

On the basis of the methods described above for the development of a computational scheme, block diagram, and algorithm of iterative multiplication of a tensor by a vector, we arrive at the final architecture of the system.

Parametrically configurable system 1' is presented in Figure 1.

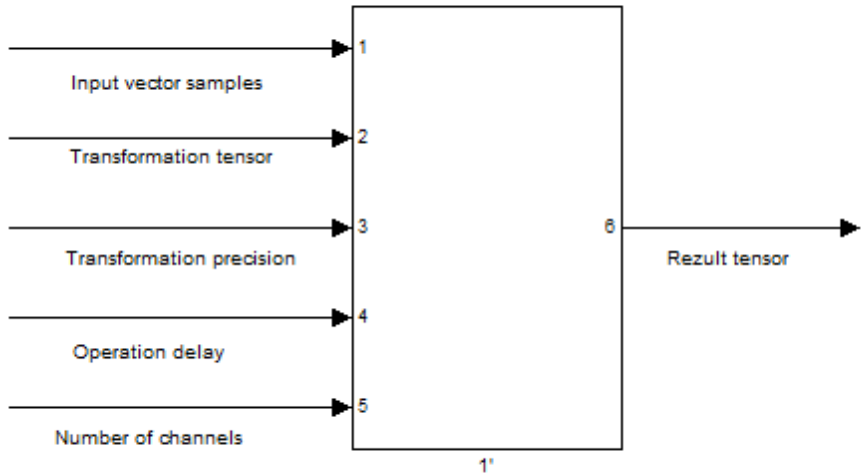

*Figure 1. Parametrically configurable system 1' for iterative multiplication of a vector by a tensor.*

The system 1' has five inputs and one output. Input 1 of system 1' receives current element $\chi$ (213) of the input vectors of each channel. Input 2 of system 1' receives current values of the transformation tensor $[\tilde{T}]_{N_1,N_2,...,N_m,...,N_M}$ (141). Input 3 of system 1' receives current value of the rounding precision $\varepsilon$ (142). Input 4 of system 1' receives current value of the operational delay $\delta$ (196). Input 5 of system 1' receives current value of the number of channels $\sigma$ (209). System 1' output 6 contains the resultant tensor $[P]_{N_1,N_2,...,N_m,...,N_{M-1}}$ (217).

The internal system 1' architecture is presented in Figure 2 and consists of the following blocks:

1 – Precision converter 1;

2 – Factorizing unit 2;

3 – Multiplier set 3;

4 – Reducer 4;

5 – Summator set 5;

6 - Indexer 6;

7 - Positioner 7;

8 – Delay component set 8;

9 – Result extractor 9.

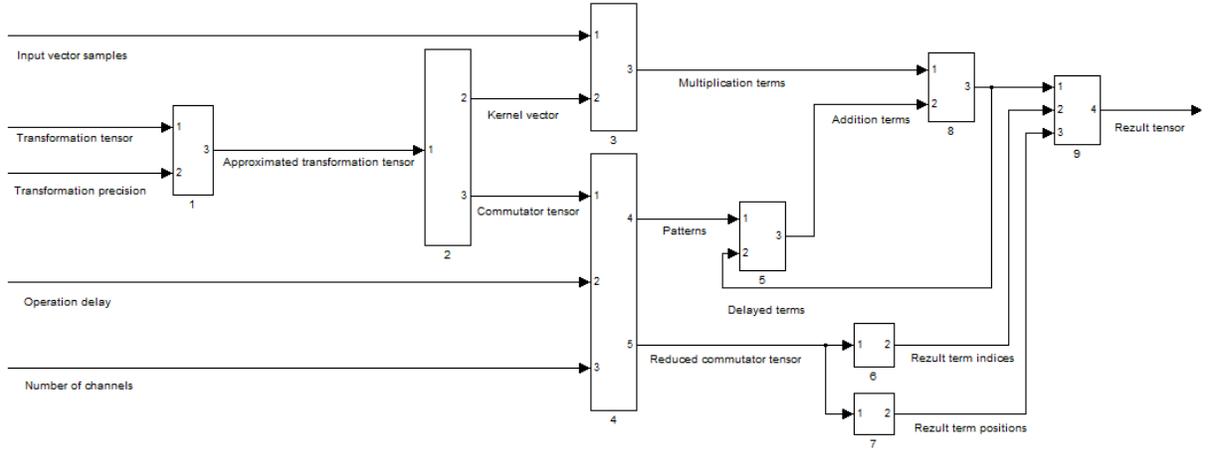

*Figure 2. Internal architecture of a parametrically configured system 1' for the iterative multiplication of a vector by a tensor.*

The blocks described above are connected in the following way. Input 1 of precision converter 1 is the system 1' input 2. It contains the transformation tensor $[\widetilde{T}]_{N_1,N_2,...,N_m,...,N_M}$ (141). Input 2 of precision converter 1 is the system 1' input 3. It contains current value of the rounding precision $\varepsilon$ (142). Input 3 of precision converter 1 contains the rounded tensor $[T]_{N_1,N_2,...,N_m,...,N_M}$ (142) and is connected to input 1 of factorizing unit 2. Output 2 of factorizing unit 2 contains the entirety of the obtained kernel vector $[U]_L$ (148) and is connected to input 2 of multiplier set 3. Output 3 of factorizing unit 2 contains the entirety of the obtained commutator in compact form $[Y]_{N_1,N_2,...,N_m,...,N_M}$ (157) and is connected to input 1 of reducer 4. Input 1 of multiplier set 3 is the system 1' input 1. It contains the elements $\chi$ (213) of the input vectors of each channel. Output 3 of multiplier set 3 contains $\varphi_{\mu,\xi}$ (213), the results of the products of the elements of the kernel and the most recently input element $\chi$ (213) of the input vector of one of the channels, and is connected to input 1 of the delay component set 8. Input 2 of reducer 4 is the system 1' input 4. It contains the operational delay $\delta$ (196). Input 3 of reducer 4 is the system 1' input 5. It contains the number of channels $\sigma$ (209). Output 4 of reducer 4 contains the entirety of the obtained matrix of combinations $[Q]_{p_1-L,5}$ (209) and is connected to input 1 of summator set 5. Output 5 of reducer 4 contains the tensor representing the reduced commutator $[Y]_{N_1,N_2,...,N_m,...,N_M}$ (208) and is connected to input 1 of indexer 6 and to input 1 of positioner 7. Output 3 of summator set 5 contains the new values of the sums of the combinations $\varphi_{\mu+\omega_{1,1}-1,\xi}$ (215) and is connected to input 2 of delay component set 8. Output 2 of indexer 6 contains the indices of the sums of the indices $[R]_{N_1,N_2,...,N_m,...,N_{M-1}}$ (208) of the sums of the combinations comprising the resultant tensor $[P]_{N_1,N_2,...,N_m,...,N_{M-1}}$ (217), and is connected to input 2 of result extractor 9. Output 2 of positioner 7 contains the positions $[D]_{N_1,N_2,...,N_m,...,N_{M-1}}$ (207) of the sums of the combinations comprising the resultant tensor $[P]_{N_1,N_2,...,N_m,...,N_{M-1}}$ (217) and is connected to input 3 of result extractor 9. Output 3 of delay component set 8 contains all the relevant values $\varphi_{\mu,\xi}$ (213), calculated previously as the products of the elements of the kernel by the elements $\chi$ (213) of the input vectors and the sums of the combinations $\varphi_{\mu+\omega_{1,1}-1,\xi}$ (215). This output is connected to input 2 of summator set 5 and to input 1 of result extractor 9. Output 4 of result extractor 9 is a system 1' output 6. It contains the resultant tensor $[P]_{N_1,N_2,...,N_m,...,N_{M-1}}$ (217).

The internal architecture of reducer 4 is presented in Figure 3 and consists of the following blocks:

1 - Pattern set builder 10;

2- Delay adjuster 11;

3 - Channel number adjuster 12.

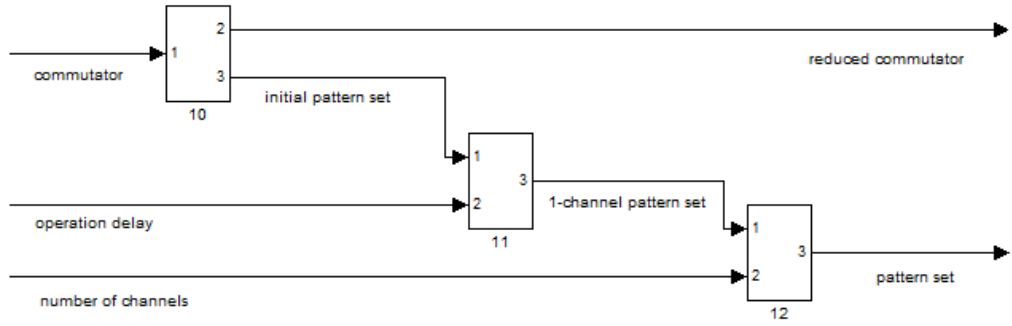

Figure 3. Internal architecture of reducer 4.

The blocks of the reducer 4 are connected in the following way. Input 1 of pattern set builder 10 is reducer 4 input 1. It contains the entirety of the obtained commutator in compact form $[Y]_{N_1,N_2,\ldots,N_m,\ldots,N_M}$ (157). Output 2 of pattern set builder 10 is reducer 4 output 5. It contains the tensor representing the reduced commutator $[Y]_{N_1,N_2,\ldots,N_m,\ldots,N_M}$ (208). Output 3 of pattern set builder 10 contains the entirety of the obtained preliminary matrix of combinations $[Q]_{p_1-L,4}$ (209) and is connected to input 1 of delay adjuster 11. Input 2 of delay adjuster 11 is reducer 4 input 2. It contains current value of the operational delay $\delta$ (196). Output 3 of delay adjuster 11 contains delay adjusted matrix of combinations $[Q]_{p_1-L,5}$ (209) and is connected to input 1 of channel number adjuster 12. Input 2 of channel number adjuster 12 is reducer 4 input 3. It contains current value of the number of channels $\sigma$ (209). Output 3 of channel number adjuster 12 is reducer 4 output 4. It contains channel number adjusted matrix of combinations $[Q]_{p_1-L,5}$ (209).

Functional algorithmic block-diagrams of precision converter 1, factorizing unit 2, multiplier set 3, summator set 5, indexer 6, positioner 7, delay component set 8, result extractor 9, pattern set builder 10, delay adjuster 11, and channel number adjuster 12 are present in Figures 4-14.

Figure 4. Block-diagram of the precision converter 1 functional algorithm.

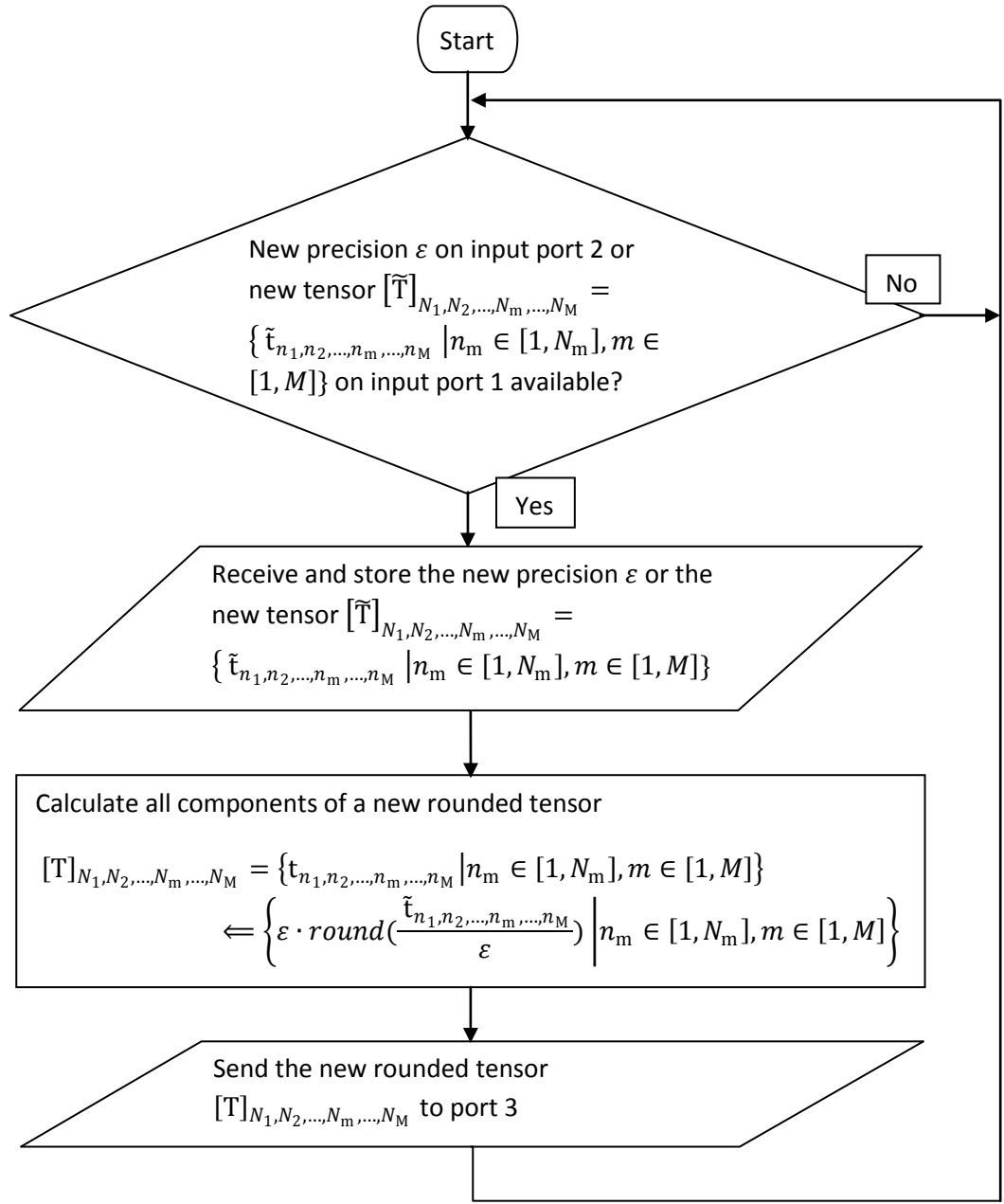

Figure 5. Block-diagram of the factorizing unit 2 functional algorithm.

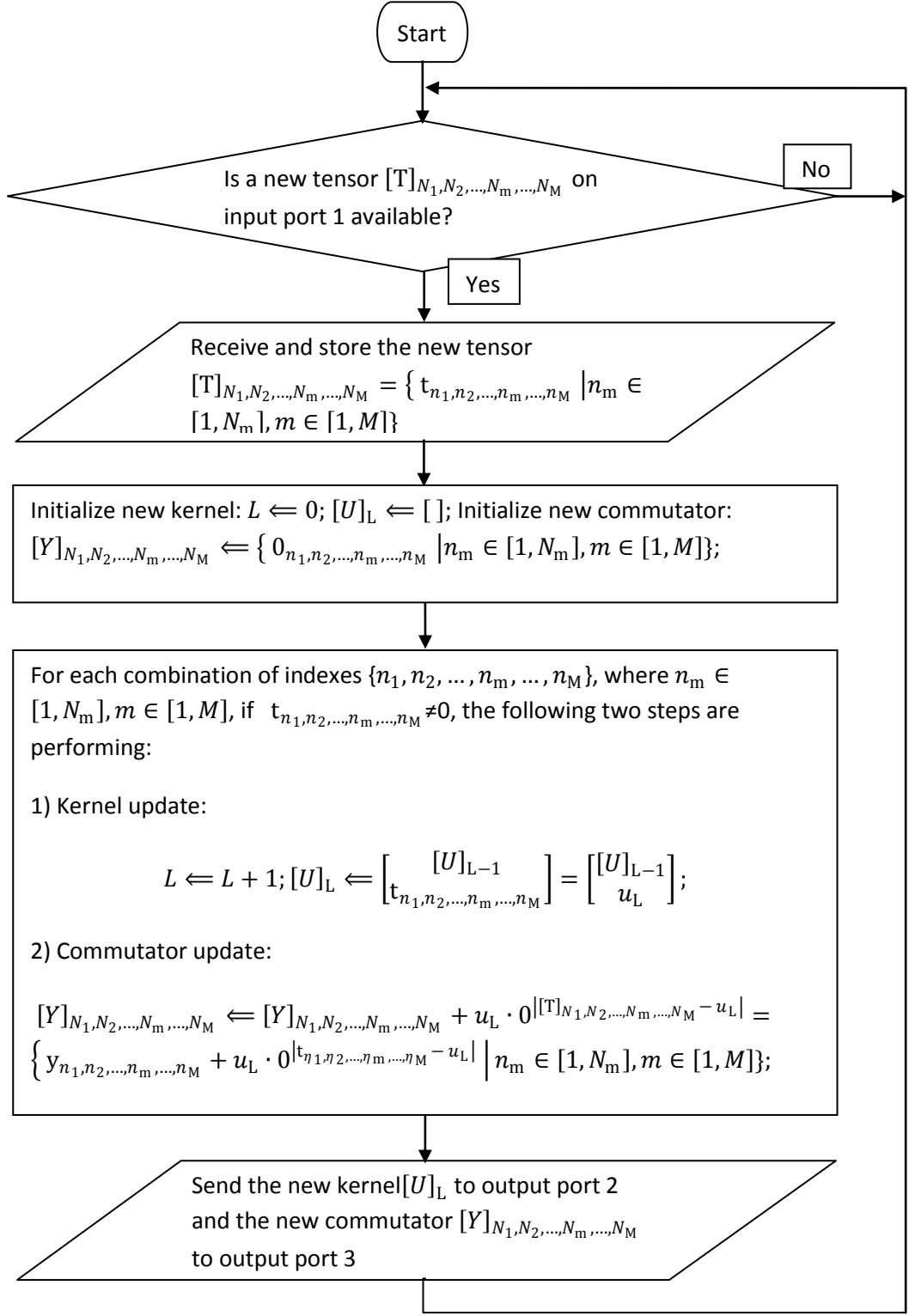

Figure 6. Functional block-diagram of multiplier set 3.

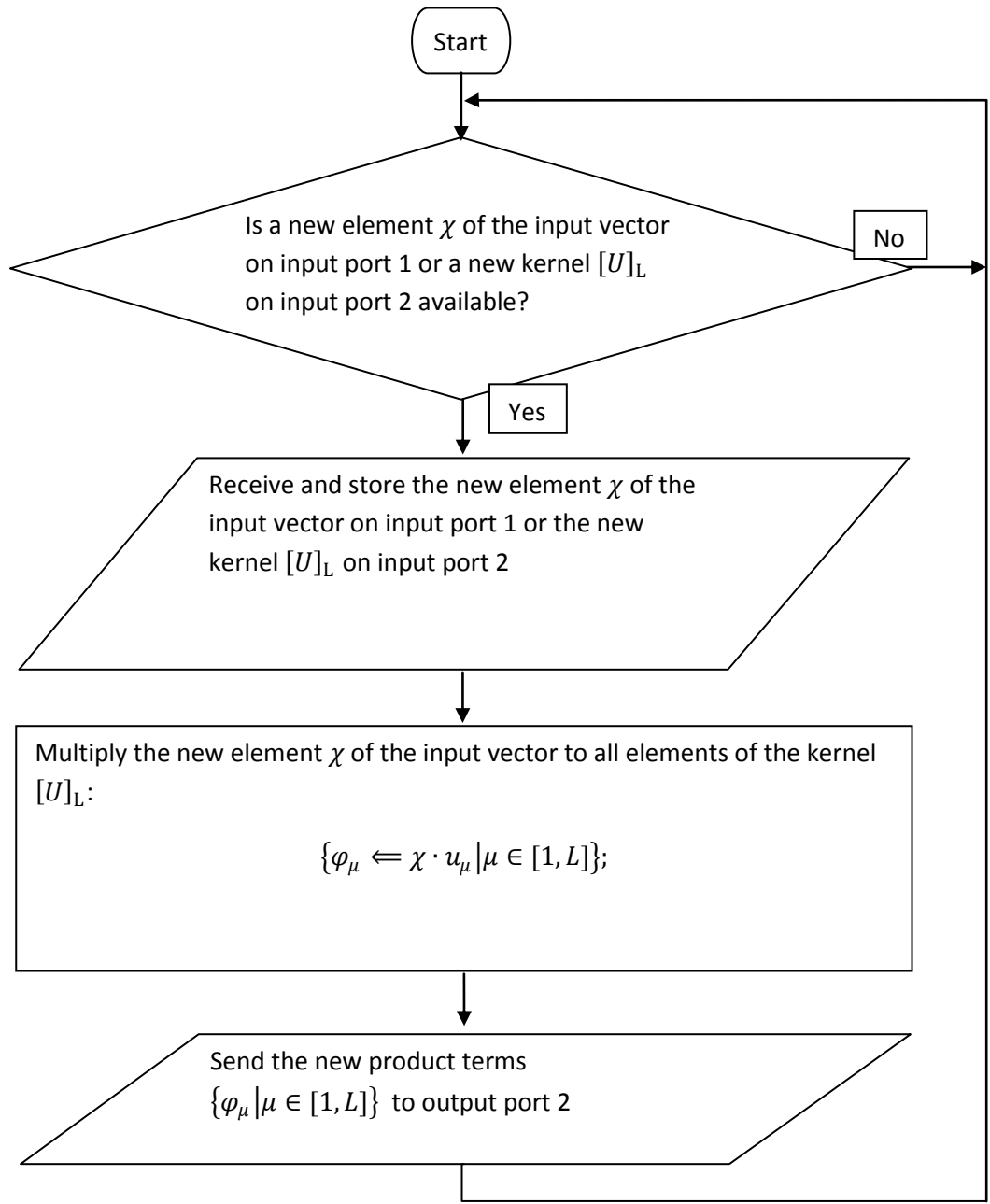

Figure 7. Functional block-diagram of summator set 5.

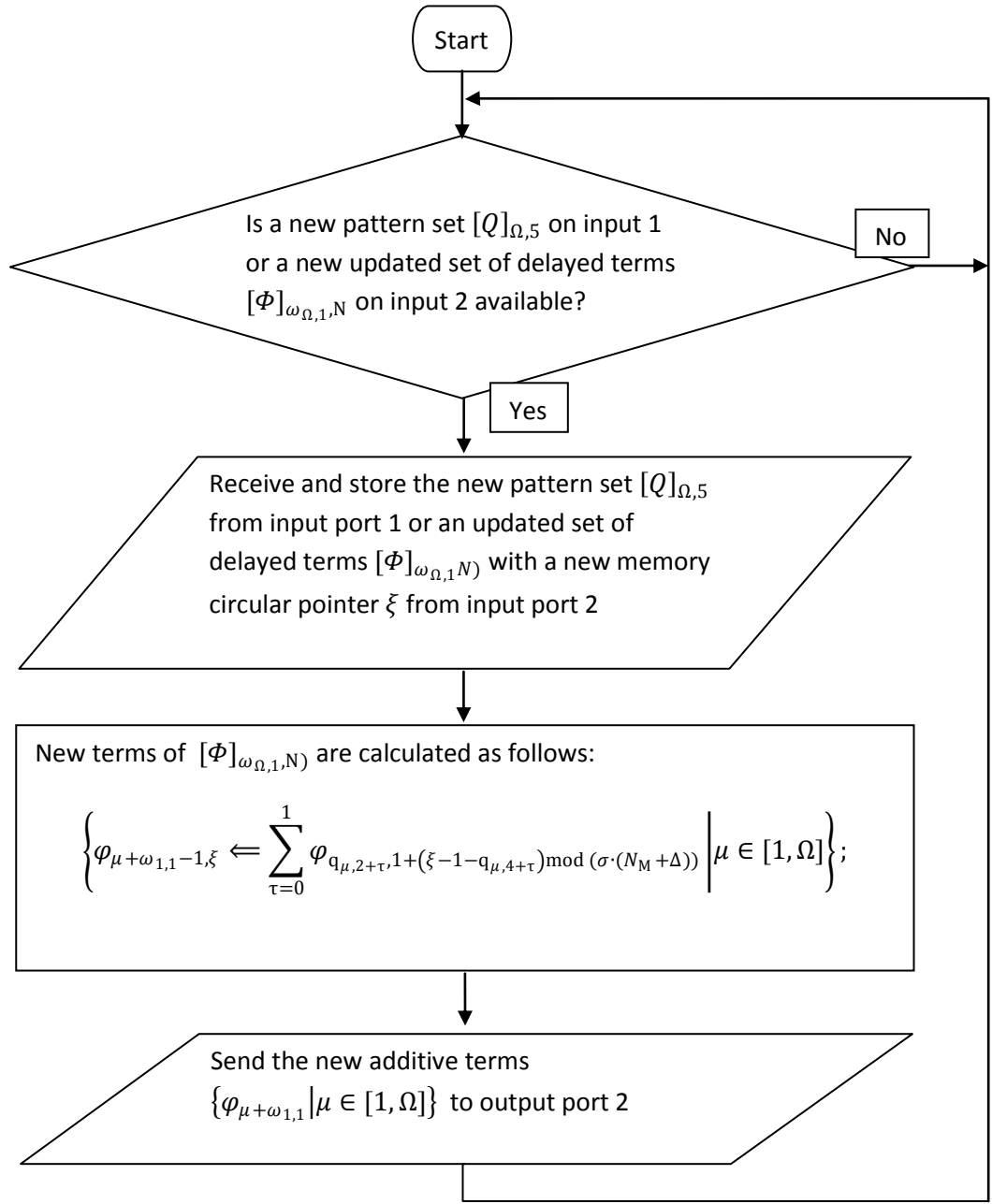

Figure 8. Functional block-diagram of indexer 6.

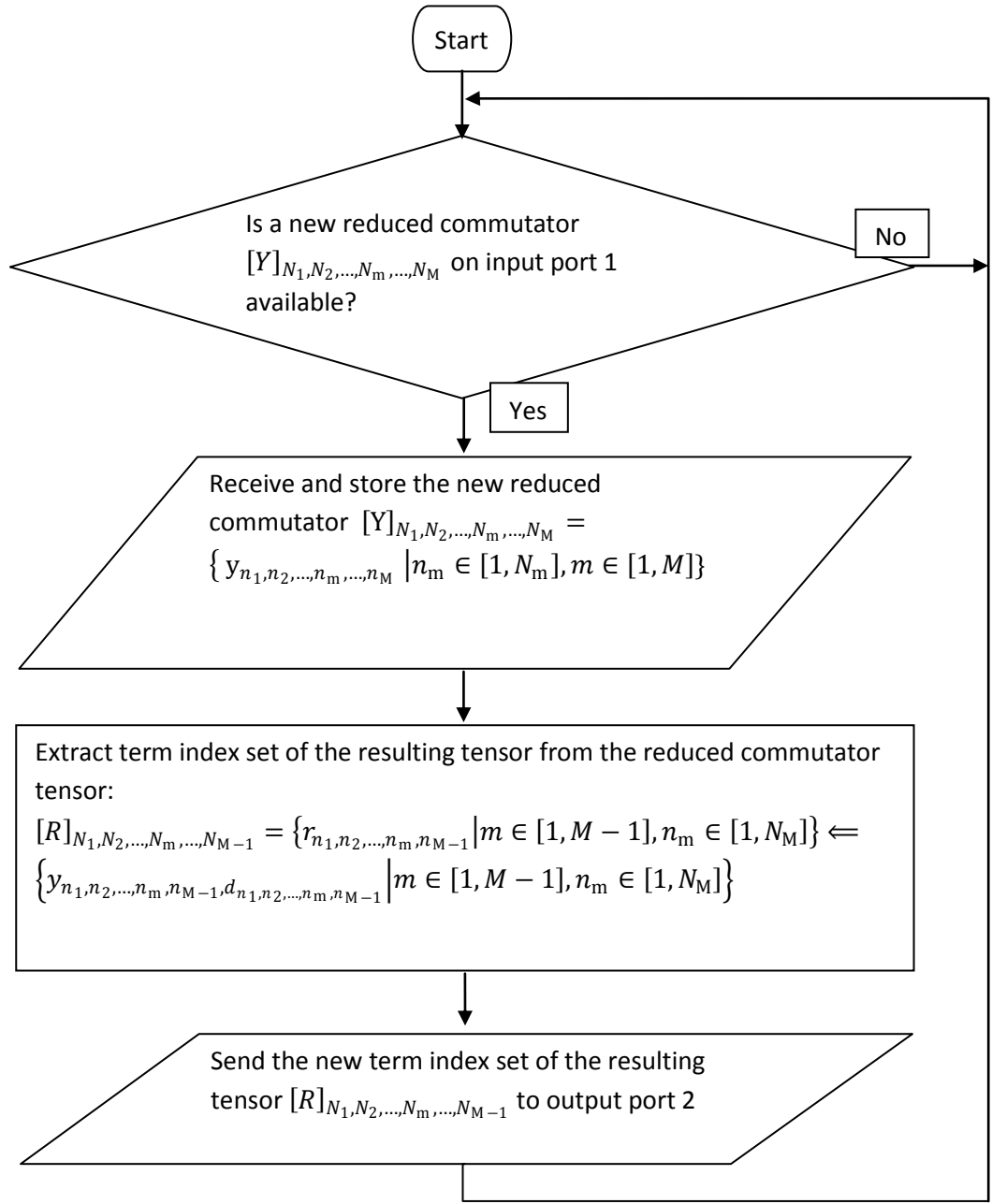

Figure 9. Functional block-diagram of positioner 7.

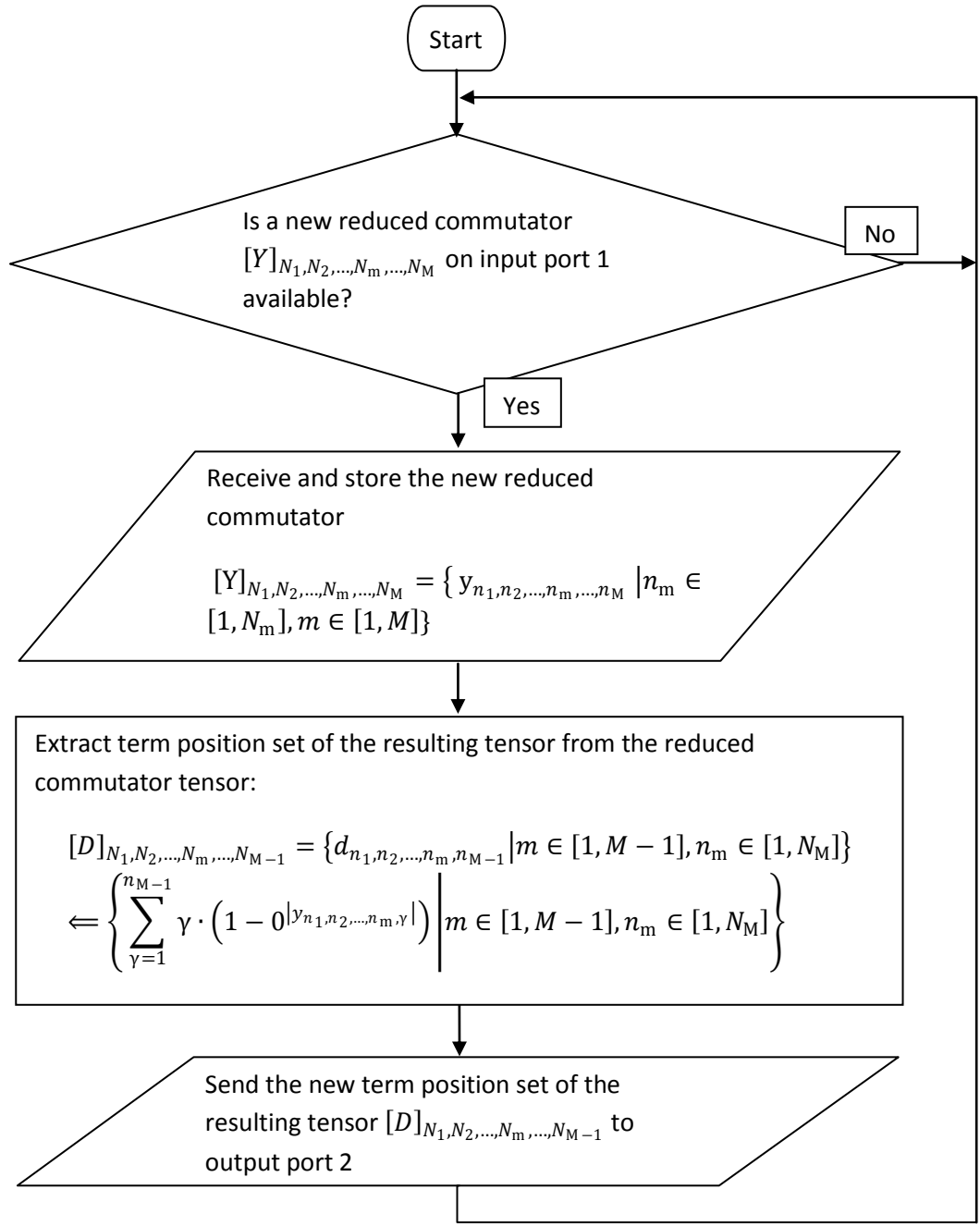

Figure 10. Functional block-diagram of delay set 8.

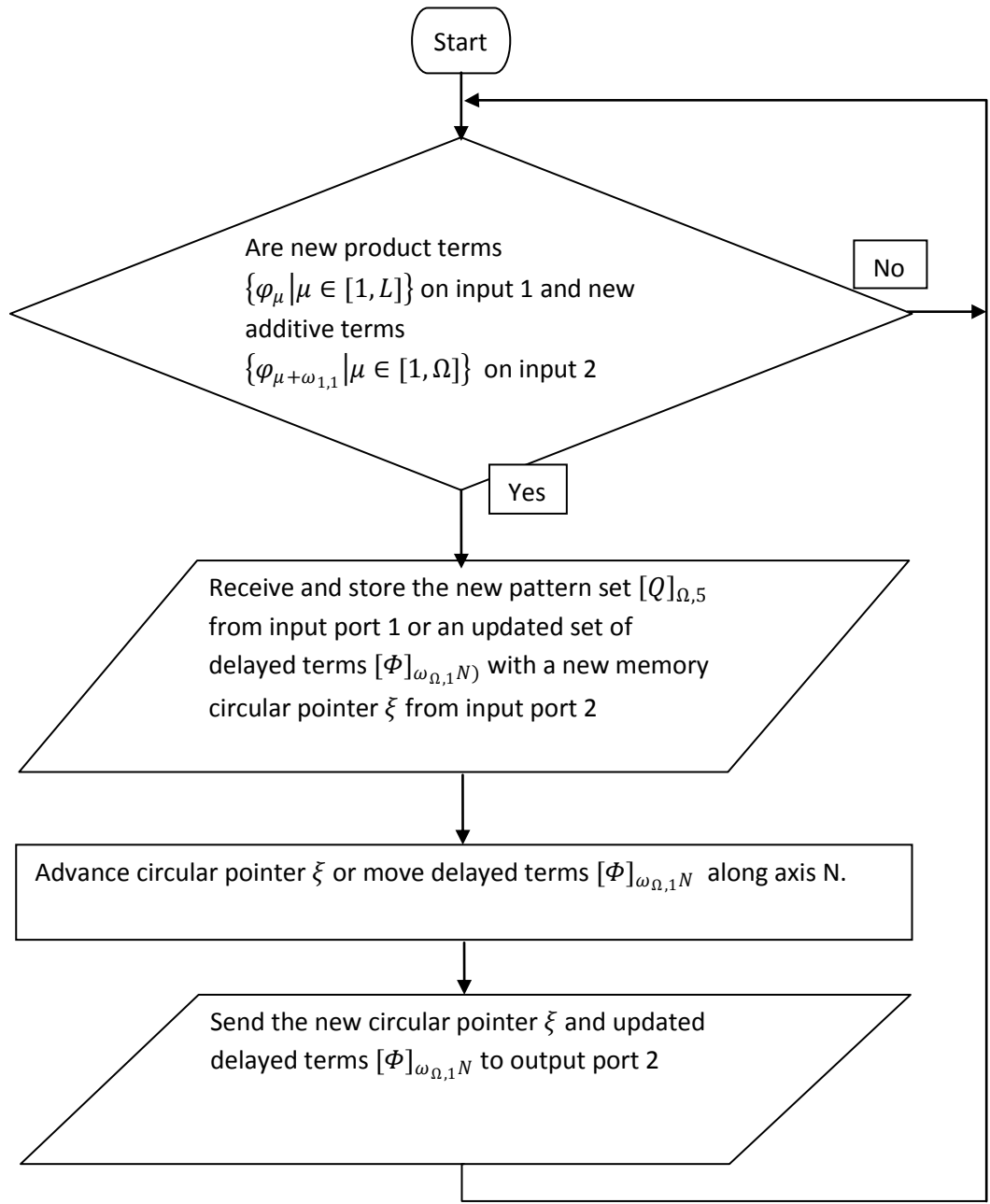

Figure 11. Functional block-diagram of result extractor 9.

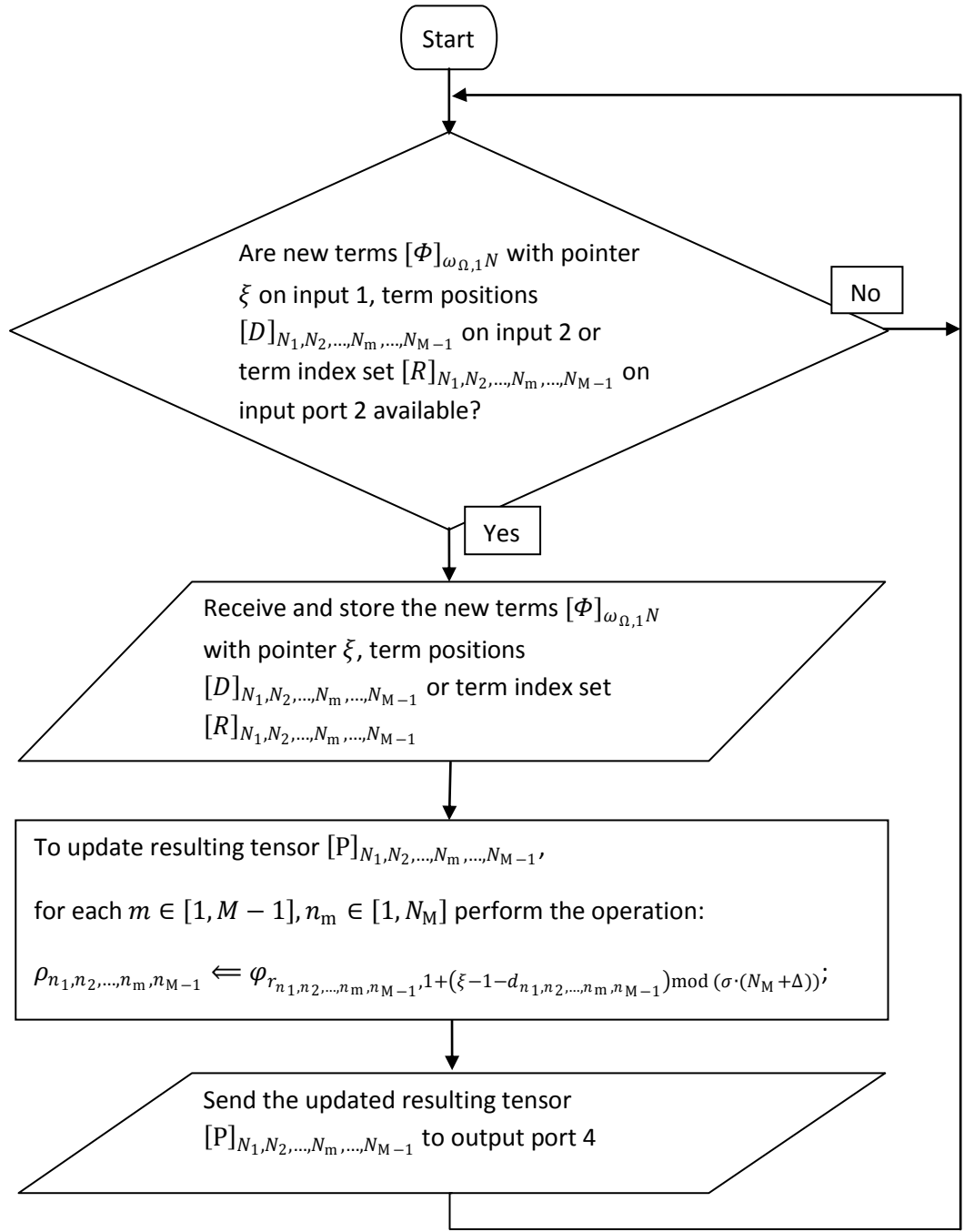

Figure 12. Functional block-diagram of pattern set builder 10 of reducer 4.

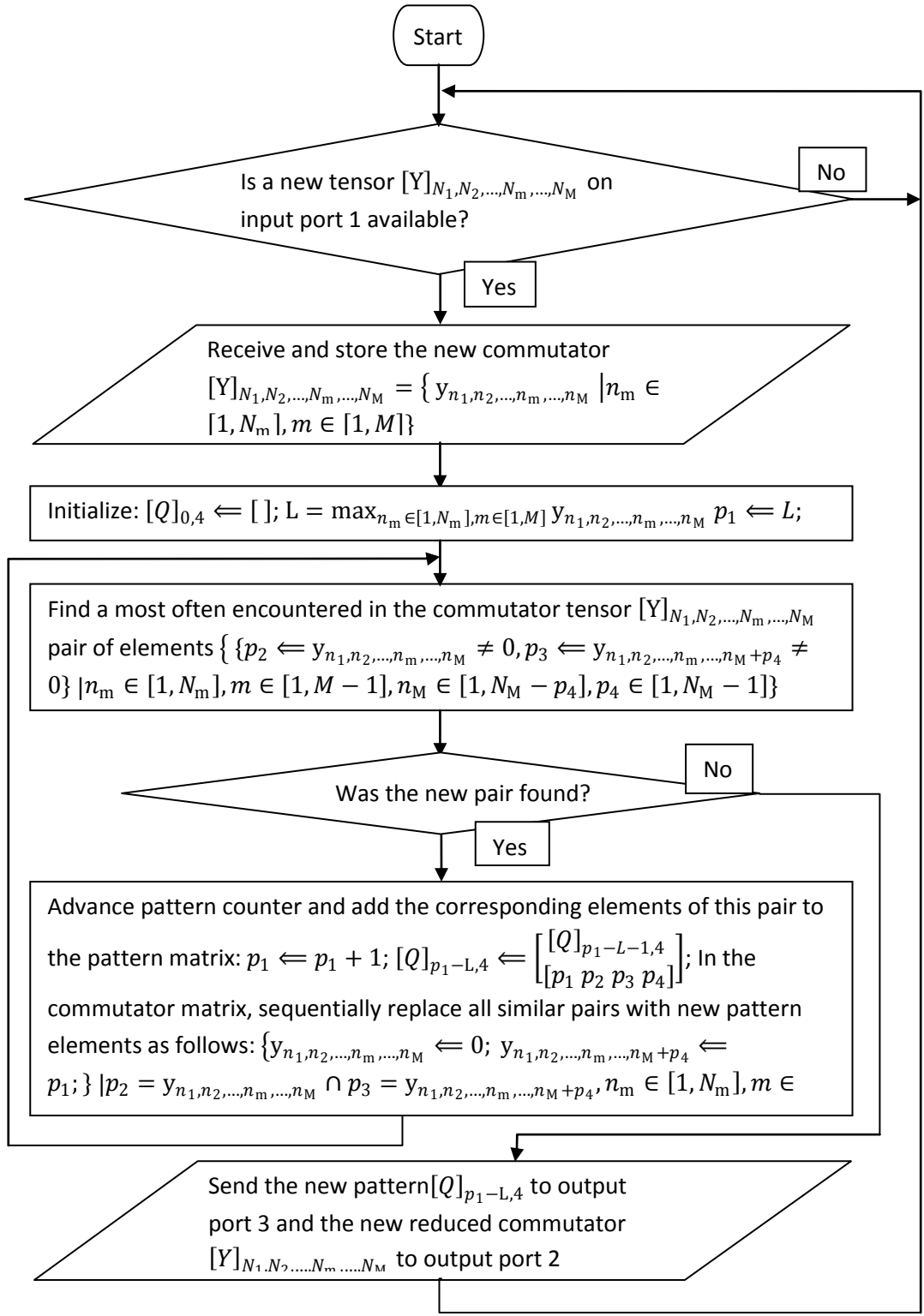

Figure 13. Functional block-diagram of a delay adjuster 11 of a reducer 4.

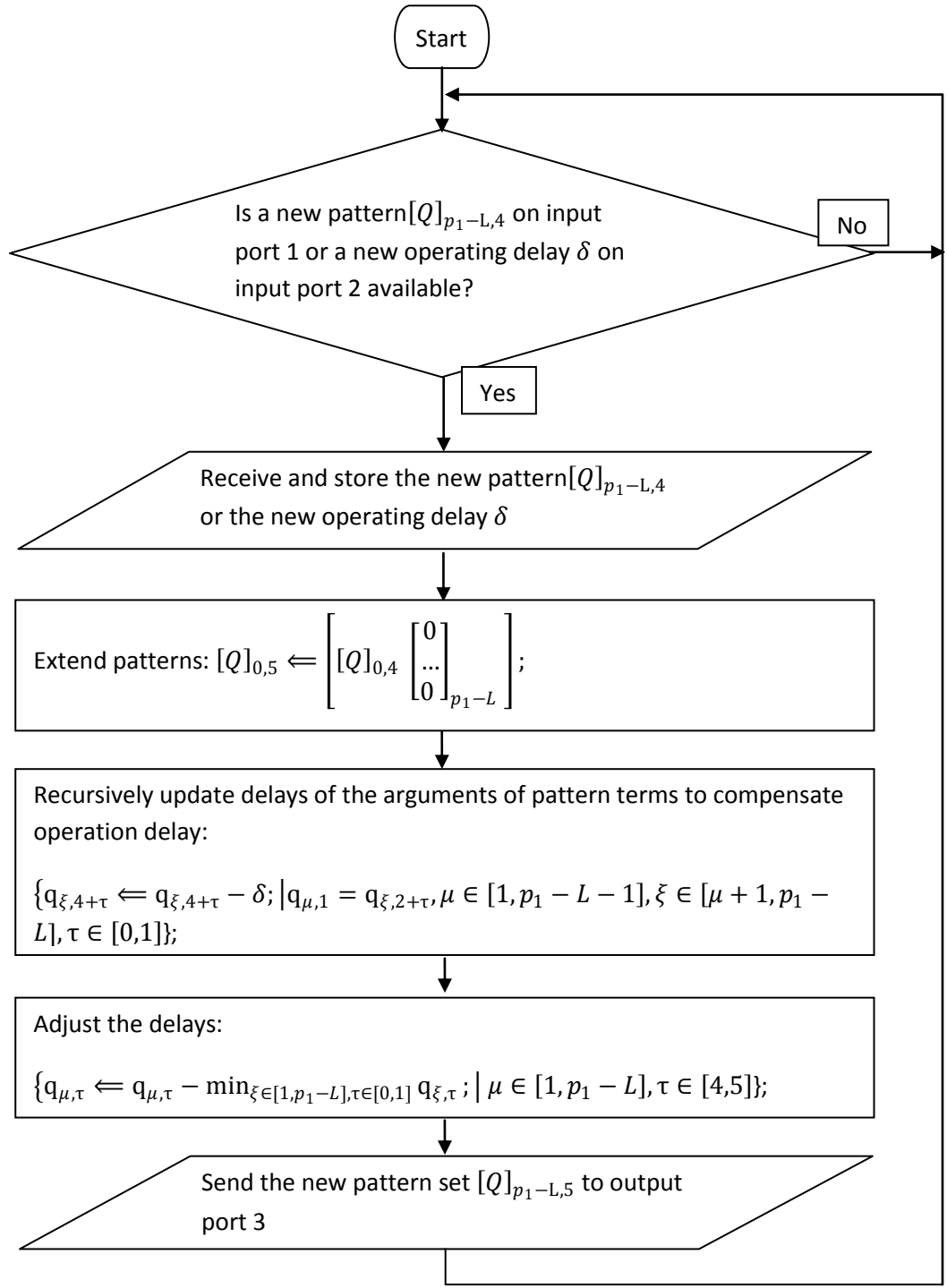

Figure 14. Functional block-diagram of channel number adjuster 12 of reducer 4.

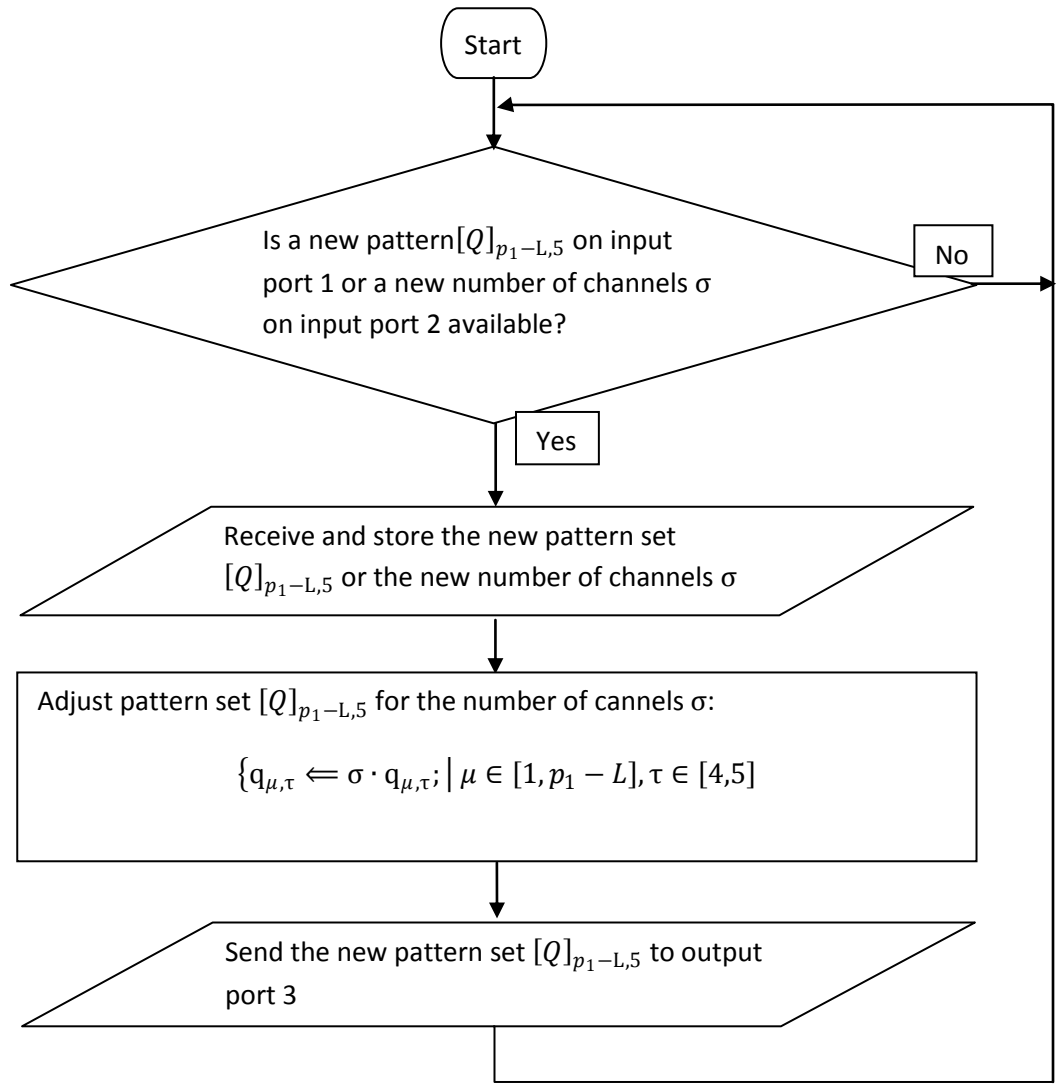

## 29. Selected applications.

The theory described herein is anticipated to be useful in the creation of a wide variety of systems in experimental physics, radio electronics, acoustics, control systems, encryption and data compression. It is difficult to give a full list of scientific fields that have some application for vector, matrix, or tensor operations.

For the sake of illustration, one of the possible applications of the developed theory to the information transmission through communication channels is presented.

The transfer of a series of information bits, or binary values, without error-correcting encoding uses information symbols. Each symbol contains

$$n_s$$

binary values, or bits. The set of all unique symbols constitutes an alphabet of length

$$N_s = 2^{n_s} \tag{233}$$

The piecewise-constant function of time formed by the succession of statistically independent symbols contains first-order discontinuities and, therefore, has an infinitely broad Fourier frequency spectrum, which makes it impossible to transfer via a physical channel. To eliminate this difficulty, the sequence is symmetrized and smoothed by a filter that limits its bandwidth and, inevitably, introduces a significant correlation between its elements, depending chiefly on the time difference between the elements and decreasing as this time difference increases. Such a smoothed function constitutes a so-called video signal, which modulates one or several parameters of the so-called carrier signal. For example, a harmonic carrier may have modulation of amplitude, phase, or frequency: that is, any one of its parameters amplitude, phase, or frequency may be a function of the video signal. Additionally, the modulated signal may be subject to some distortion in amplifiers or other points in its transmission. Such a modulated signal is transmitted on a connection channel where many other factors may exist that add noise or distortion to the transmitted signal.

The signal may be represented as a sequence of vectors

$$[V]_N = \begin{bmatrix} v_1 \\ \dots \\ v_n \\ \dots \\ v_N \end{bmatrix} \tag{234}$$

each of which is formed from the preceding vector by a linear shift of all its elements one position up. Each element $v_k$ of this vector is a succeeding complex sample of the received signal with a carrier frequency of 0. At each successive iteration the lowest position of this vector is filled with a new element and element in the highest position is lost. For the demodulation of such a signal it is most efficient to use a set of matched filters. As the length of the observed signal increases, so does the probability of correctly identifying a received symbol as equal to the central symbol of the sequence corresponding to that matched filter whose response power is maximal. Each such filter is matched to a signal formed by one of the possible sequences of length

$$A_l \geq 1 \tag{235}$$

and is the complex conjugate of the corresponding signal. The periodic variation in the modulation parameters, such as the degree of amplitude modulation, the index of phase modulation, the combination of these parameters or any other parameters, leads to the fact that each unique sequence of symbols may require

$$A_m \geq 1 \tag{236}$$

distinct matched filters.

Thus the total number of matched filters is

$$M \leq A_m \cdot N_s^{A_l} = A_m \cdot 2^{n_s \cdot A_l} \tag{237}$$

The received signal is represented by discrete samples where the sampling rate is

$$A_s \geq 1 \tag{238}$$

samples per symbol. Then the formation of matched filters of length $A_l$ symbols requires vectors containing

$$N = A_l \cdot A_s \tag{239}$$

samples.

All matched filters may be represented as rows of the matrix $[\tilde{T}]_{M,N}$ of dimension $(M, N)$. To minimize the number of distinct elements of the matrix $[\tilde{T}]_{M,N}$ each of its rows may be normalized by one of its elements, for example by complex division of each of its elements by the chosen element.

$$[T]_{M,N} = \begin{bmatrix} t_{1,1} & \cdots & t_{1,N} \\ \vdots & t_{m,n} & \vdots \\ t_{M,1} & \cdots & t_{M,N} \end{bmatrix} = \begin{bmatrix} \frac{\tilde{t}_{1,1}}{\tilde{t}_{1,n}} & \cdots & \frac{\tilde{t}_{1,N}}{\tilde{t}_{1,n}} \\ \vdots & \frac{\tilde{t}_{m,n}}{\tilde{t}_{m,n}} & \vdots \\ \frac{\tilde{t}_{M,1}}{\tilde{t}_{M,n}} & \cdots & \frac{\tilde{t}_{M,N}}{\tilde{t}_{M,n}} \end{bmatrix} = \begin{bmatrix} \frac{\tilde{t}_{1,1}}{\tilde{t}_{1,n}} & \cdots & \frac{\tilde{t}_{1,N}}{\tilde{t}_{1,n}} \\ \vdots & 1 & \vdots \\ \frac{\tilde{t}_{M,1}}{\tilde{t}_{M,n}} & \cdots & \frac{\tilde{t}_{M,N}}{\tilde{t}_{M,n}} \end{bmatrix} \tag{240}$$

It is likewise possible to round the elements of the matrix to a given precision, which may diminish the number of distinct elements as a result of the permitted decrease in precision of the matrix multiplication operation.

The operation of optimal filtering is equivalent to the multiplication of the matrix $[T]_{M,N}$ by the vector $[V]_N$ of length $N$, yielding a vector of length $M$ containing the responses of all $M$ matched filters. This multiplication may be performed without intermediate steps or recursively or iteratively with preliminary factorization of the matrix $[T]_{M,N}$ as the product of a commutator and a kernel.

The result of optimal filtering is a vector of length $M$

$$[R]_M = [T]_{M,N} \cdot [V]_N \tag{241}$$

This vector consists of
$$N_s = 2^{n_s} \tag{242}$$

groups of elements that can form the vectors $[R_n]_K$, each containing

$$K = \frac{M}{N_s} = \frac{A_m \cdot N_s^{A_l}}{N_s} = A_m \cdot N_s^{A_l - 1} = A_m \cdot 2^{n_s \cdot (A_l - 1)} \tag{243}$$

elements. Each element of such a vector $[R_n]_K$ is the result of the multiplication of the received signal by those rows of the matrix $[T]_{M,N}$, that correspond to only one specific alphabet symbol. Thus the number $N_s$ of such groups or vectors is equal to the length of the alphabet. The position of the element of the vector $[R]_M$ with maximal absolute value is selected.

$$\mathrm{argmax}\bigl(\mathrm{abs}([R]_M)\bigr) = \mathrm{argmax}\bigl(\mathrm{abs}(\{r_n | n \in [1, M]\})\bigr) \tag{244}$$

The transmitted symbol is identified by the number of that element. For example, the matrix $[T]_{M,N}$ may be constructed so that a row corresponding to a symbol with smaller number may not be located lower than another row corresponding to another symbol with a higher number. In this case the symbol number is defined by

$$m = 1 + ceil\left(\frac{\text{argmax}(\text{abs}(\{r_n | n \in [1,M]\}))-1}{K}\right) = 1 + ceil\left(\frac{\text{argmax}(\text{abs}(\{r_n | n \in [1,M]\}))-1}{A_m \cdot 2^{n_S \cdot (A_l - 1)}}\right) \quad (245)$$

An alternative and equivalent approach is, rather than multiplying the signal by the matrix $[T]_{M,N}$ and subsequently extracting the vectors $[R_n]_K$ from the result of multiplication $[R]_M$, to synthesize, factor, and use the $K$ matrices $[T_n]_{K,N}$, each of which contains all those rows of the matrix $[T]_{M,N}$ that correspond only to one specific symbol of the alphabet. Then the process of optimal filtering is equivalent to the multiplication of each such matrix by the signal vector:

$$\{[R_n]_K = [T_n]_{K,N} \cdot [V]_N \, | n \in [1, N_s]\} \quad (246)$$

From each resulting vector $[R_n]_K$ the element with the largest magnitude is selected:

$$\left\{r_{max_n} = \max(\text{abs}([R_n]_K)) = \max\left(\text{abs}(\{r_{n_k} | k \in [1, K]\})\right) = \max\left(\text{abs}(\{r_{n_k} | k \in [1, K]\})\right) \, | n \in [1, N_s]\right\} \quad (247)$$

The obtained values $r_{max_n}$ are compared and the largest of these indicates the symbol which is to be transmitted.

$$m = argmax\{r_{max_n} | n \in [1, N_s]\} \quad (248)$$

Moreover, information about phase difference in the signal is contained in the argument of the complex correlation of the signal and the corresponding matched filter:

$$\varphi = atan\left(\frac{imag\left(r_{\text{argmax}(\text{abs}(\{r_n | n \in [1,M]\}))}\right)}{real\left(r_{\text{argmax}(\text{abs}(\{r_n | n \in [1,M]\}))}\right)}\right) \quad (248)$$

A measure of the received signal power is obtained from the square of the absolute value of the complex correlation between the signal and the filter matched with it:

$$p = abs\left(r^2_{\text{argmax}(\text{abs}(\{r_n | n \in [1,M]\}))}\right) = \left(real\left(r_{\text{argmax}(\text{abs}(\{r_n | n \in [1,M]\}))}\right)\right)^2 + \left(imag\left(r_{\text{argmax}(\text{abs}(\{r_n | n \in [1,M]\}))}\right)\right)^2 \quad (249)$$

The quality of the received signal as expressed as the signal-to-noise ratio is equal to the ratio of the square of the absolute value of the complex correlation of the signal and the filter matched with it, to the difference between the product of the autocorrelation of the signal and the autocorrelation of the matched filter and the square of this quantity:

$$S/N = \frac{p}{\sqrt{[V]_N^t \cdot [V]_N} \cdot \sqrt{[t_{\text{argmax}(\text{abs}(\{r_n | n \in [1,M]\})),1}, \dots, t_{\text{argmax}(\text{abs}(\{r_n | n \in [1,M]\})),N}]^t \cdot [t_{\text{argmax}(\text{abs}(\{r_n | n \in [1,M]\})),1}, \dots, t_{\text{argmax}(\text{abs}(\{r_n | n \in [1,M]\})),N}] - p}} = \frac{p}{\sqrt{(\sum_{n=1}^N v_n^2) \cdot (\sum_{n=1}^N t^2_{\text{argmax}(\text{abs}(\{r_n | n \in [1,M]\})),n})} - p} \quad (250)$$

The application of the method to fast numerical solutions of partial differential equations is described in [5].